\documentclass{article}

\usepackage{arxiv}

\usepackage[utf8]{inputenc} 
\usepackage[T1]{fontenc}    
\usepackage{hyperref}       
\usepackage{url}            
\usepackage{booktabs}       
\usepackage{amsfonts}       
\usepackage{nicefrac}       
\usepackage{microtype}      
\usepackage{lipsum}		
\usepackage{graphicx}
\usepackage[square,sort,comma,numbers]{natbib}
\usepackage{doi}
\usepackage{amsmath}
\usepackage{bm}
\title{Kinematic fitting of Neutral current events in deep inelastic \emph{ep} collisions.}

\author{ \href{https://orcid.org/0000-0003-2755-5682}{\includegraphics[scale=0.06]{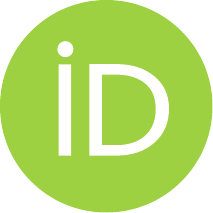}\hspace{1mm}Ritu Aggarwal}\thanks{This research work was carried out under the INSPIRE Faculty research grant awarded to the author by the Department of Science and Technology, Government of India.} \\
	Department of Technology\\
	Savitribai Phule Pune University\\
	India 411007 \\
	\texttt{ritu.aggarwal1@gmail.com} \\
	\And
	\href{https://orcid.org/0000-0003-0244-5129}{\includegraphics[scale=0.06]{orcid.pdf}\hspace{1mm}Allen Caldwell} \\
	Max Plank Institute for Physics\\
	Munich\\
	Germany, 80805 \\
	\texttt{caldwell@mpp.mpg.de} \\
}

\renewcommand{\shorttitle}{\textit{arXiv} Template}

\hypersetup{
pdftitle={A template for the arxiv style},
pdfsubject={q-bio.NC, q-bio.QM},
pdfauthor={David S.~Hippocampus, Elias D.~Striatum},
pdfkeywords={First keyword, Second keyword, More},
}

\begin{document}
\maketitle

\begin{abstract}

	In this paper we present a technique to reconstruct the scaling variables defining $ep$ deep inelastic scattering by performing a kinematic fit. This reconstruction technique makes use of the full potential of the data collected.  It is based on Bayes’ Theorem and involves the use of informative priors.  The kinematic fit method has been tested using a simulated sample of $ep$  neutral current events at a center of mass energy of 318~GeV with $Q^2$ > 400 GeV$^2$. In addition to the scaling variables, this method is able to estimate the energy of possible initial state radiation
(E$_\gamma$ ) which otherwise goes undetected.  A better
resolution than standard electron and double angle techniques in the reconstruction of scaling variables
is achieved using a kinematic fit. 
\end{abstract}

\keywords{Kinematic fit \and Deep inelastic scattering \and Resolution of scaling variables}

\section{Introduction}
Deep inelastic scattering (DIS) of leptons on  hadrons is one of the fundamental experimental methods to probe the internal structure of hadrons. A precise knowledge of the structure of hadrons is important in the quest to uncover phenomena beyond the Standard Model of particle Physics in various high energy collider experiments that are running and also those which are planned in the future.

The next generation of lepton-hadron/ion colliders~\cite{ref:LHeC}-~\cite{ref:Awake} will extend the study of hadronic matter at higher energies and higher luminosities. While preparing for the next generation of updated high energy colliders, it is appropriate to study the analysis methods which can harness the full potential of the future colliders.
A DIS event can be categorised with the Lorentz invariants  $Q^2$, $x$ and $y$~\cite{ref:DIS}.  $Q^2$ is the negative of the square of the four momentum transferred, Bjorken-$x$ is interpreted as the momentum fraction of the proton taken by the struck quark in the Breit frame and $y$ is interpreted as the fraction of energy transferred from the electron to the proton in the frame where the proton is at rest. 

\begin{figure}
\centering
\includegraphics[scale=.3]{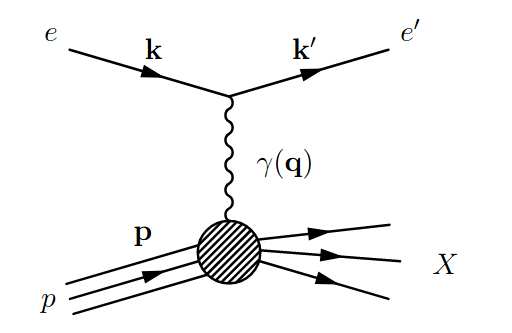}
\caption{Deep Inelastic Scattering of an electron on a proton in a Neutral Current Channel.}
\label{fig:Dis}

\end{figure}

In this paper we present a kinematic  fit technique to reconstruct the event kinematics using Bayesian inference methods~\cite{ref:bayes1}. The leptonic and hadronic information of the final state from the detector is used to reconstruct the kinematic variables, and, as we show, provides a better resolution compared to conventional methods. The algorithm for performing a kinematic fit is discussed in Section 4 in detail. In addition to the extraction of the kinematic variables with improved resolution, this method can be used to infer if Initial State Radiation (ISR, E$_\gamma$) is present in the interaction. A further advantage of the kinematic fit method using Bayesian Analysis is that it provides the uncertainty on the reconstructed kinematic variables.

\section{Reconstruction of kinematic variables}
Figure~\ref{fig:Dis} shows the Feynman diagram representing neutral current (NC) DIS of electrons on protons mediated by the exchange of a virtual photon.
The three kinematic scaling variables describing the interaction are defined in terms of the incoming proton 4-momentum, ${\bf p}$ and the incoming and scattered lepton 4-momenta ${\bf k, k'}$ as
	
\begin{equation}
\label{eq:q2}
Q^{2} = -(\bf{q} \cdot \bf{q}) 
\end{equation}
where, $\bf{q} = (\bf{k}-\bf{k}^{'})$\begin{equation}
\label{eq:x}
x=Q^{2}/2(\bf{p} \cdot \bf{q})
\end{equation}
and
\begin{equation}
\label{eq:y}
y=(\bf{p} \cdot \bf{q})/(\bf{p} \cdot \bf{k}) \; .
\end{equation}
$Q^2$ is the virtuality of the exchanged boson and gives the scale of the interaction, with $Q=1$~GeV corresponding to a transverse distance scale of approximately $0.2$~fm.
In a frame where the proton has very large momentum
the Bjorken-$x$ variable has an intuitive interpretation  
as the fractional momentum carried by the struck parton in the scattering process.
The scaling variable $y$ gives the energy transferred from the electron to the proton in the frame where proton is at rest. The three kinematic variables are related to the center of mass energy of the interaction,  $s$, as 
	
\begin{equation}
\label{eq:q2_2}
Q^2 \approx sxy.
\end{equation}

\begin{figure}
\centering
\includegraphics[scale=.35]{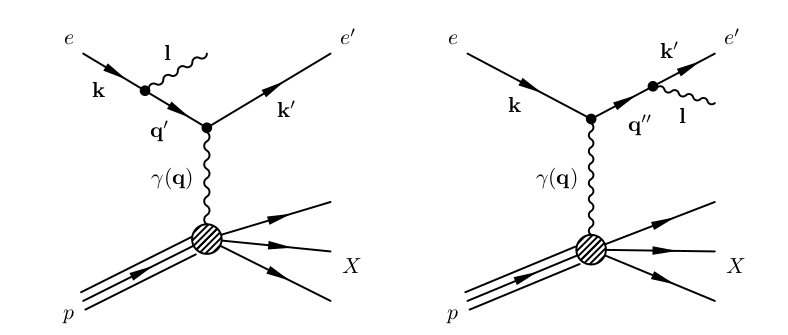}
\caption{ISR (left) and FSR (right) from the incoming and outgoing electrons respectively.}
\label{fig:ISR}

\end{figure}

There are QED processes that can accompany the  NC DIS interaction in the form of ISR and Final State Radiation (FSR)  as shown in Figure ~\ref{fig:ISR}. The ISR typically goes unrecorded in the detector as it is generated at small angles in the direction of the incoming electron. An event with ISR has a smaller value of the incoming electron energy participating in the scattering and hence the center of mass energy is reduced. The reconstruction of kinematic variables in case of unrecorded ISR can have a strong bias. The FSR is, however, typically recorded, and added to the outgoing electron.  
   The information from the detector is processed and provided for further analysis in the form of the following four independent quantities

\begin{itemize}
    \item$E_e$, scattered electron energy 
    \item $\theta_e$, scattered electron angle
    \item $\delta_h =\sum\limits_{i} E_{i}(1-\cos\theta_{i})$, sum over energy deposits $E_i's$ at angle $\theta_i's$ in the `Calorimeter' of the detector, which are not assigned to the scattered lepton 
    \item $P_{T,h} =\sqrt{(\sum\limits_{i}  Px_i)^2 +(\sum\limits_{i} Py_i)^2}$, transverse momentum of the hadronic final state (HFS),\\ where $Px_i$ = $E_i$ $\sin\theta_i$ $ \cos\phi_i$ and  $Py_i$ = $E_i$ $\sin\theta_i$ $\sin\phi_i$. 
\end{itemize}

Using $\delta_h$ to summarize the HFS has an advantage as it removes the contribution from the spectator quarks very elegantly and was first used by Jacquet and Blondel~\cite{ref:JB}. The total $E-P_Z(\sim\delta  = \delta_h +\delta_e$) in the interaction remains conserved as both $E$ and $P_Z$ are conserved. Here, $\delta_e  = E_e (1-\cos\theta_e)$.  If there is no energy loss down the beam pipe in the direction of the electron, then the $\delta$ measured in the detector is very close to 2A, where A is the electron beam energy. Therefore, in the case of no ISR, $\delta_h +\delta_e $ = 2A.

Ideally any two of the the above four quantities are required to calculate the three unknown scaling variables $x$, $y$ and $Q^2$. 
 The kinematic variables at small $x$ and low $Q^2$ can be calculated with a good resolution using the electron energy and angular information in what is called as electron-only method (EL), while at the large $x$ and high $Q^2$, the double angle method (DA) is often used. The use of a kinematic fit for a better reconstruction of the kinematic variables is contemplated as discussed in~\cite{ref:KF1},~\cite{ref:Recon1},~\cite{ref:highx}. The electron and double angle methods are briefly discussed below. A more complete summary of reconstruction methods can be found~\cite{ref:Recon1},~\cite{ref:Recon2}.    
\subsection{Electron-only method}
In the electron method~\cite{ref:el}, the information from the scattered electron in the event final state is used to reconstruct the kinematic variables.  The kinematic variables $x$, $y$ and $Q^2$ are reconstructed from   E$_{e}$ and $\theta_{e}$ as follows:

\begin{equation}
\label{elm}
 Q^{2}_{EL} = 2AE_{e}(1+\cos\theta_{e})=\delta E_{e}(1+\cos\theta_{e})
\end{equation}
\begin{equation}
\label{yel}
 y_{EL} = 1 - \frac{E_{e}}{2A}(1-\cos\theta_{e}) = 1 - \frac{\delta_e}{\delta}
\end{equation}
\begin{equation}
 x_{EL} = \frac{Q^{2}_{EL}}{sy_{EL}} =
\frac{E_{e}\cos^{2}\frac{\theta}{2}}{P(1-\frac{E_{e}}{A}\sin^{2}
\frac{\theta}{2})}
\end{equation}

where, $P$ is the incident proton beam energy.
\subsection{Double Angle method}
\label{sec:DA_meth}
 In the double angle method~\cite{ref:el,ref:DA}, the kinematic variables are reconstructed using $\theta_e$ and the scattered hadron angle, $\gamma_{h}$. 
 $\gamma_{h}$ is calculated  as 

\begin{equation}
\cos\gamma_h = \frac{P_{T,h}^2 - \delta_h^2}{P_{T,h}^2 + \delta_h^2} \ ,
\label{eqn:gamma_h}
\end{equation}
The kinematic variables in the double angle method are then  calculated as follows:
\begin{equation}
 Q^{2}_{DA} =
4A^{2}.\frac{\sin\gamma_{h}(1+\cos\theta_{e})}{\sin\gamma_{h}+\sin\theta_{e
}-\sin(\gamma_{h}+\theta_{e})}
\end{equation}

\begin{equation}
 x_{DA} =
\frac{A}{P}.\frac{\sin\gamma_{h}+\sin\theta_{e}+\sin(\gamma_{h}
+\theta_{e})}{\sin\gamma_{h}+\sin\theta_{e}-\sin(\gamma_{h}+\theta_{e})}
\end{equation}

\begin{equation}
 y_{DA} =
\frac{\sin\theta_{e}(1-\cos\gamma_{h})}{\sin\gamma_{h}+\sin\theta_{e}
-\sin(\gamma_{h}+\theta_{e})}
\end{equation}

One of the advantages of the double angle method is that it is not sensitive to the electron or jet energy calibrations as the kinematic variables are reconstructed using the angular information.  However this method is sensitive to the initial and final state radiations and simulation of the color flow.

\section{Kinematic Fit}
\label{sec:headings}
The goal of the kinematic fit technique (KF) is to infer the kinematic variables along with the possible ISR energy, $E_\gamma$. These quantities here are grouped as $\bm{\lambda}$ = ($ \textit{x},\textit{y},E_{\gamma} $) and  the measured quantities as $\bf{D}$ = $(E_e, \theta_e, \delta_h, P_{T,h})$. Using Bayes theorem~\cite{ref:bayes1}, the probability distribution for the parameter set
$\bf{\lambda}$ given the set of measured quantities $\bf{D}$ can be written as

\begin{equation}
\label{eq:Baysian1}
 P(\boldsymbol{\lambda}|\boldsymbol{D}) \propto P(\boldsymbol{D}|\boldsymbol{\lambda}) P_{o}(\boldsymbol{\lambda}).
\end{equation}

Here $P(\bf{D}|\bm{\lambda})$ is the likelihood function  which gives the probability of making a measurement $\bf{D}$ given true values  $\bm{\lambda}$ and $P_{o}(\bm{\lambda})$ is the prior information on $\bm{\lambda}$.
The $P(\boldsymbol{D}|\bm{\lambda})$ distribution can be written as
\begin{equation}
\label{eq:Baysian2a}
 P(\boldsymbol{D}|\boldsymbol{\lambda})=P(\boldsymbol{D}|x,y,E_\gamma)=P(E_e, \theta_e, \delta_h, P_{T,h}|x,y,E_\gamma) \newline
\end{equation}

\begin{equation}
\label{eq:Baysian2b}
 =P(E_e, \theta_e|x,y,E_\gamma)P(\delta_h, P_{T,h}|x,y,E_\gamma).
\end{equation}
In a first attempt, we further factorize the likelihood as
\begin{equation}
\label{eq:Baysian2c}
 \approx P(E_e|x,y,E_\gamma)P(\theta_e|x,y,E_\gamma)P(\delta_h|x,y,E_\gamma)P( P_{T,h}|x,y,E_\gamma).
\end{equation}

In our KF code, we use
\begin{equation}
\label{eq:Baysian2}
 P(\boldsymbol{D}|\boldsymbol{\lambda})=P(E_e|E_e^{\lambda})P(\theta_e|\theta_e^{\lambda})P(\delta_{h}|\delta_{h}^{\lambda})P(P_{T,h}|P_{T,h}^{\lambda})
\end{equation}
 where the values of $E_e^{\lambda}$, $\theta_e^{\lambda}$, $\delta_h^{\lambda}$ and $P_{T,h}^{\lambda}$ are obtained for a given set $\boldsymbol{\lambda}$.
 
We choose E, F, $\theta$ and $\gamma$ to represent the true values of the generated electron and quark final state energies and scattering angles respectively. These can be calculated from the true\footnote{$Q^2$ is defined by the exchanged Boson, $x =\frac{Q^2}{2\boldsymbol{p} \cdot \boldsymbol{q}}$,  $y = \frac{Q^2}{s^\prime x}$ and $s^\prime=(\mathbf{k}+\mathbf{p}-\mathbf{l})^2$ in the presence of ISR with four momentum $\mathbf{l}$ $\approx$ $(E_\gamma,0,0,-E_\gamma)$} $Q^2$, $x$ and $y$ of the event as

\begin{equation}
\label{eq:true1}
E = xyP + A_r(1-y)                
\end{equation}

\begin{equation}
\label{eq:true2}
F = x(1-y)P + y A_r                
\end{equation}

\begin{equation}
\label{eq:true3}
\cos\theta = \frac {xyP - A_r(1-y)} {xyP + A_r(1-y)}                
\end{equation}

\begin{equation}
\label{eq:true4}
\cos\gamma = \frac {x(1-y)P - yA_r} {x(1-y)P + yA_r}.                
\end{equation}
Here, $$ A_r  = A - E_{\gamma}$$

where, $E_{\gamma}$ is the energy of an ISR photon.

  In case of ISR, the  effective lepton energy participating in the interaction is reduced to $A_r$. 
  For the HFS, the true $\delta_h$ and transverse momentum are given as

\begin{equation}
\label{eq:true5}
\delta_h^{gen} = F(1-\cos\gamma)                
\end{equation}

\begin{equation}
\label{eq:true6}
P_{T,h}^{gen} = F\sin\gamma                
\end{equation}

 Each factor on the right hand side  of Equation~\ref{eq:Baysian2} is initially assumed to be a Gaussian PDF with width defined by the smearing factor applied to incorporate the detector effects (Equations ~\ref{eq:param1e}-~\ref{eq:param2h}).  
  The likelihood can therefore be written as

\begin{equation}
\label{eq:Pos}
P(\boldsymbol{D}|\boldsymbol{\lambda})  \propto   \frac{1}{\sqrt{2\pi}\sigma_E}e^{-\frac{(E_e-E_e^{\lambda})^2}{2\sigma_E^2}}\frac{1}{\sqrt{2\pi}\sigma_\theta}e^{-\frac{(\theta_e-\theta_{e}^{\lambda})^2}{2\sigma_\theta^2}}\frac{1}{\sqrt{2\pi}\sigma_{\delta_{h}}}e^{-\frac{(\delta_h-\delta_{h}^{\lambda})^2}{2\sigma_{\delta_{h}}^2}}\frac{1}{\sqrt{2\pi}\sigma_{P_{T,h}}}e^{-\frac{(P_{T,h}-P_{T,h}^{\lambda})^2}{2\sigma_{P_{T,h}}^2}}. 
\end{equation}

 The prior distribution $P_{o}(\boldsymbol{\lambda})$ used in this analysis   reflects the basic features of the DIS cross section on $x$ and $y$.

\section{Simulated data}
For this analysis, the sample of $10^6$ DIS NC ep events with center of mass energy 318~GeV and $Q^2 >$400 GeV$^2$ was generated using the Rapgap-3.303~\cite{ref:rapgap} Monte Carlo generator interfaced with HERACLES~\cite{ref:Heraceles}, where the latter is used to apply $O(\alpha)$ QED corrections. 
The detector simulation effects are introduced as Gaussian smearing on the true generated quantities E, F, $\delta_h^{gen}$ and $P_{T,h}^{gen}$ .

The electron energy and electron angle resolutions are taken from the ZEUS detector performance as reported in~\cite{ref:aggarwal}.
 
\begin{equation}
\label{eq:param1e}
\sigma_E/E = 19.59\%/ \sqrt{E} \oplus 0.825\%               
\end{equation}
 
\begin{equation}
\label{eq:param2e}
\sigma_\theta/\theta = 0.25\%/ \sqrt{\theta}              
\end{equation}

\begin{figure}
\centering

    \includegraphics[width=0.48\linewidth]{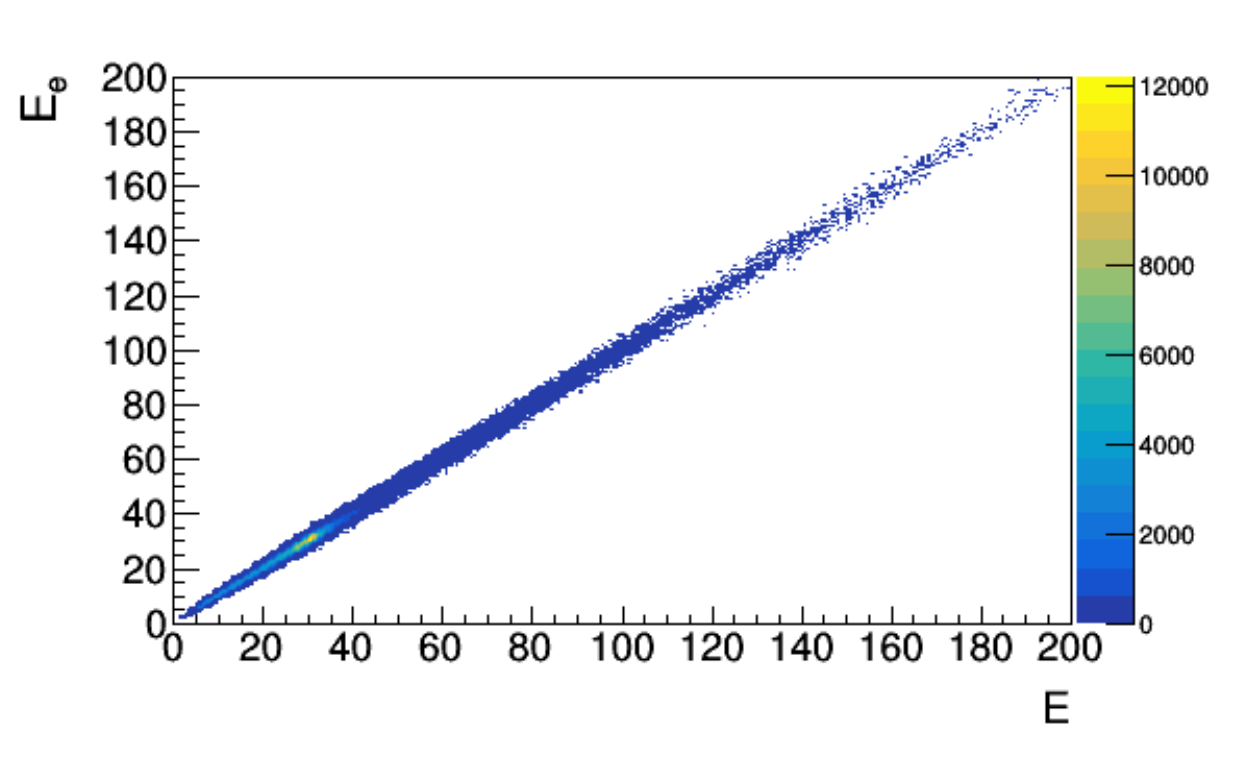}
     \includegraphics[width=0.48\linewidth]{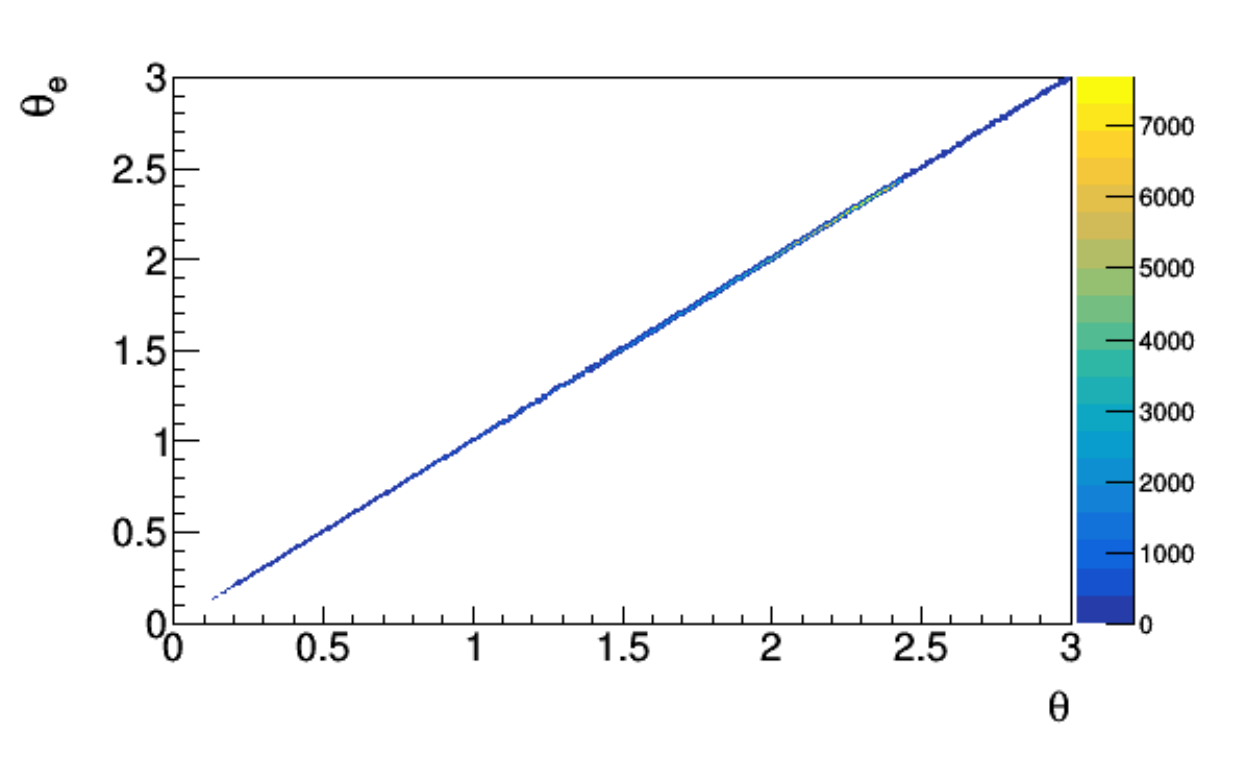}\\
     
     \includegraphics[width=0.48\linewidth]{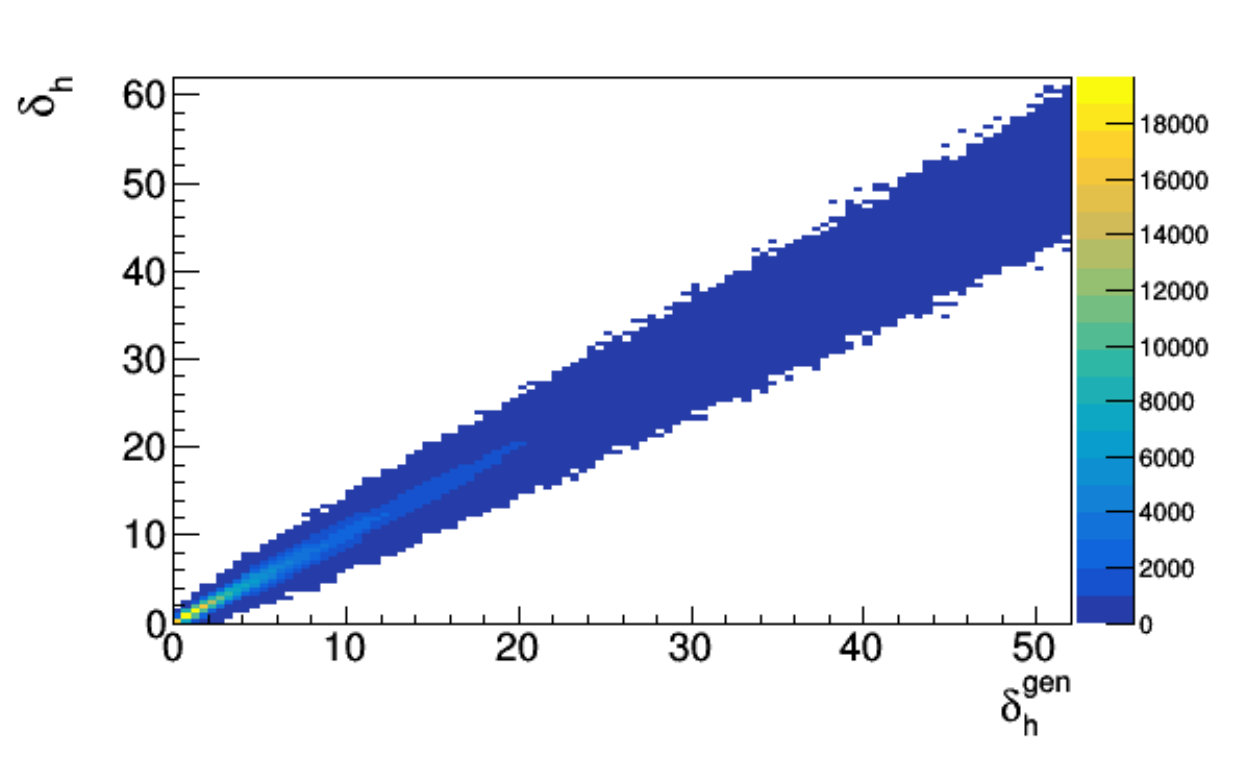}
    \includegraphics[width=0.48\linewidth]{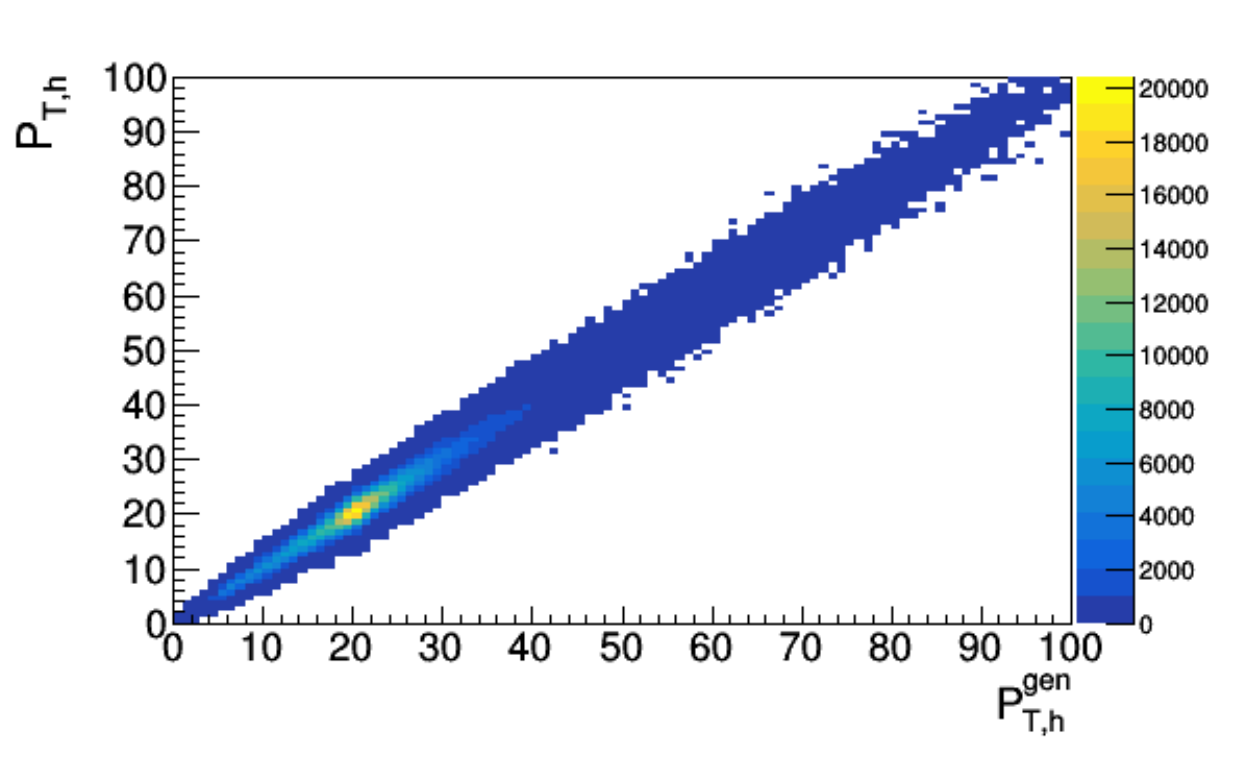}
\caption{Correlation of smeared $E_e$, $\theta_e$, $\delta_{h}$ and $P_{T,h}$ in the simulated  to their respective true generated values.}
\label{fig:controlplots}
\end{figure}

The simulated HFS is obtained by smearing the  $\delta_h^{gen}$ and $P_{T,h}^{gen}$  using

\begin{equation}
\label{eq:param1h}
\sigma_{\delta_h} / \delta_h^{gen} = 35\%/ \sqrt{\delta_h^{gen}}              
\end{equation}

\begin{equation}
\label{eq:param2h}
\sigma_{P_{T,h}}/P_{T,h}^{gen} = 35\% / \sqrt{P_{T,h}^{gen}}.             
\end{equation}
These values are motivated by detailed study of resolution  of $\delta_h$ and $P_{T,h}$ in the ZEUS detector~\cite{ref:tuning}.
The correlation of the four smeared final state properties, (E$_e$,$\theta_e$, $\delta_h$ and $P_{T,h}$), in the simulated data to their respective true generated values are plotted and shown in Figure~\ref{fig:controlplots}.
Using uncorrelated Gaussian smearing for the four measured quantities is not strictly appropriate, and we expect correlations between ($E$, $\theta$) and ($P_{T,h}$, $\delta_{h}$) to be present. A detector simulation will be needed to include this.

\begin{figure}
\centering
\includegraphics[scale=.35]{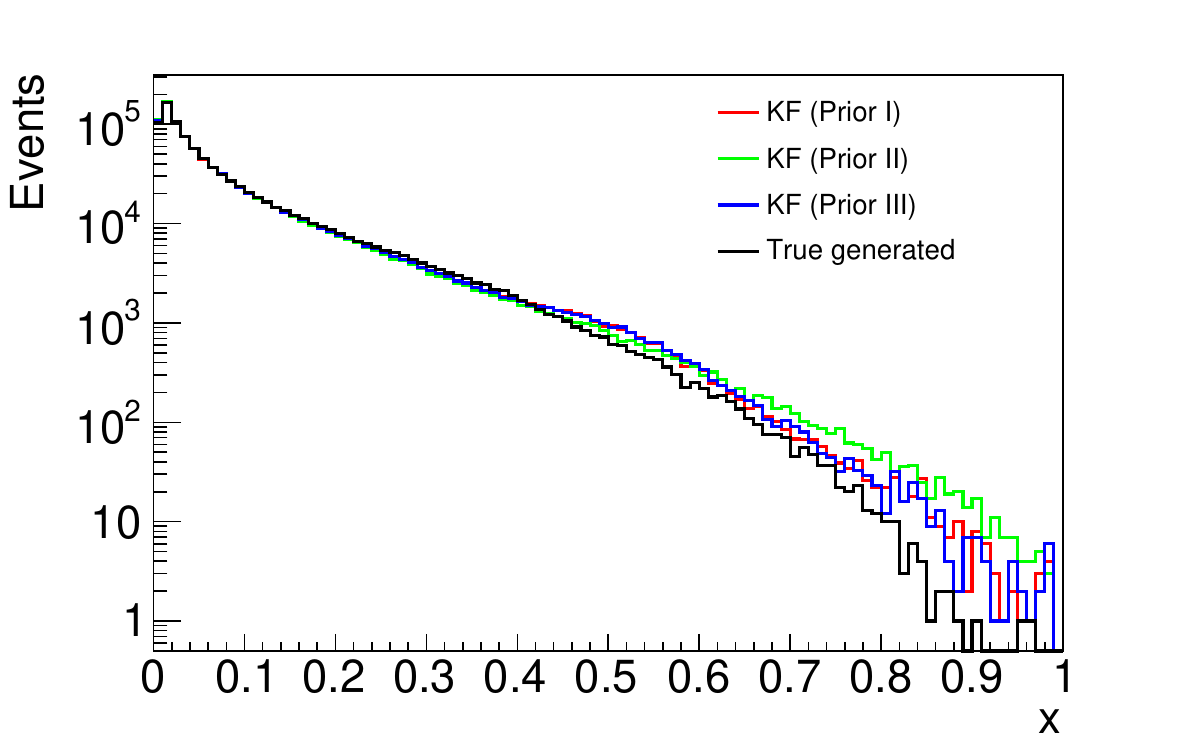}
\includegraphics[scale=.35]{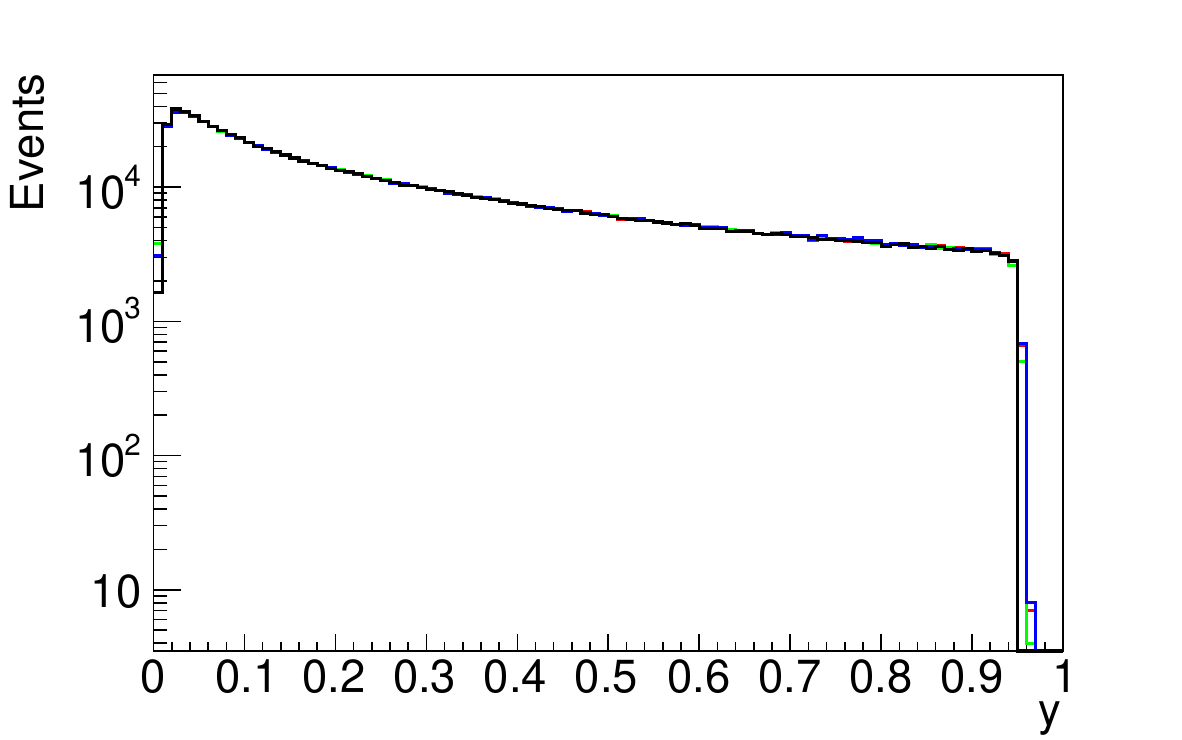}
\includegraphics[scale=.35]{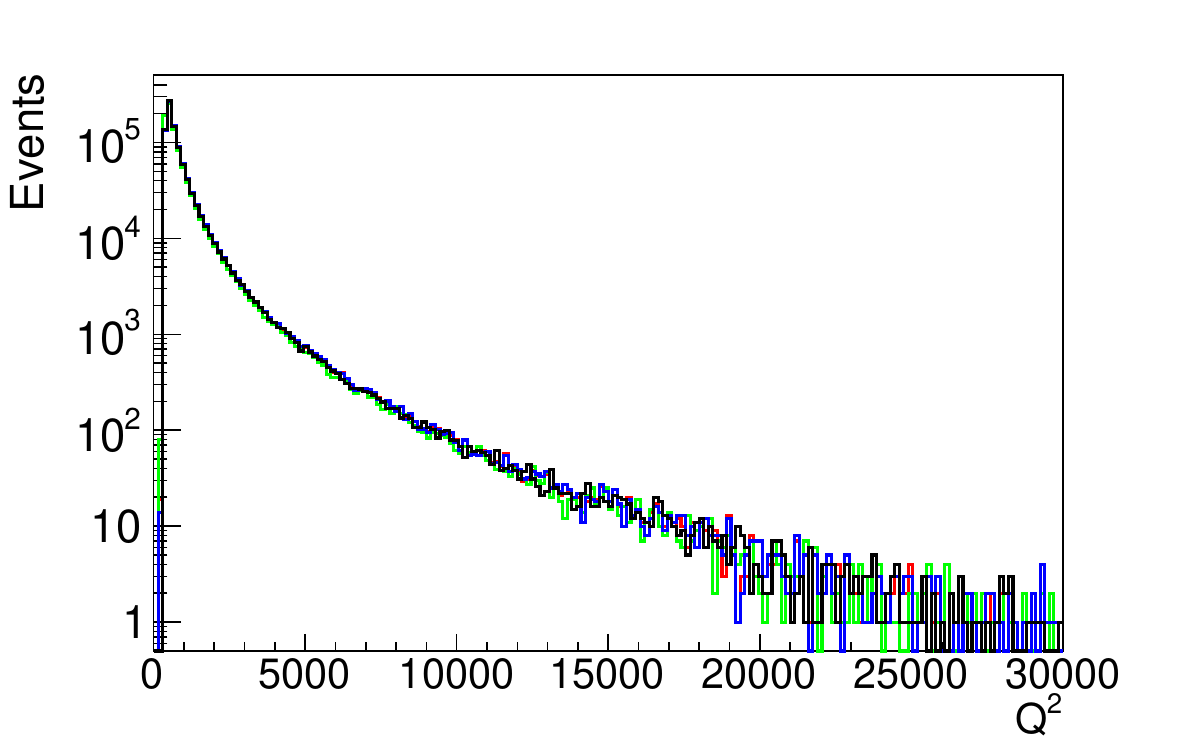}
\includegraphics[scale=.35]{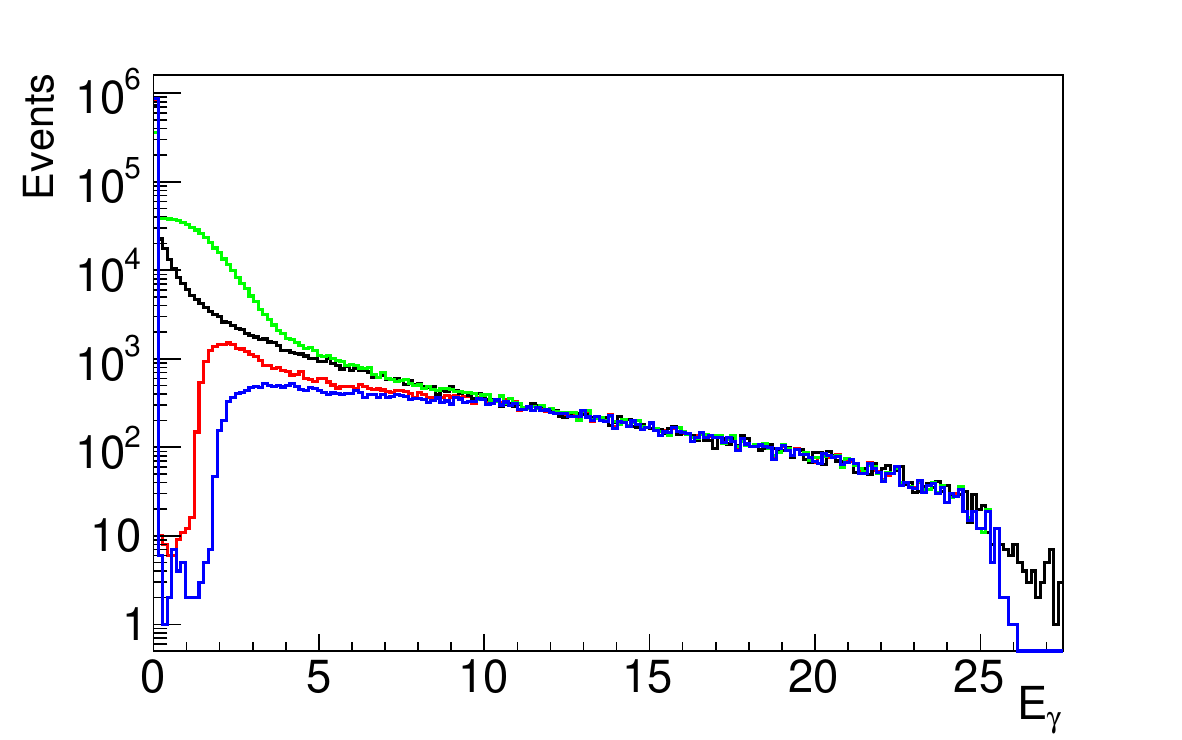}
\caption{x, y, Q$^2$ and $E_{\gamma}$ from the kinematic fit as compared to the true quantities using different priors.}
\label{fig:xyE}

\end{figure}

\begin{figure}
\centering
\includegraphics[scale=.4]{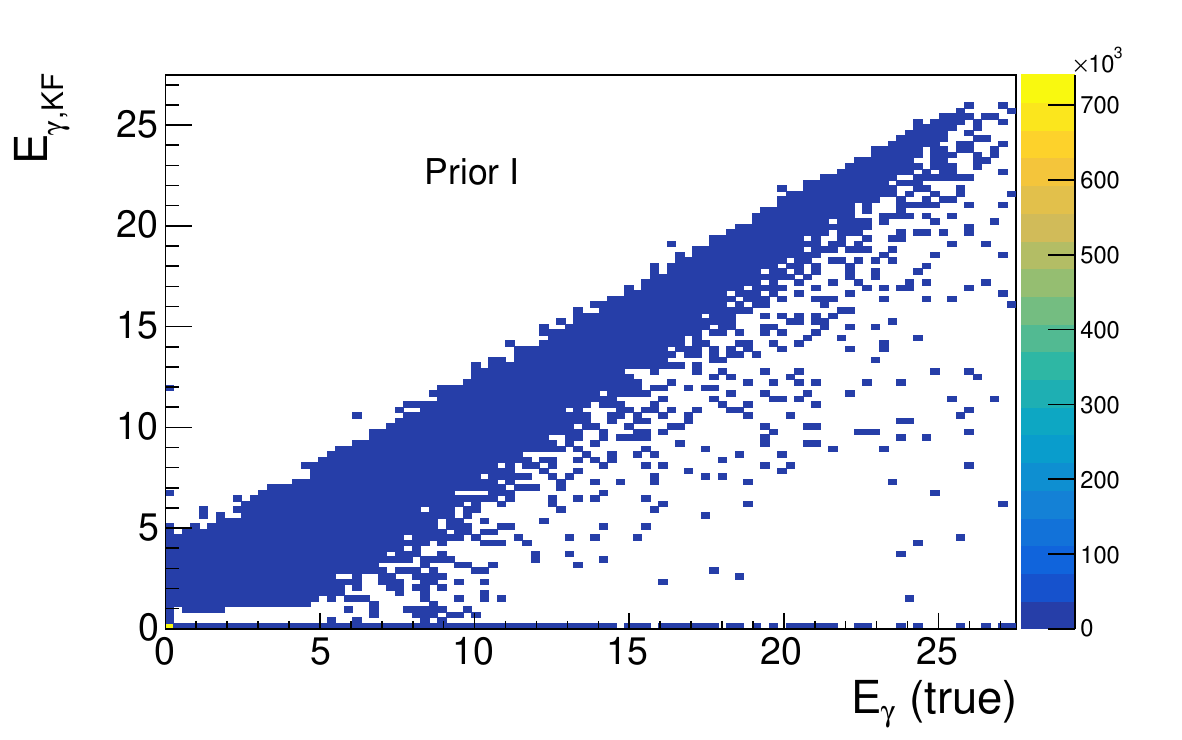}
\includegraphics[scale=.4]{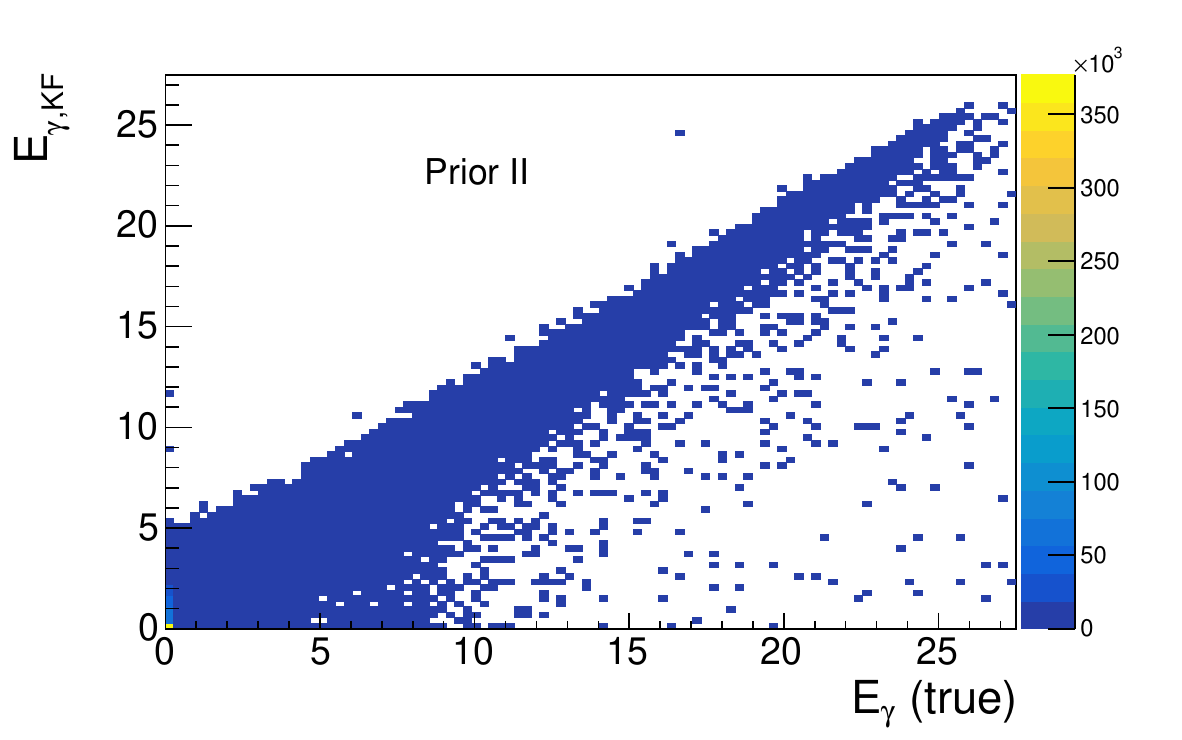}
\includegraphics[scale=.4]{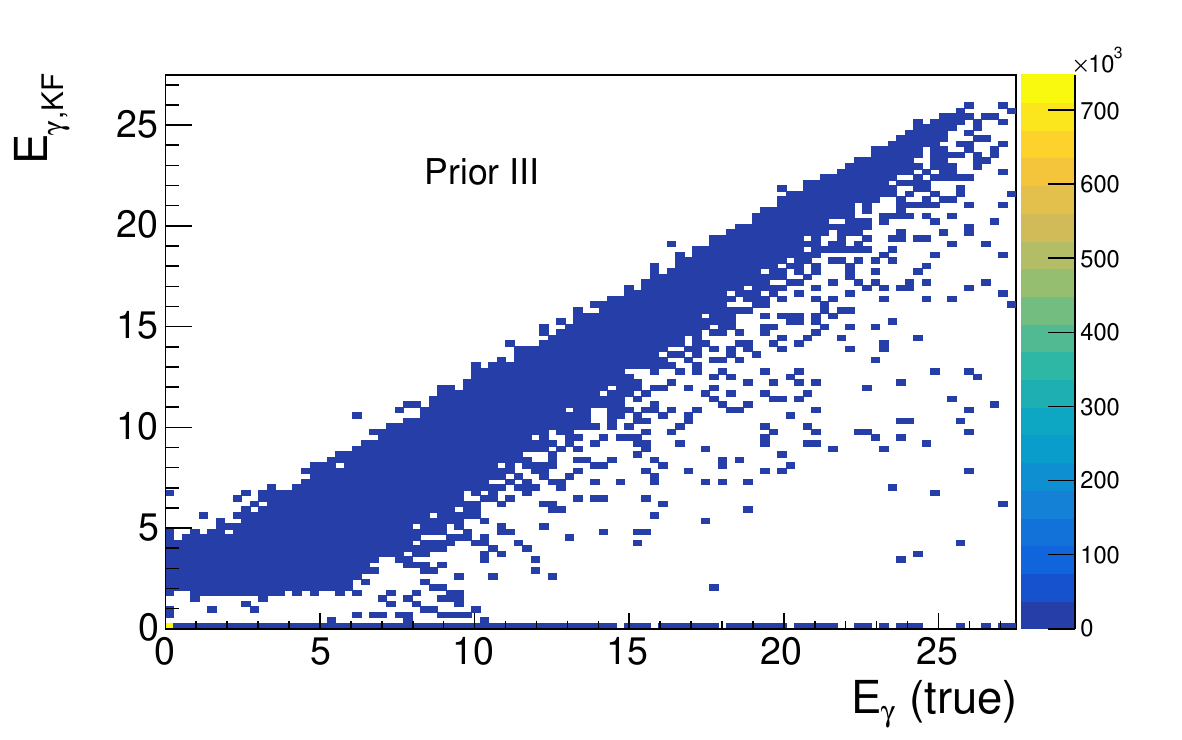}
\caption{ Correlation of $E_\gamma$ from the kinematic fit to it's true value for three different priors. }
\label{fig:ISR2D}
\end{figure}

\begin{figure}
\centering
\includegraphics[width=0.44\linewidth]{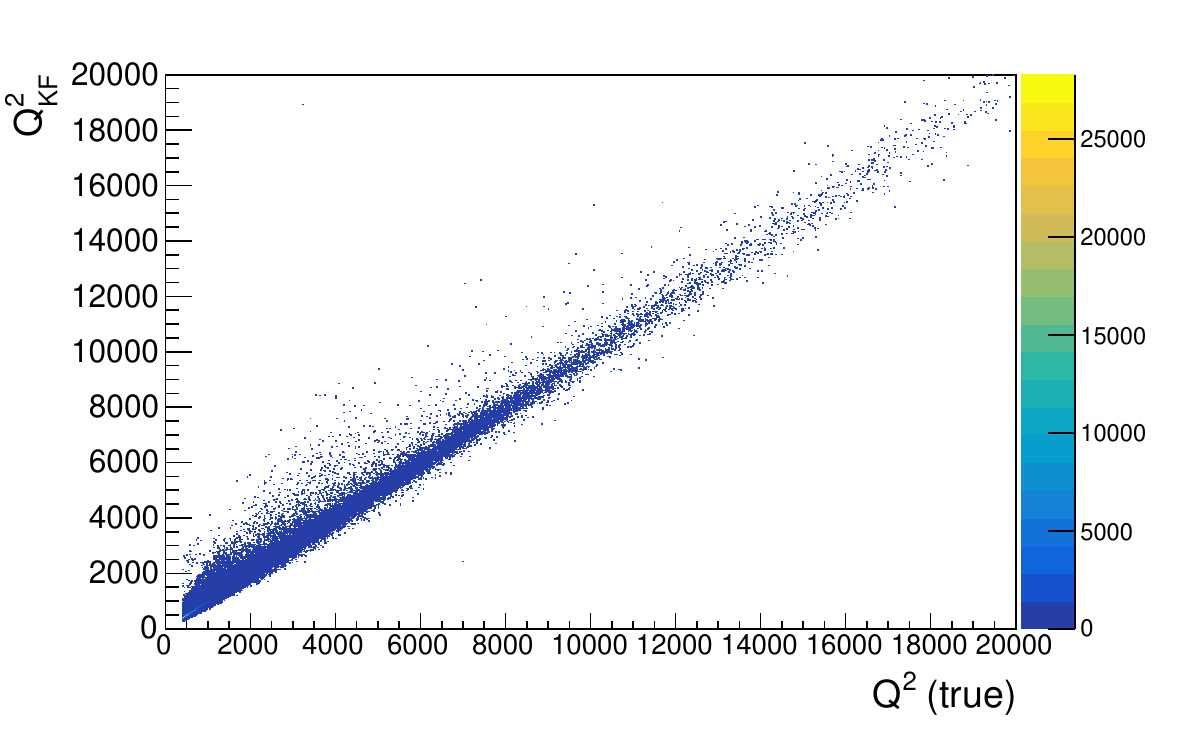}
\includegraphics[width=0.44\linewidth]{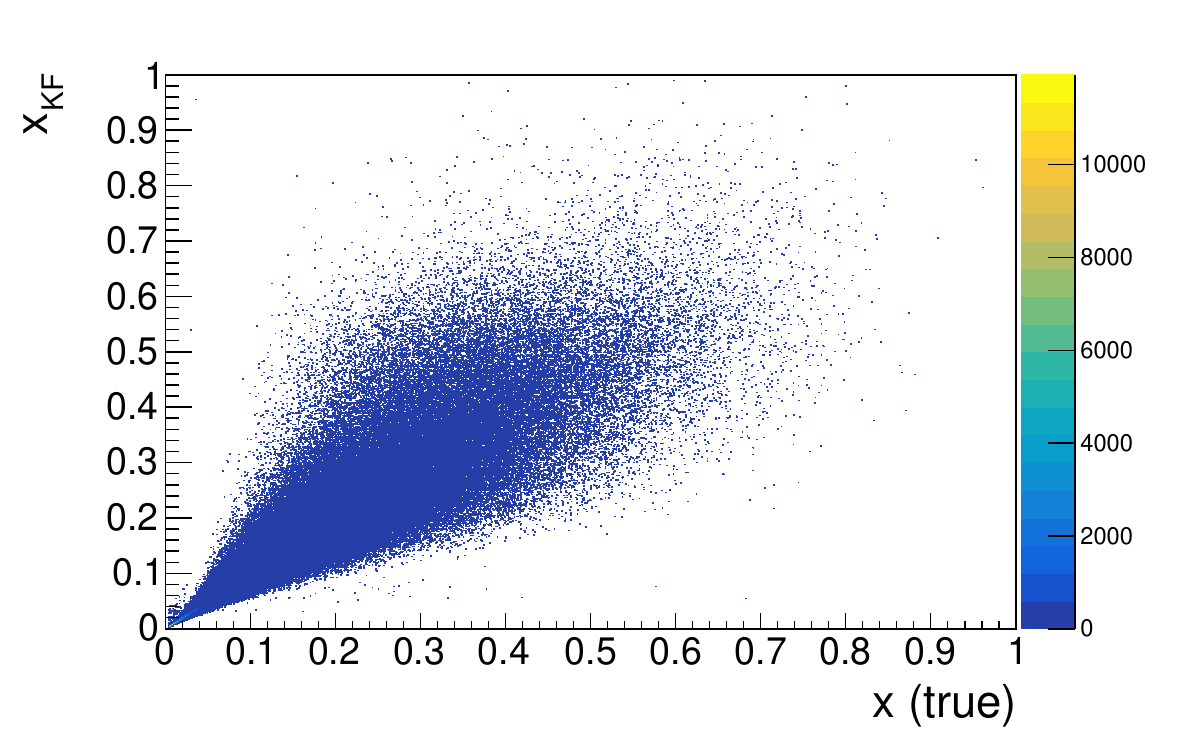}

\includegraphics[width=0.44\linewidth]{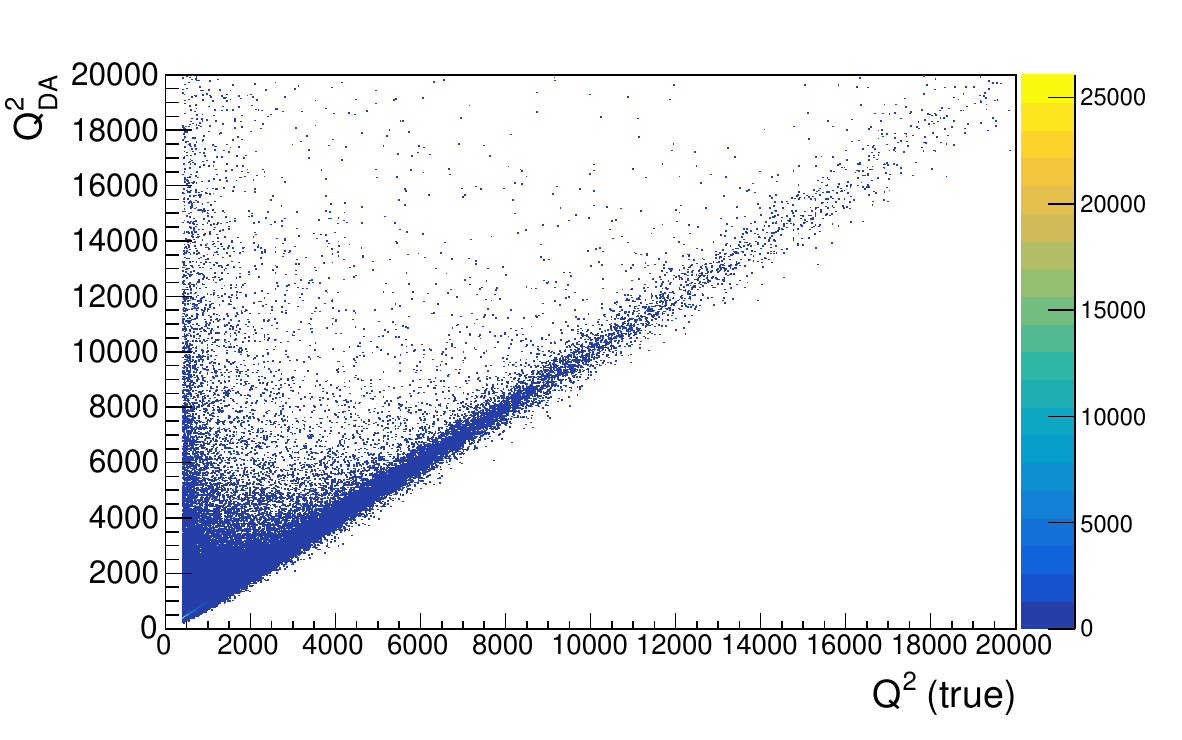}
\includegraphics[width=0.44\linewidth]{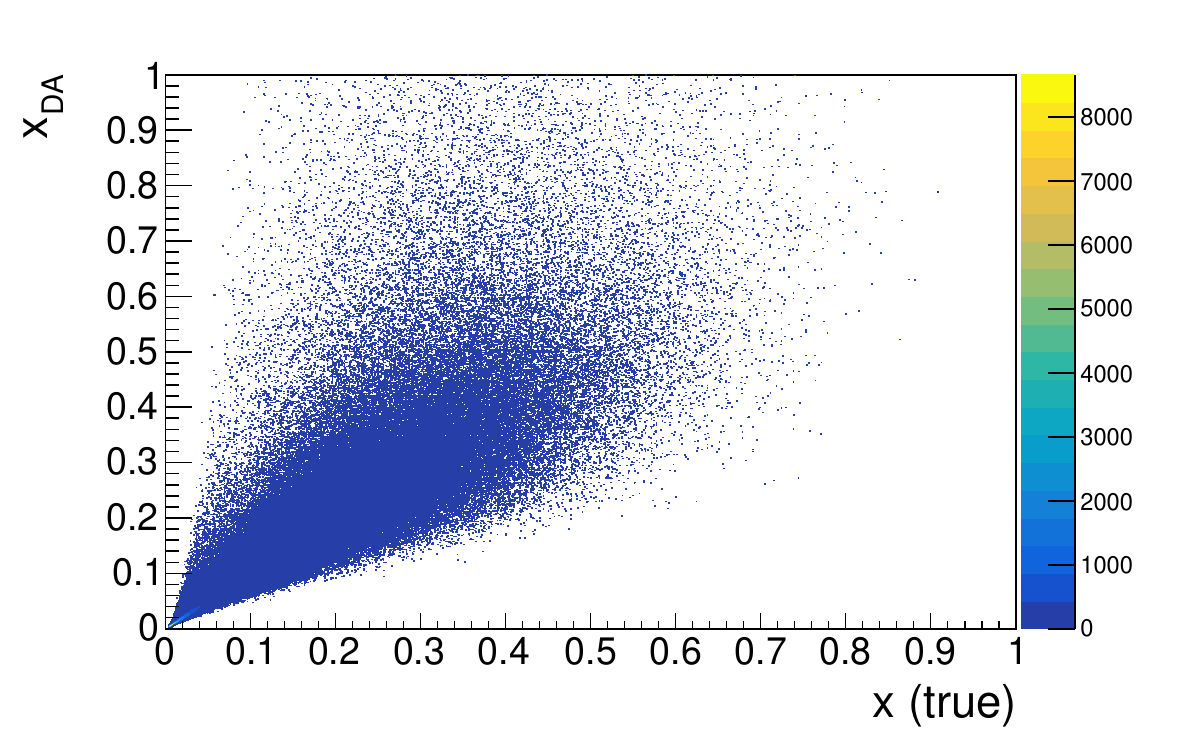}
\includegraphics[width=0.44\linewidth]{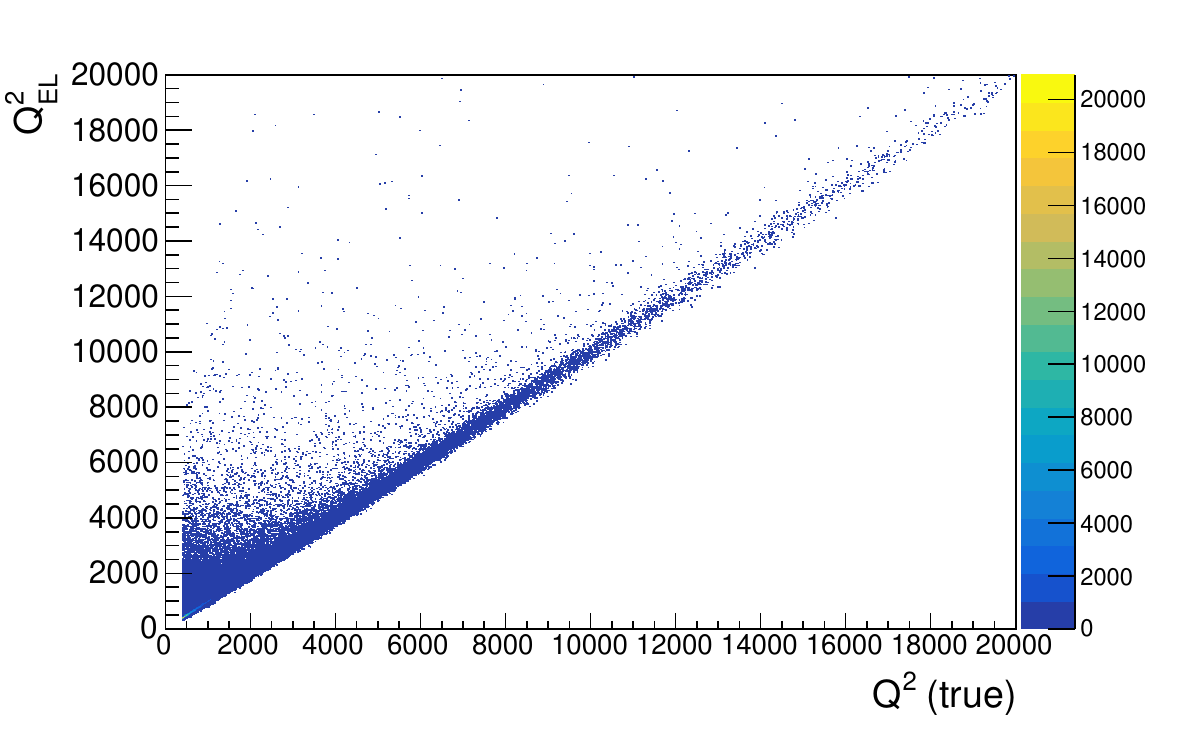}
\includegraphics[width=0.44\linewidth]{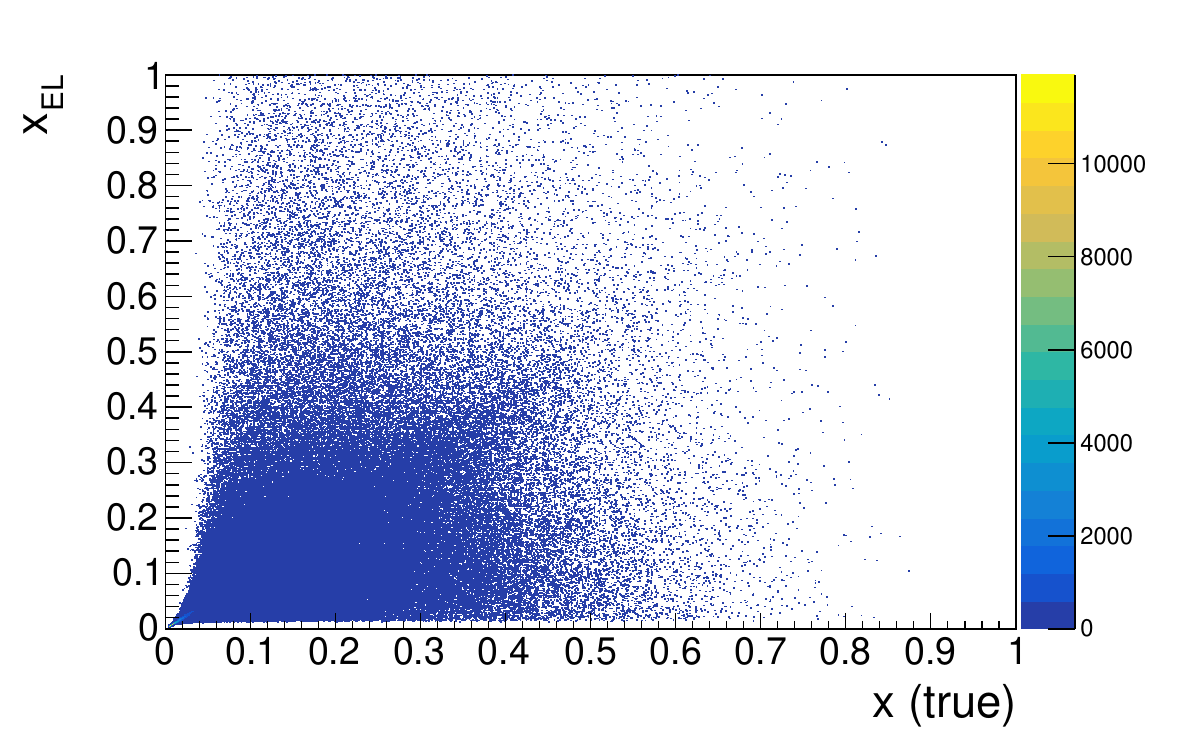}

\caption{Correlation of $x$ and $Q^2$ from the kinematic fit (KF), double angle (DA) and electron (EL) reconstruction methods (upper, middle and lower rows respectively) to the true generated values.}
\label{fig:xq2_corr}

\end{figure}

\begin{figure}
\centering
\includegraphics[scale=.4]{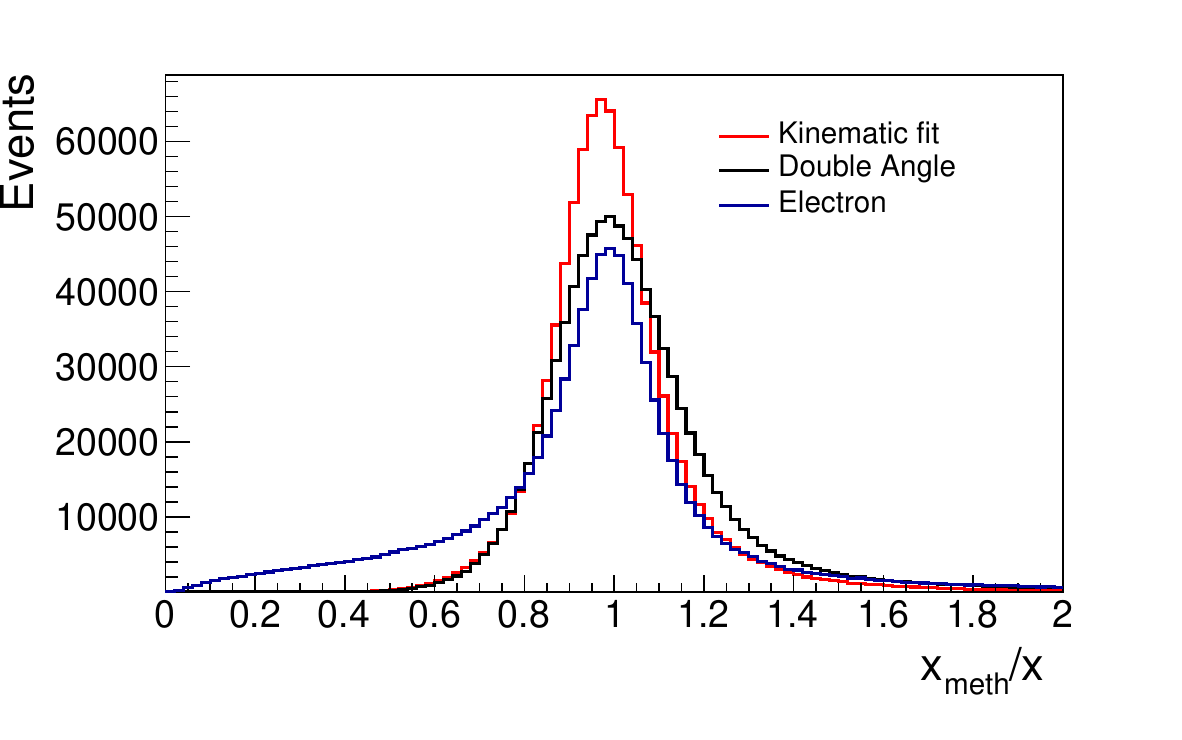}
\includegraphics[scale=.4]{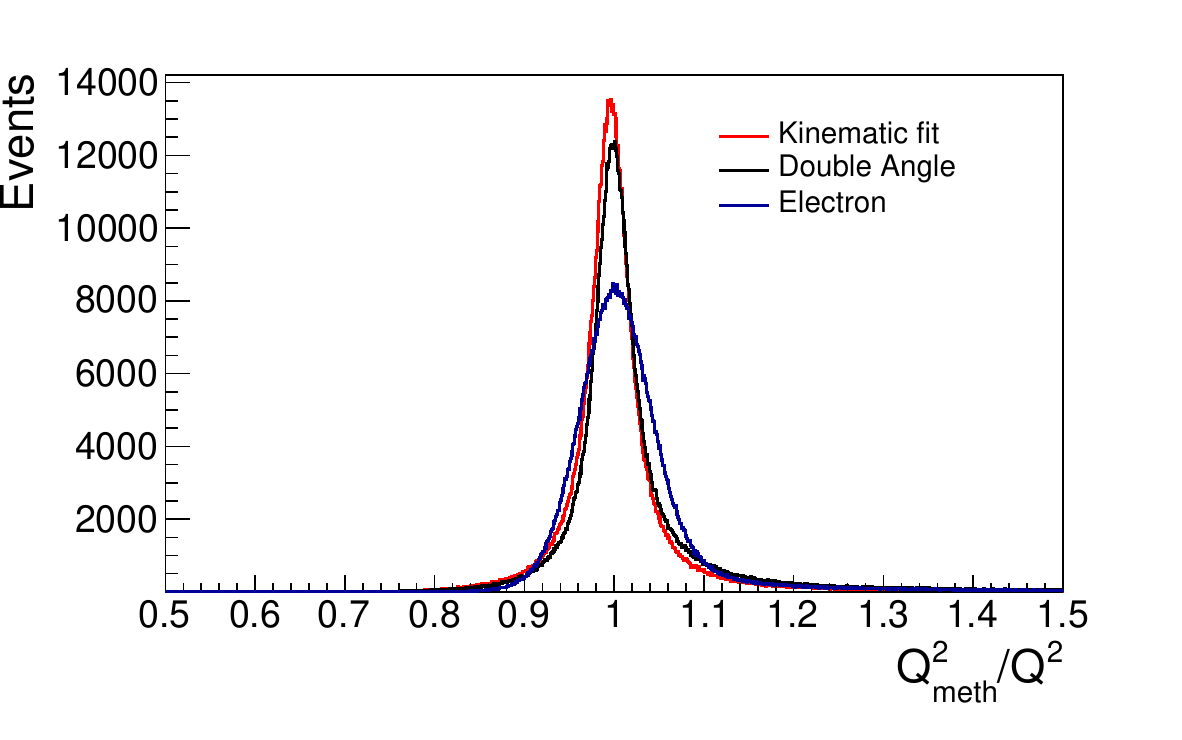}
\caption{Ratio of $x$ and $Q^2$ from the  kinematic fit, double angle and electron reconstruction methods as compared to the true quantities for all generated events.}
\label{fig:xq2_main}

\end{figure}

\section{Results from the Kinematic Fit}
The kinematic fit is performed using Bayes theorem on the simulated sample with  $Q^2 >$ 400 GeV$^2$ and after filtering out the generated QED Compton~\cite{ref:Heraceles} events\footnote{The QED Compton events serve as background to the DIS NC data, and at large $x$ and high $Q^2$ are expected to be negligible~\cite{ref:highx}.}. The likelihood function is taken from Equation ~\ref{eq:Pos}. For the main results shown in this paper, we do not extract the full posterior probability distribution but only the values of the parameters at the mode of the posterior probability. To extract the uncertainties and correlations in the parameters  $\boldsymbol{\lambda}$, the BAT software~\cite{ref:bat}, e.g., can be used.
Two example fits of the kinematic variables using BAT are shown in Appendix A. 

The Bayesian inference to get the set of most probable quantities $\boldsymbol{\lambda}=(x,y, E_\gamma)$ given the measurement $\boldsymbol{D} = (E, \theta_e, \delta_h, P_{T,h})$, is calculated using the Bayesian analysis toolkit, BAT. $Q^2$ is obtained from the KF method  as 
$$Q^2_{KF} = s^{\prime}x_{KF} y_{KF},$$ where, $s^{\prime}$ is the center of mass energy which gets reduced when an ISR is reconstructed from kinematic fit and is given as
$$s^{\prime} = 4(A-E_\gamma)P.$$
The comparison of the distribution for $x$, $y$ $Q^2$ and  $E_\gamma$  obtained from the kinematic fit method to the generated values are shown in Figure~\ref{fig:xyE}.

  Three different prior distributions, $P_{o}(\boldsymbol{\lambda})$, were studied and are listed below. The kinematic fit was performed for each one of the priors separately. The comparison of  $x$, $y$ and  $E_\gamma$ distributions obtained from the kinematic fit method  using different prior choices are also shown in Figure~\ref{fig:xyE}. The results are extracted using three different prior choices.

\begin{itemize}
	\item Prior I : the Bremsstrahlung cross section on E$_\gamma$
	
\begin{equation}
\label{eq:prior1}
         P_{o}(\boldsymbol{\lambda}) = \frac{1+(1-y)^{2}}{x^{3}y^{2}} \frac{[1+(1-E_{\gamma}/A)^{2}]}{E_{\gamma}/A}
\end{equation}
     
	\item Prior II : steeply falling factor for $E_\gamma$
		
\begin{equation}
\label{eq:prior2}
    	P_{o}(\boldsymbol{\lambda}) = \frac{1+(1-y)^{2}}{x^{3}y^{2}} \frac{1}{E_{\gamma}^2}.
\end{equation}
    	
	\item Prior III : flat prior for $E_\gamma$
	
\begin{equation}
\label{eq:prior3}
    	P_{o}(\boldsymbol{\lambda}) = \frac{1+(1-y)^{2}}{x^{3}y^{2}}.
\end{equation}
	\end{itemize}
	
	There is no significant difference observed in the results from the three priors. However, some differences are observed in the $E_\gamma$ distribution for low values of $E_\gamma$. Priors I and III  underestimated the ISR with very small values of $E_\gamma$, whereas,  Prior II overestimated the ISR. 

One of the advantages of the kinematic fit approach is the estimation of the energy of the ISR in the event. Figure~\ref{fig:ISR2D} shows the correlation of  $E_\gamma$ estimated from the kinematic fit to it's true value in the event using three different priors.  As observed from the comparison of different Priors shown in  Figures~\ref{fig:xyE} and~\ref{fig:ISR2D}, the $E_\gamma$ is estimated effectively using the Prior I (Equation ~\ref{eq:prior1}) and this is used in the subsequent analysis.

\subsection{Comparison to other methods}

For the comparison to other methods, the kinematic fit is performed with the likelihood function as given in Equation~\ref{eq:Pos} and prior I with Bremsstrahlung cross section for $E_\gamma$. Figure~\ref{fig:xq2_corr} shows the correlation of $x$ and $Q^2$  reconstructed from the kinematic fit method to the true generated value. The correlation of $x$ and $Q^2$ reconstructed from the double angle and electron methods to the true generated value are also shown in Figure~\ref{fig:xq2_corr}. It is observed that the $x$ and $Q^2$ reconstructed from the kinematic fit method has a smaller spread in the correlation plots as compared to the double angle and electron methods.

The bias and resolution in the $x$ and $Q^2$ reconstruction from the kinematic fit method is compared to the electron and the double angle method in Figure~\ref{fig:xq2_main}. 
The Figure shows the comparison of ratios\footnote{Here in the ratios, $x_{meth}$ and $Q^2_{meth}$ will refer to the variables reconstructed from any of the el, double angle and KF methods, and the $x$ and $Q^2$ without subscript would represent their true generated values obtained using the exchanged Boson information.} $x_{meth}/x$ and  $Q^2_{meth}/Q^2$ from different methods for the full simulated data set. The ratios obtained from the kinematic fit method are observed to have minimum width implying a better resolution which can be attributed to the maximum information of the final state being used in the analysis as compared to any of the double angle and electron methods.

\begin{figure}
\centering
\includegraphics[scale=.27]{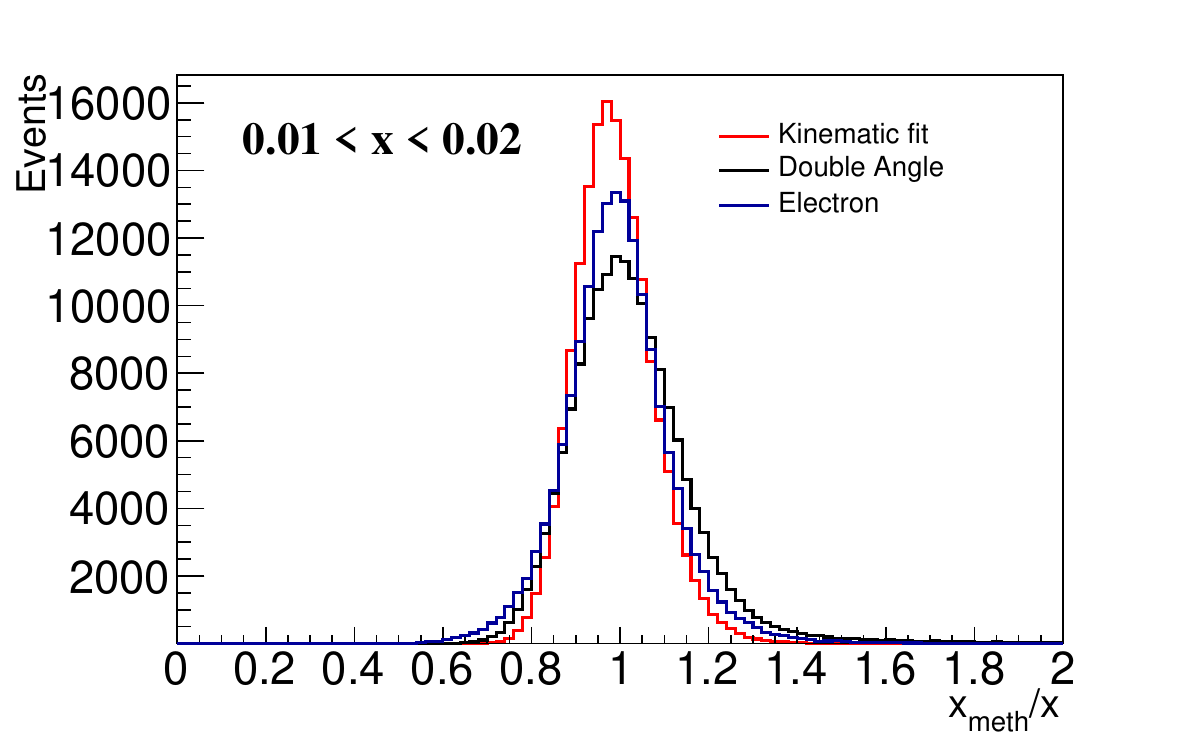}
\includegraphics[scale=.27]{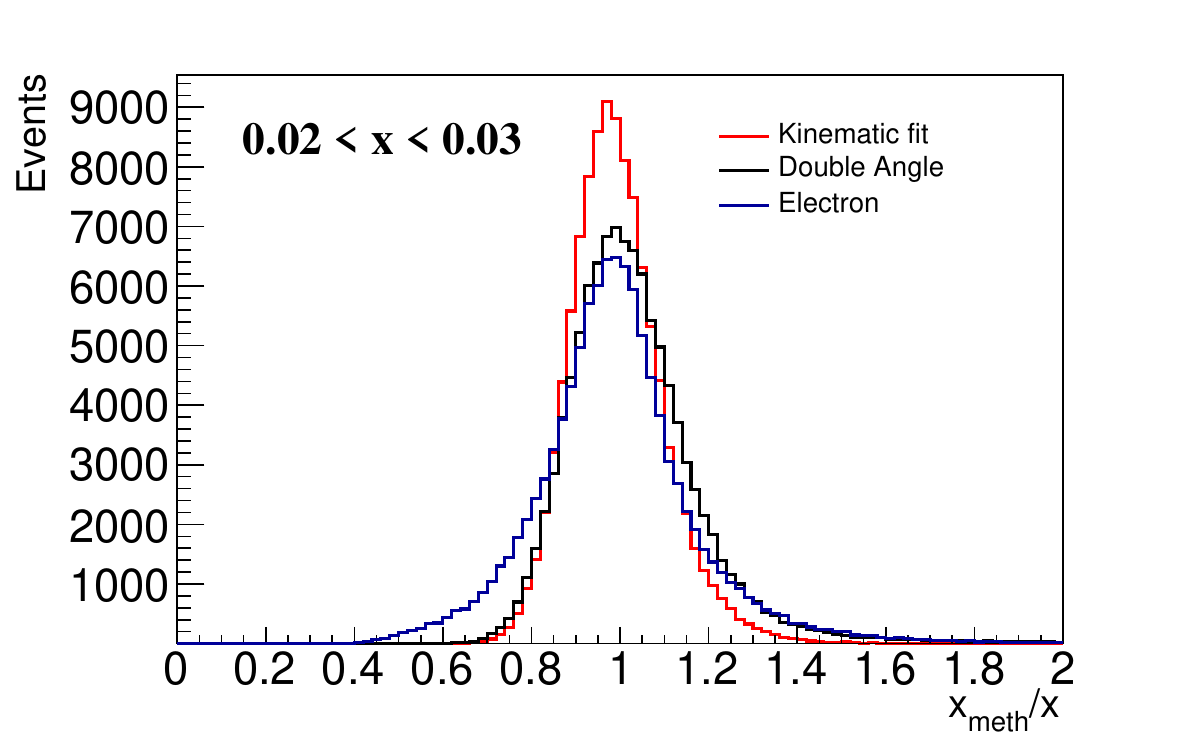}
\includegraphics[scale=.27]{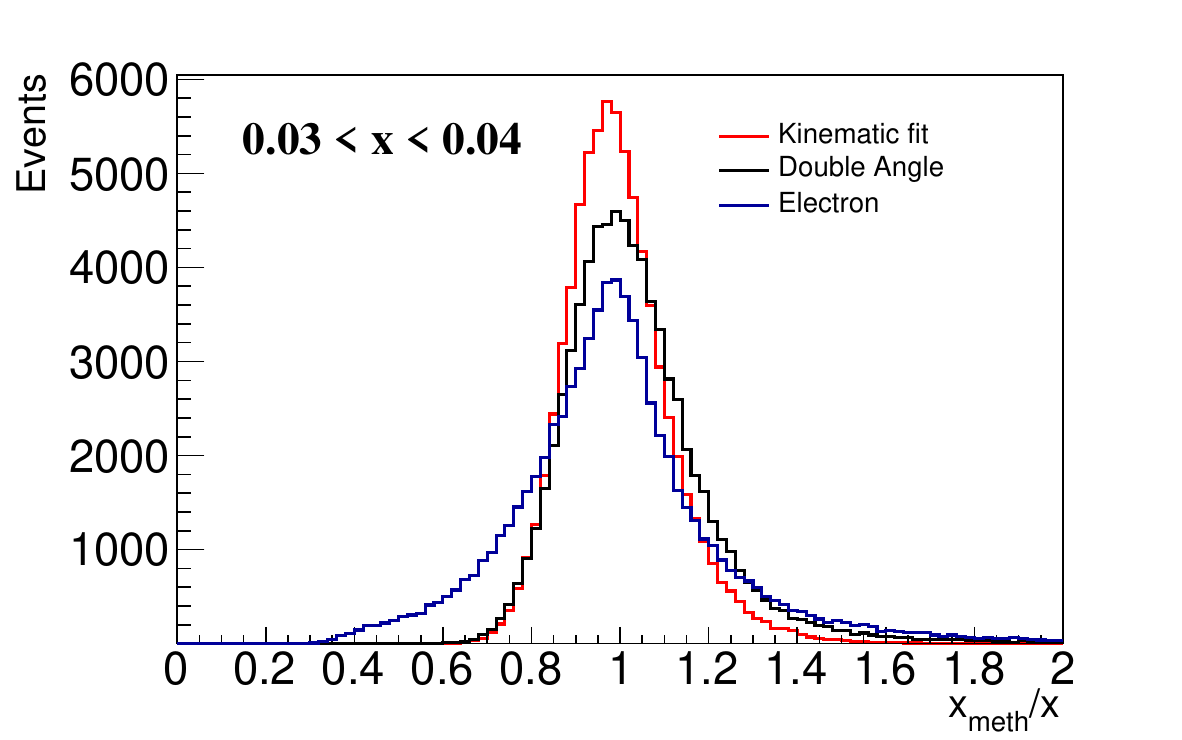}
\includegraphics[scale=.27]{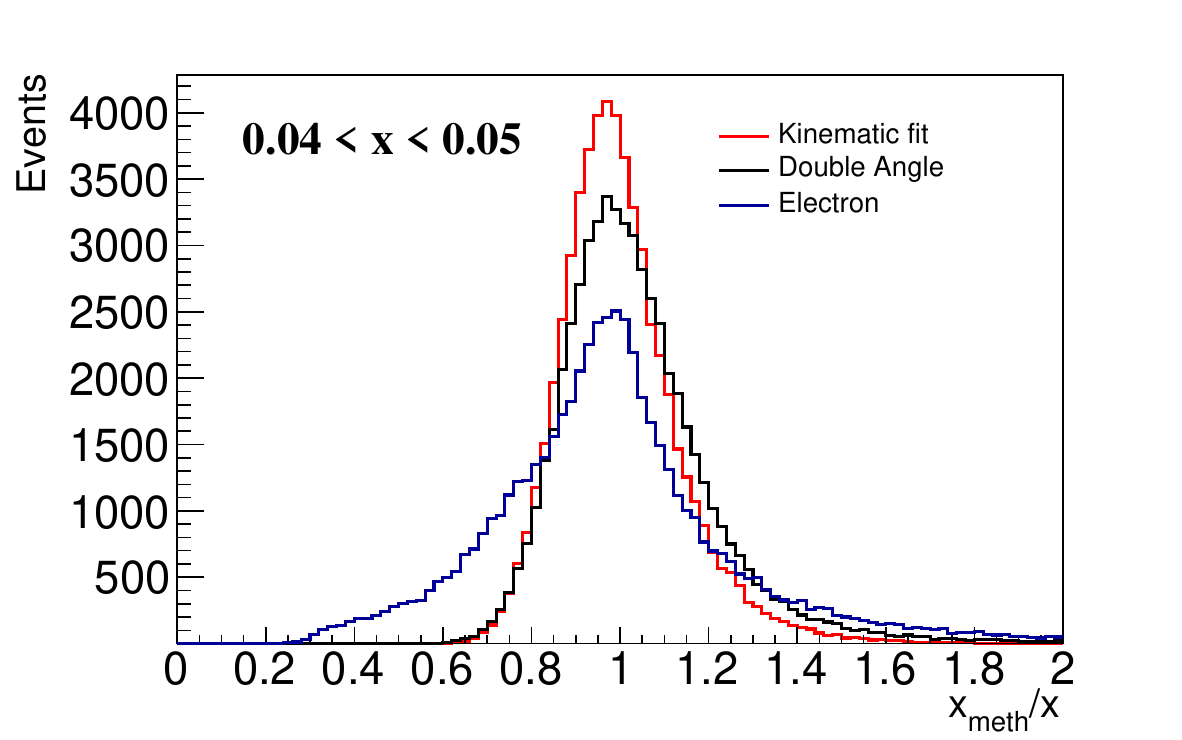}
\includegraphics[scale=.27]{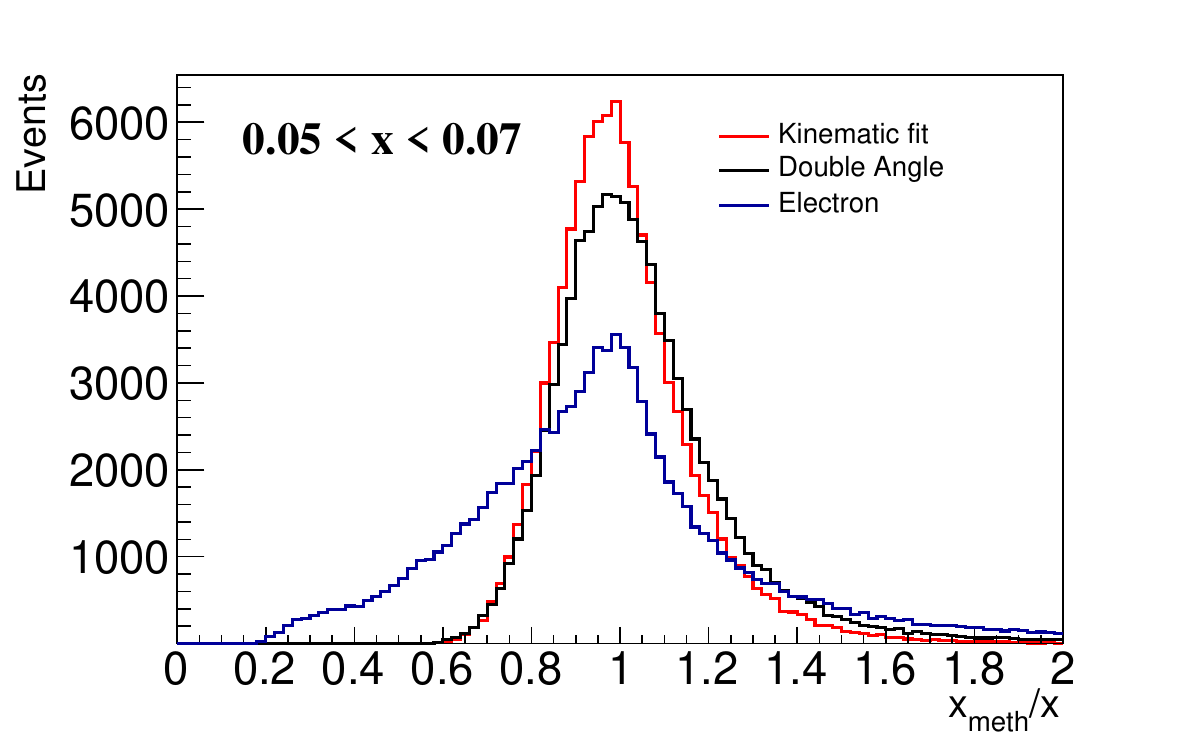}
\includegraphics[scale=.27]{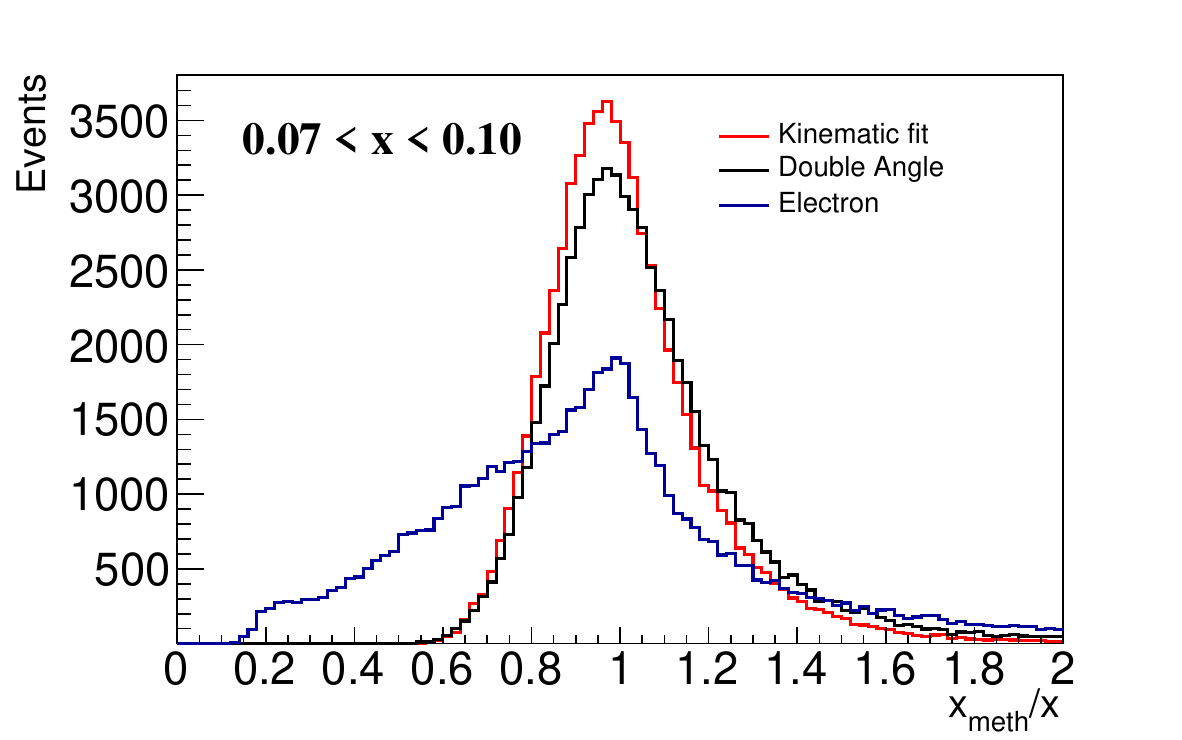}
\includegraphics[scale=.27]{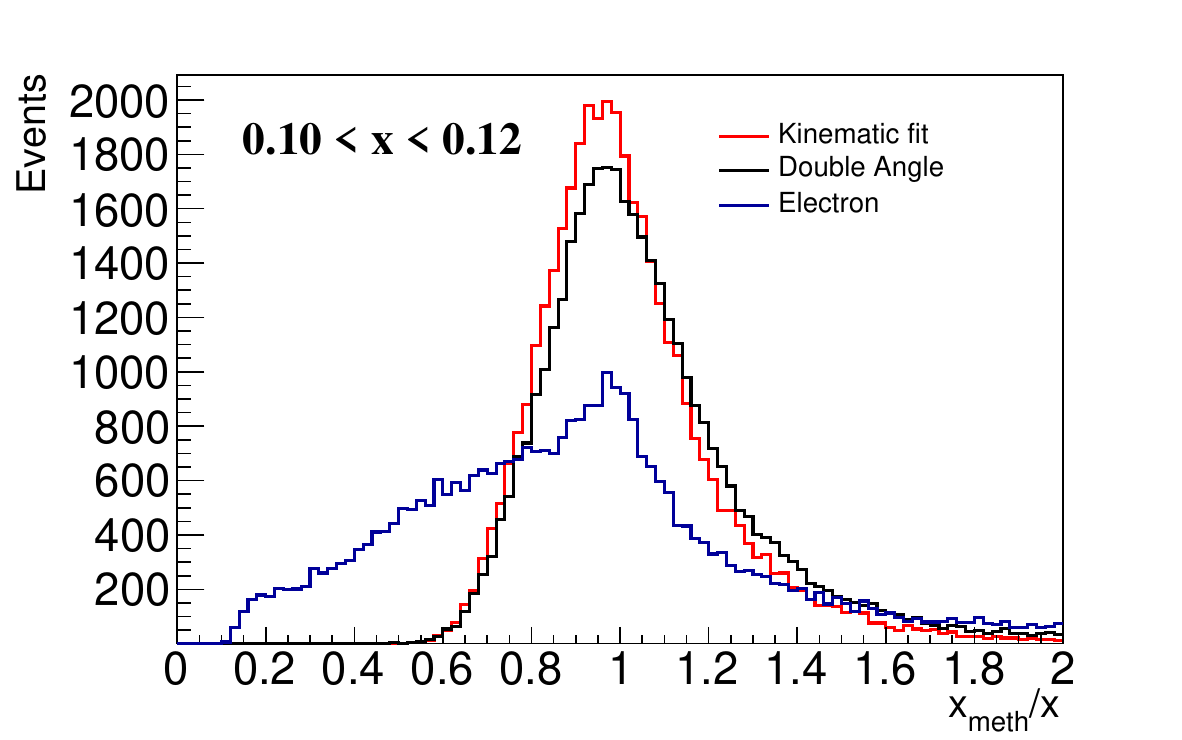}
\includegraphics[scale=.27]{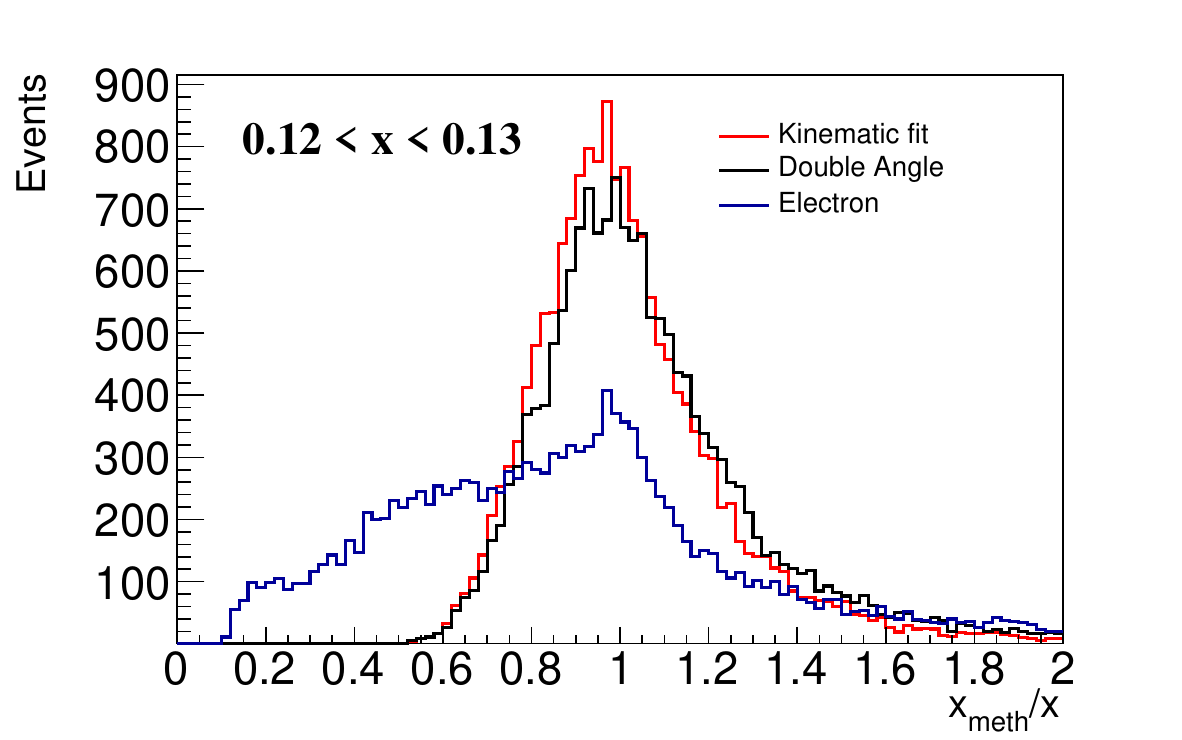}
\includegraphics[scale=.27]{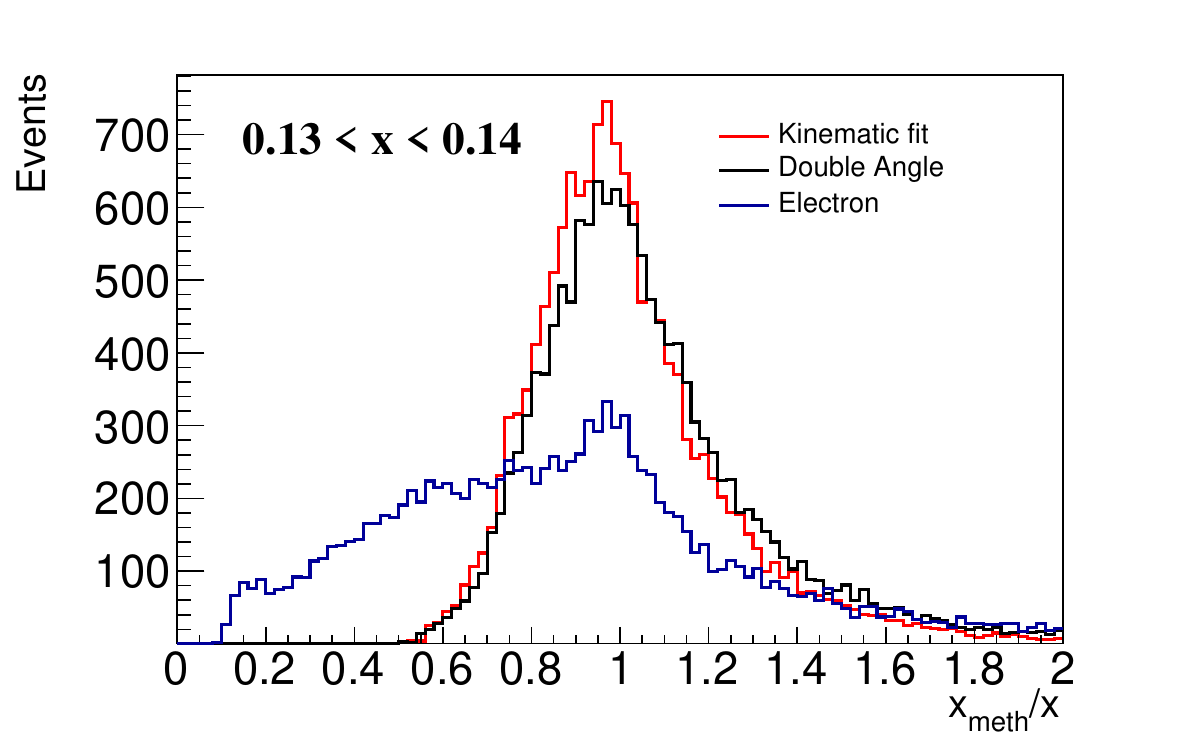}
\includegraphics[scale=.27]{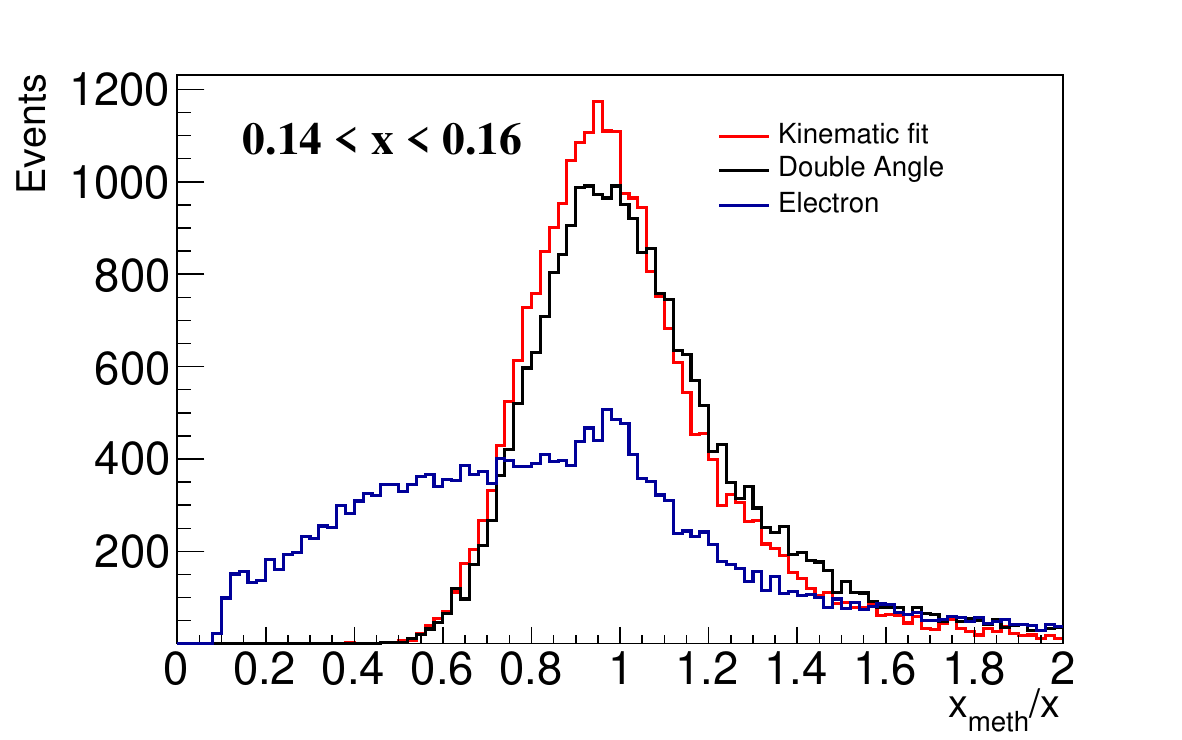}
\includegraphics[scale=.27]{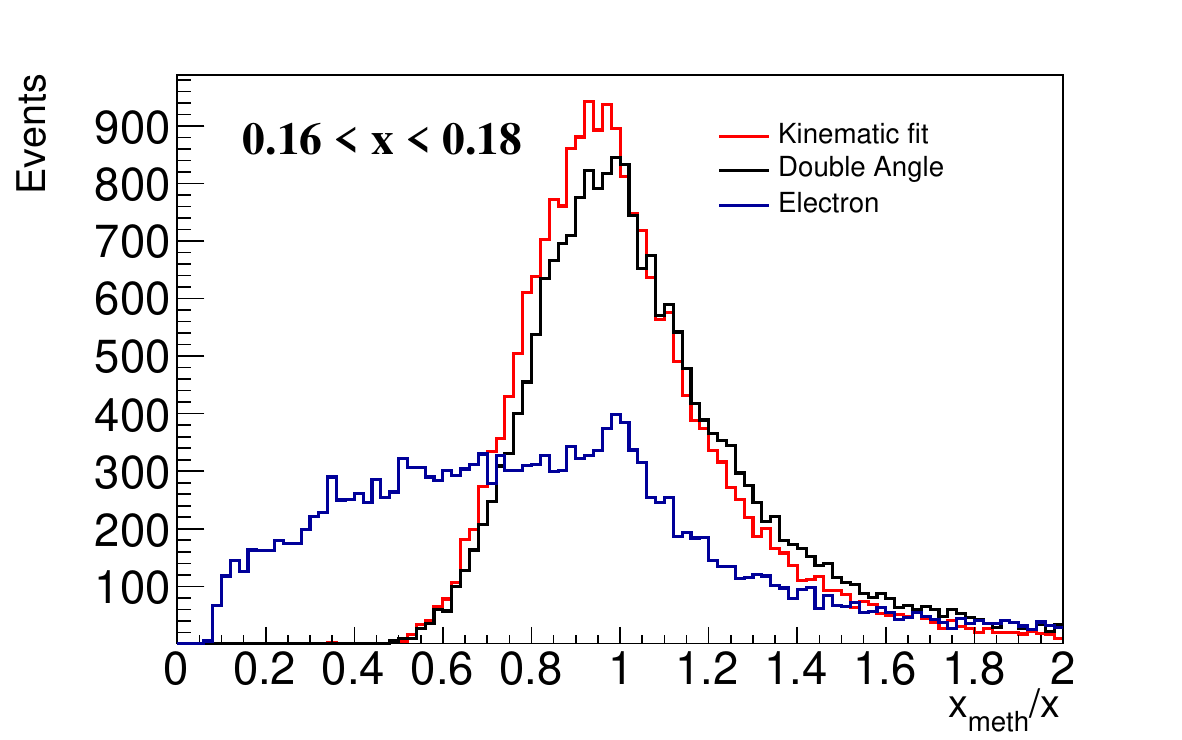}
\includegraphics[scale=.27]{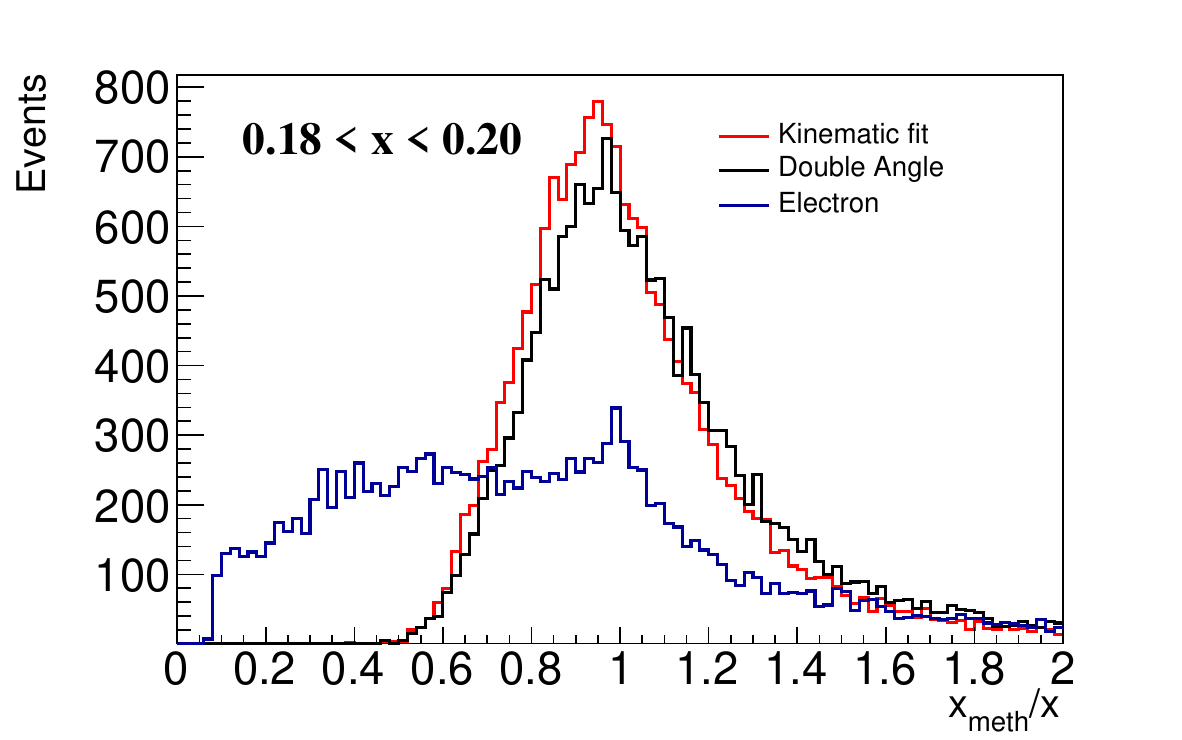}
\includegraphics[scale=.27]{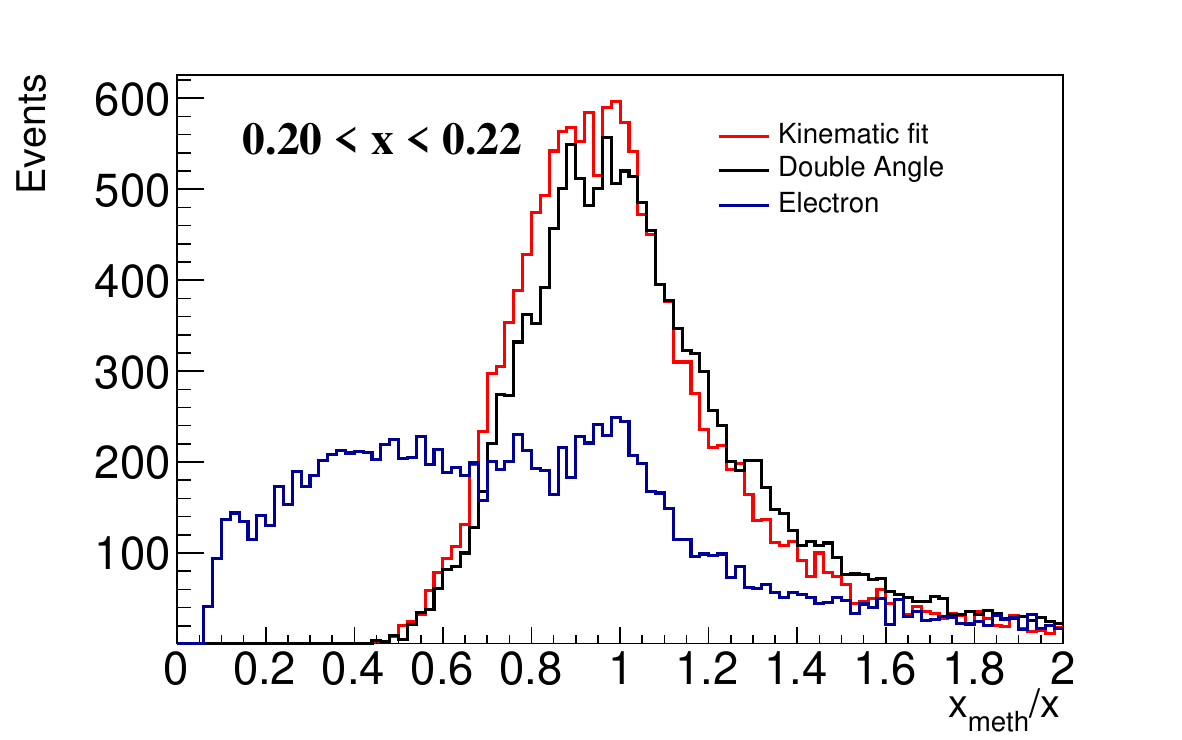}
\includegraphics[scale=.27]{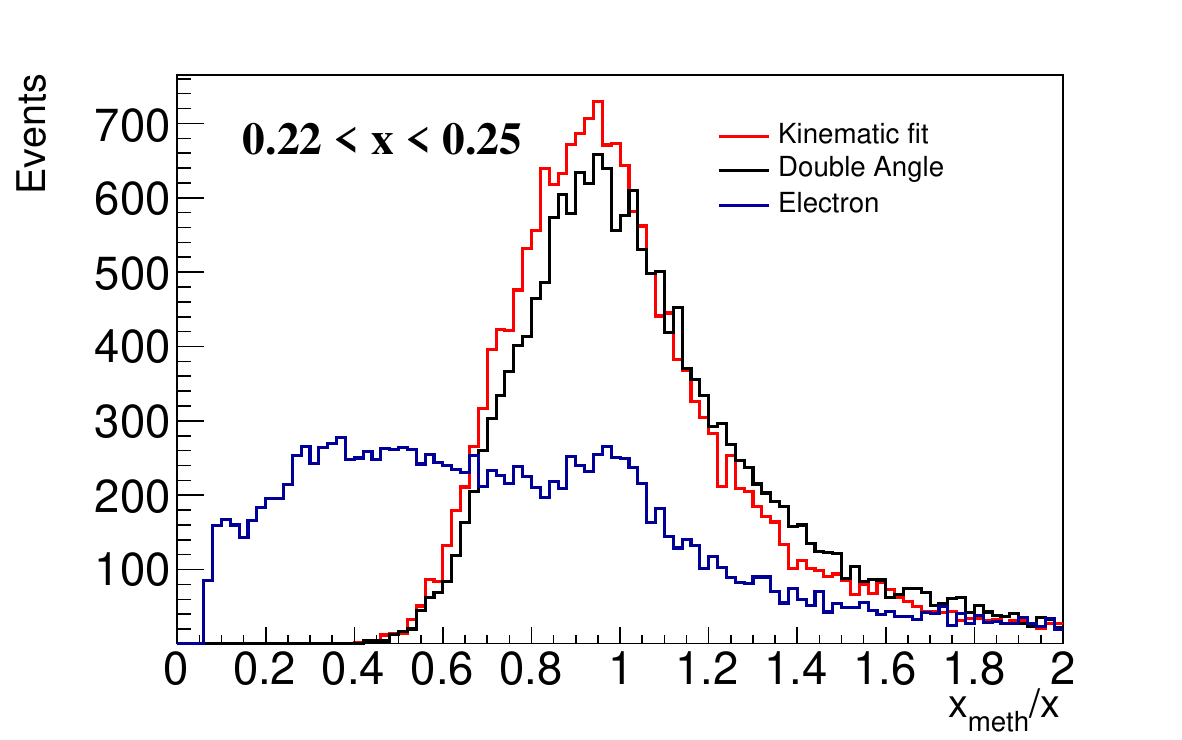}
\includegraphics[scale=.27]{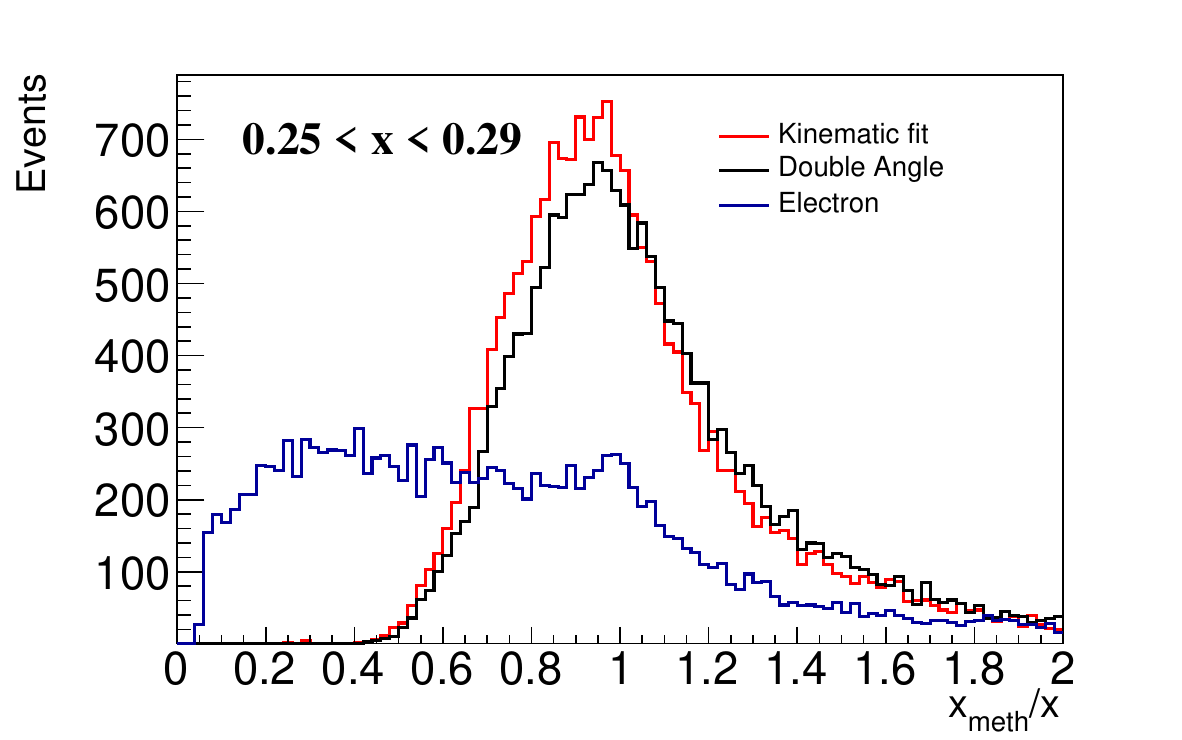}
\includegraphics[scale=.27]{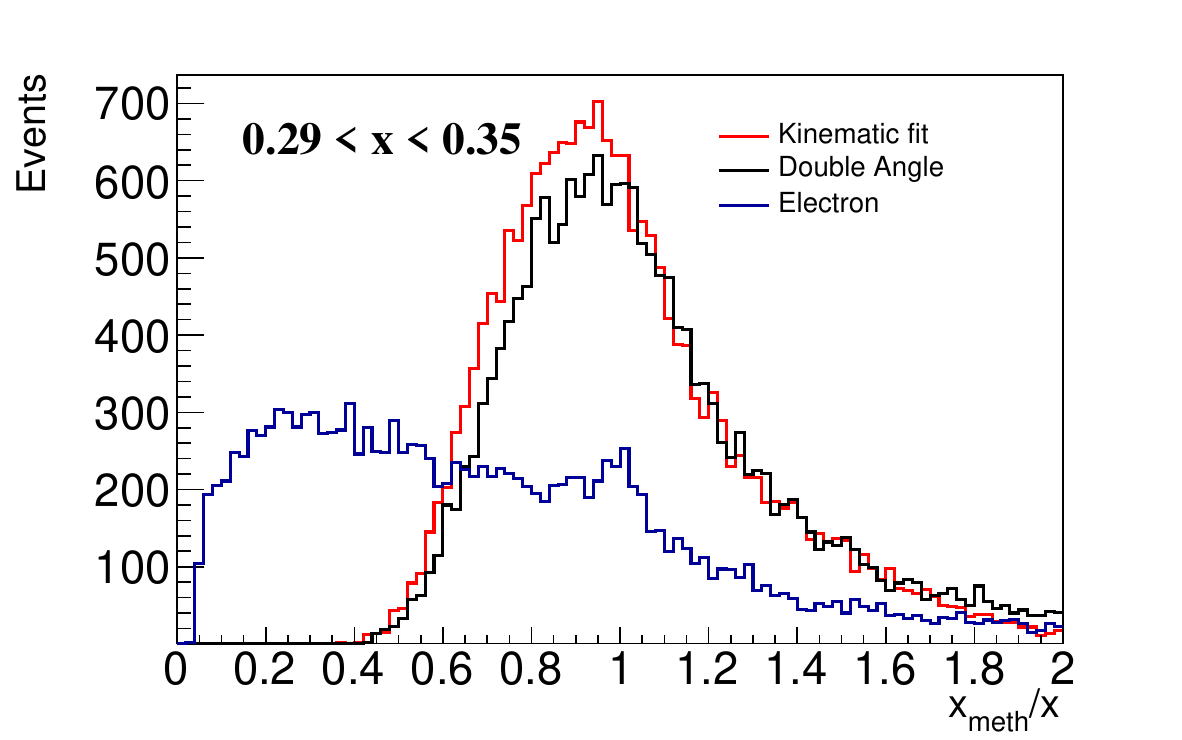}
\includegraphics[scale=.27]{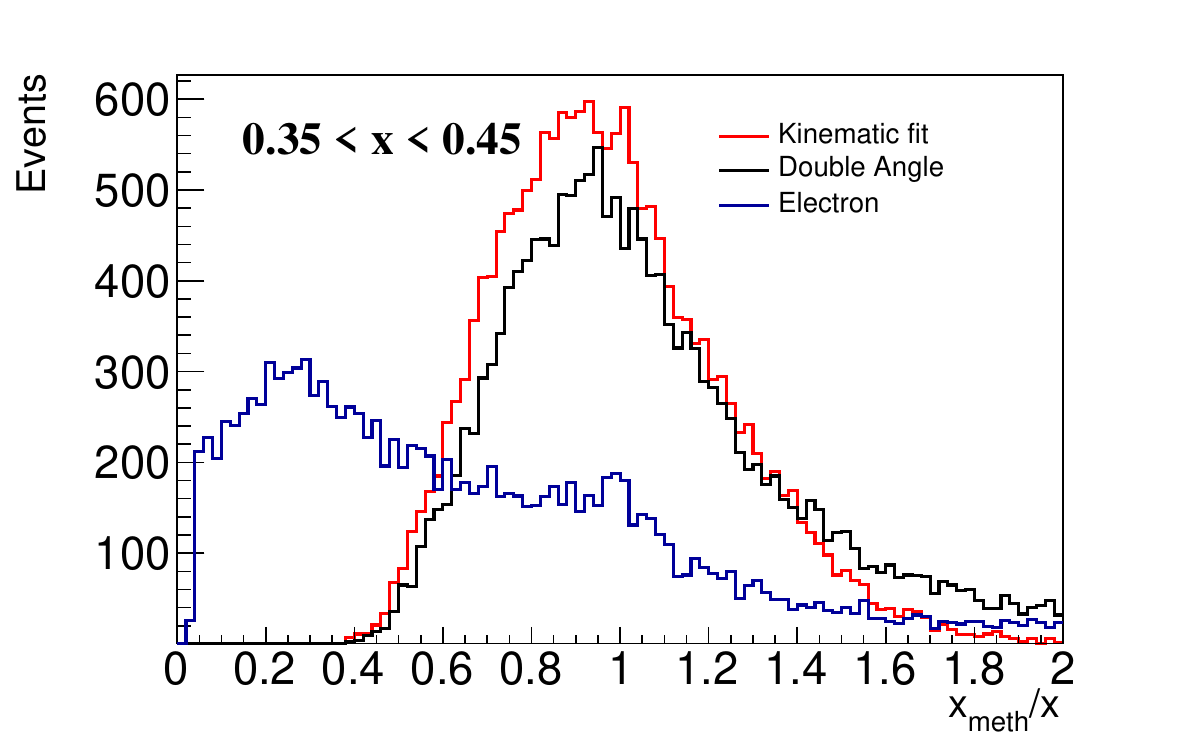}
\includegraphics[scale=.27]{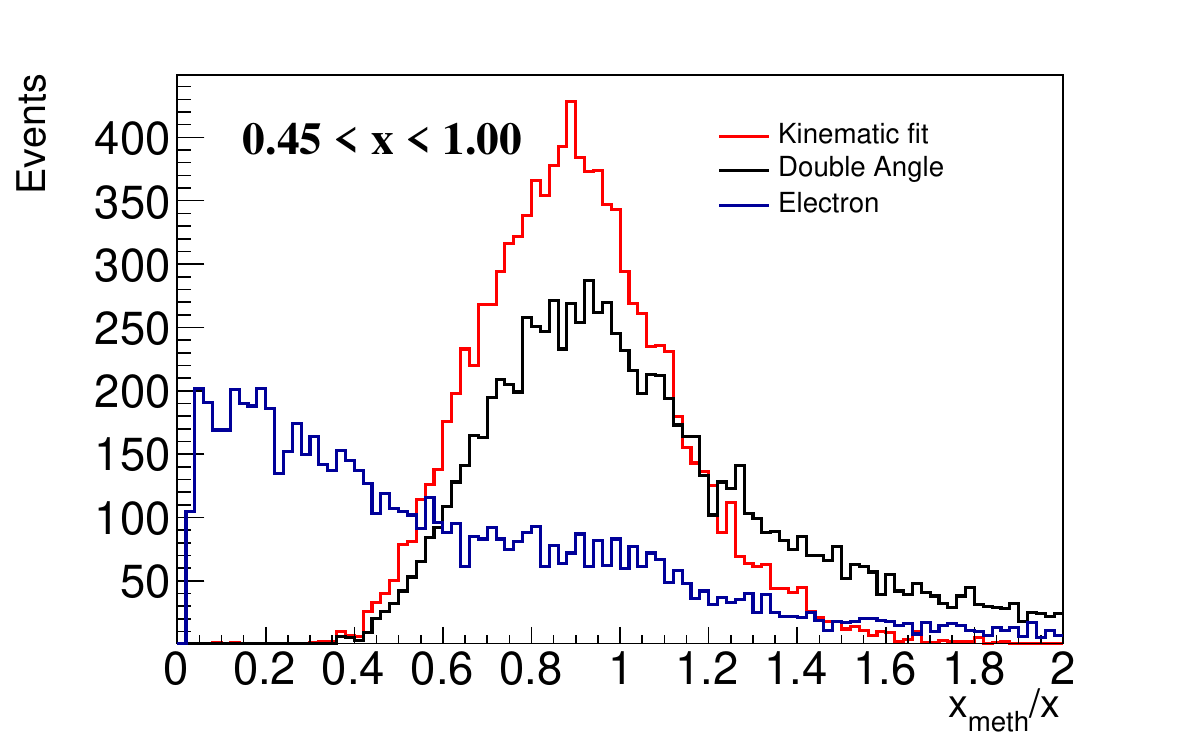}
\caption{Ratio of $x$ from the kinematic fit, double angle and electron reconstruction methods to the true value in different bins of $x$.}
\label{fig:x_bins}

\end{figure}

\begin{figure}
\centering
\includegraphics[scale=.30]{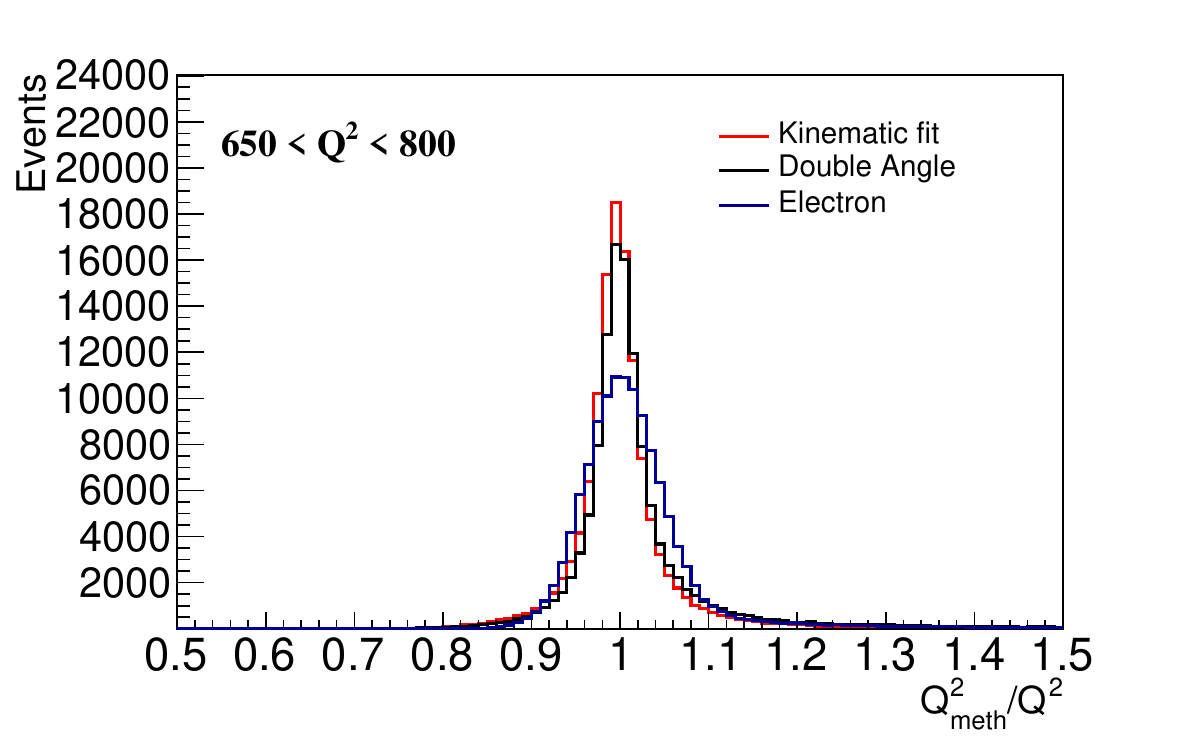}
\includegraphics[scale=.30]{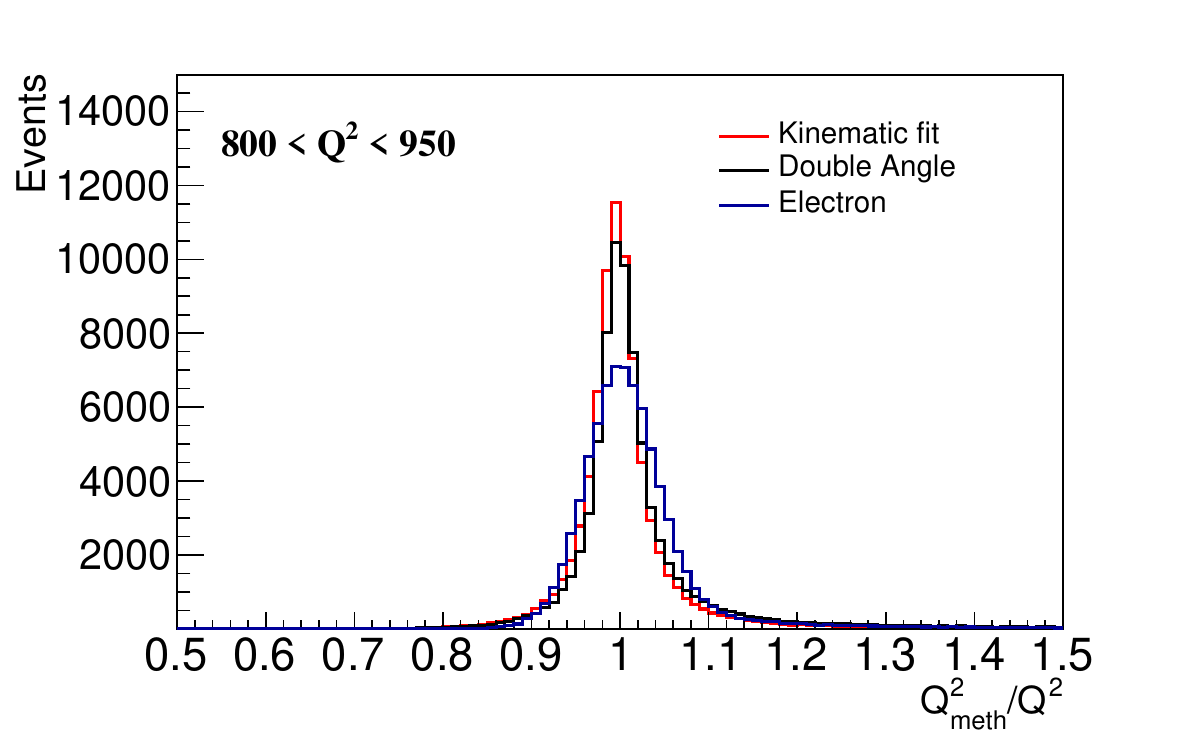}
\includegraphics[scale=.30]{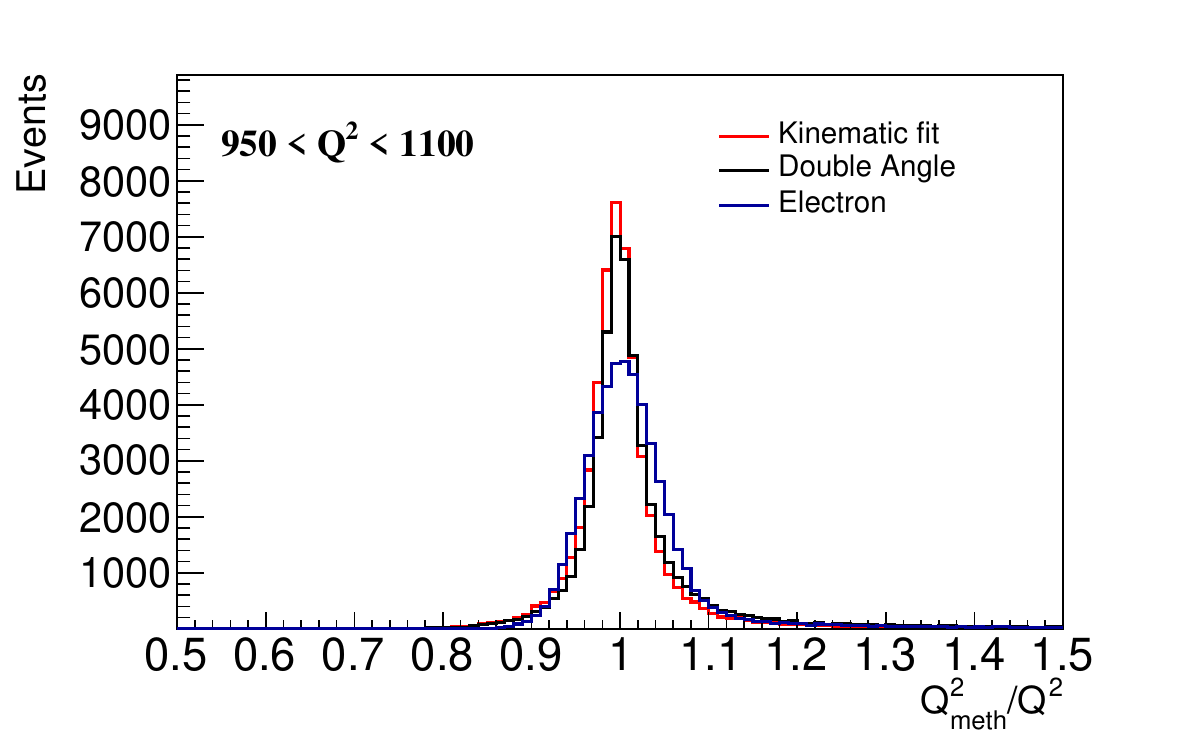}
\includegraphics[scale=.30]{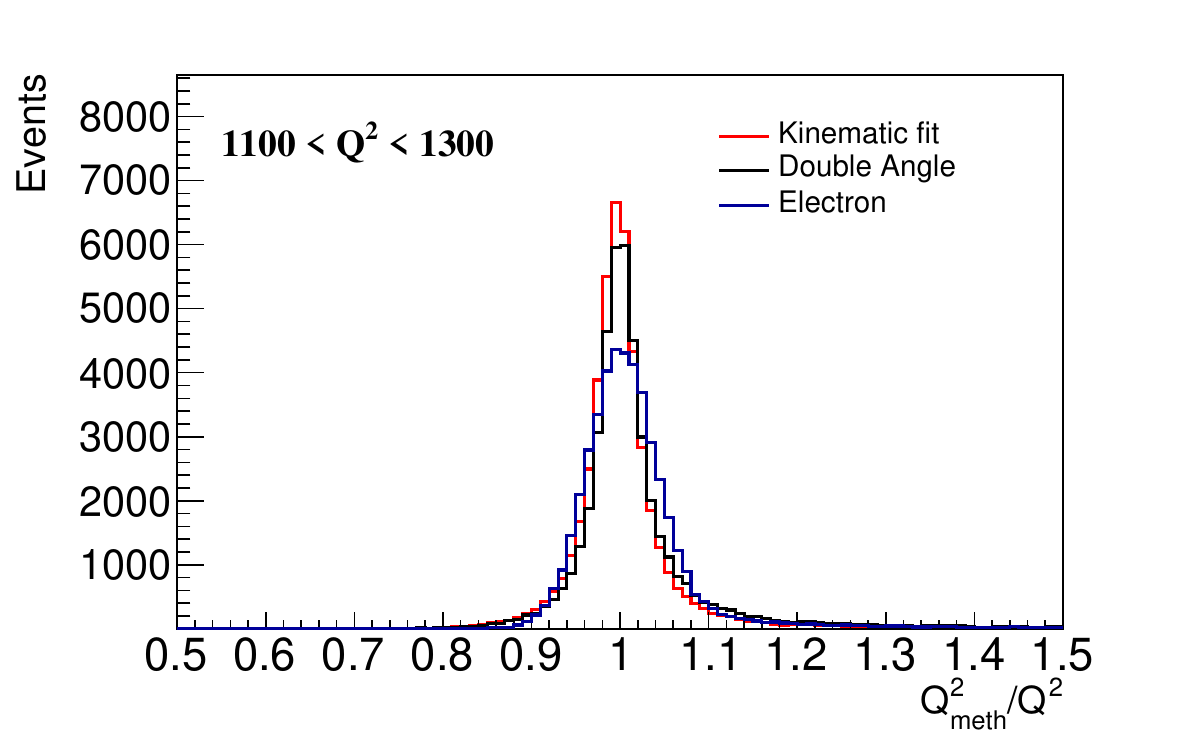}
\includegraphics[scale=.30]{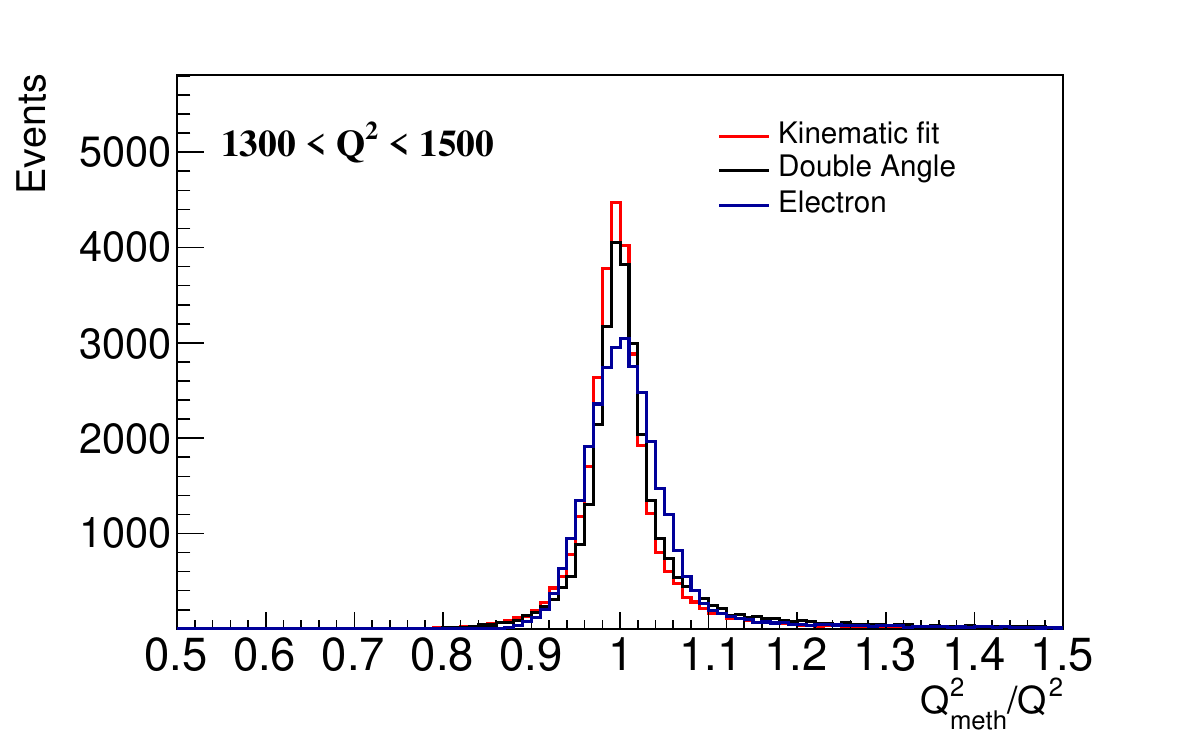}
\includegraphics[scale=.30]{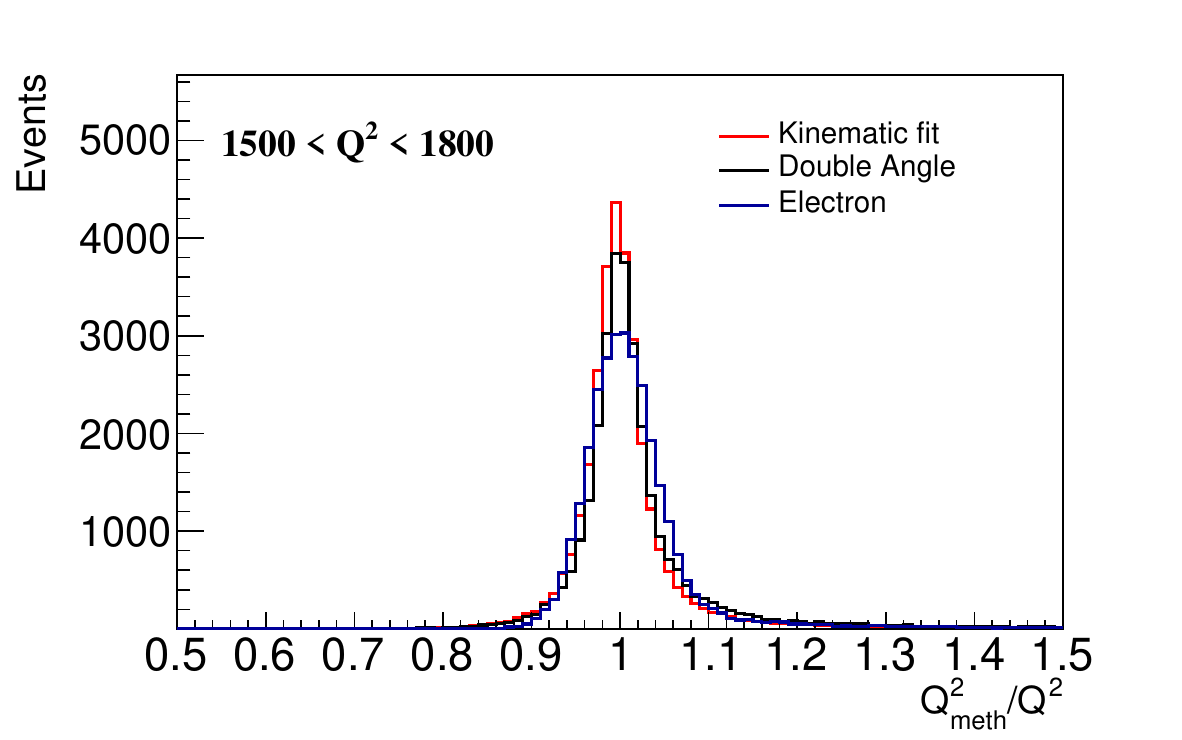}
\includegraphics[scale=.30]{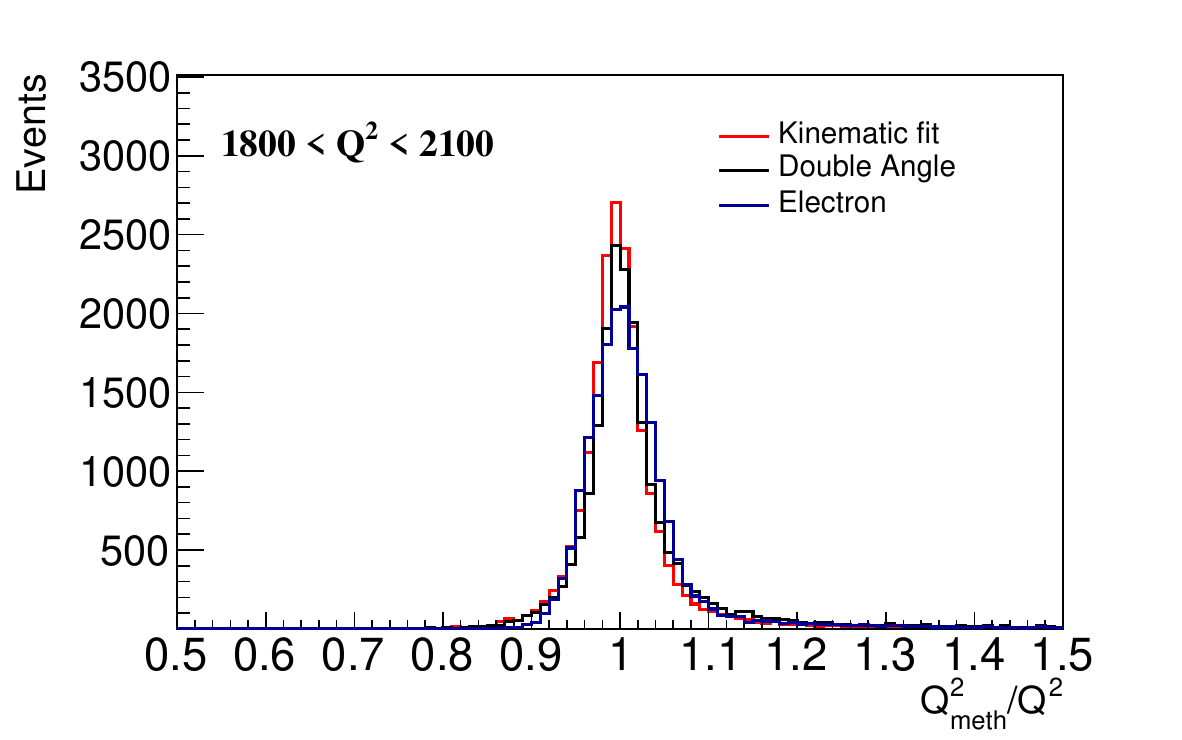}
\includegraphics[scale=.30]{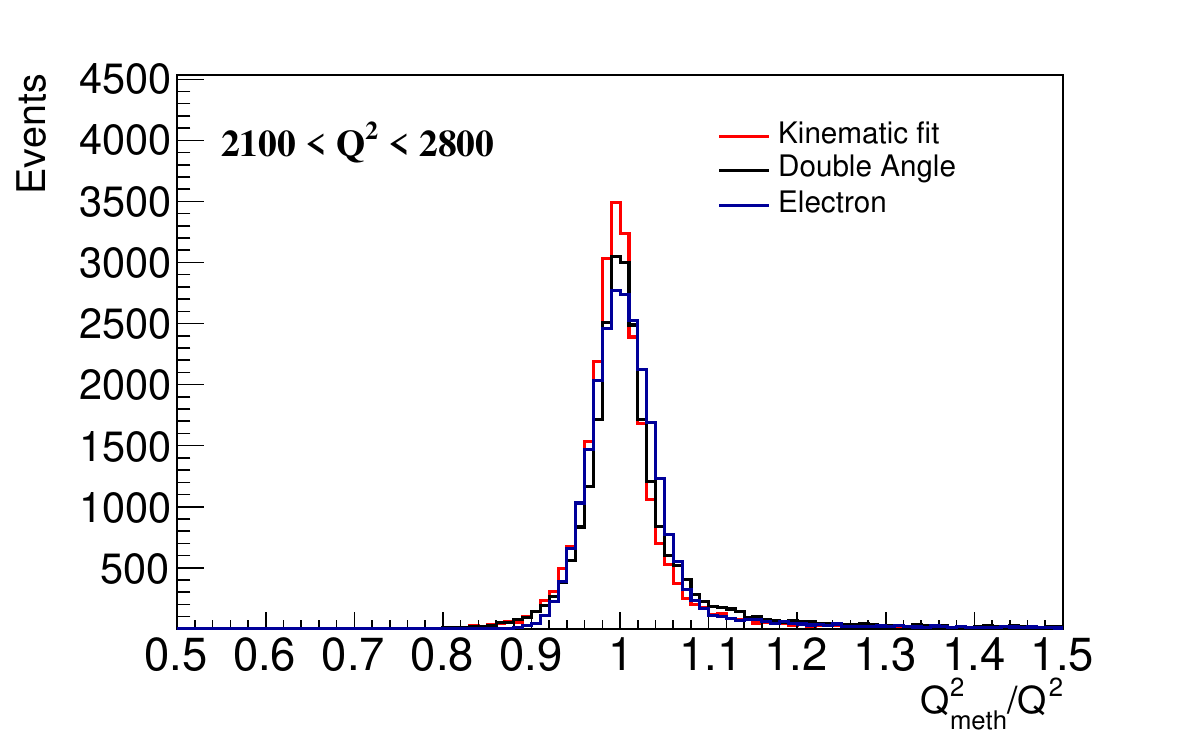}
\includegraphics[scale=.30]{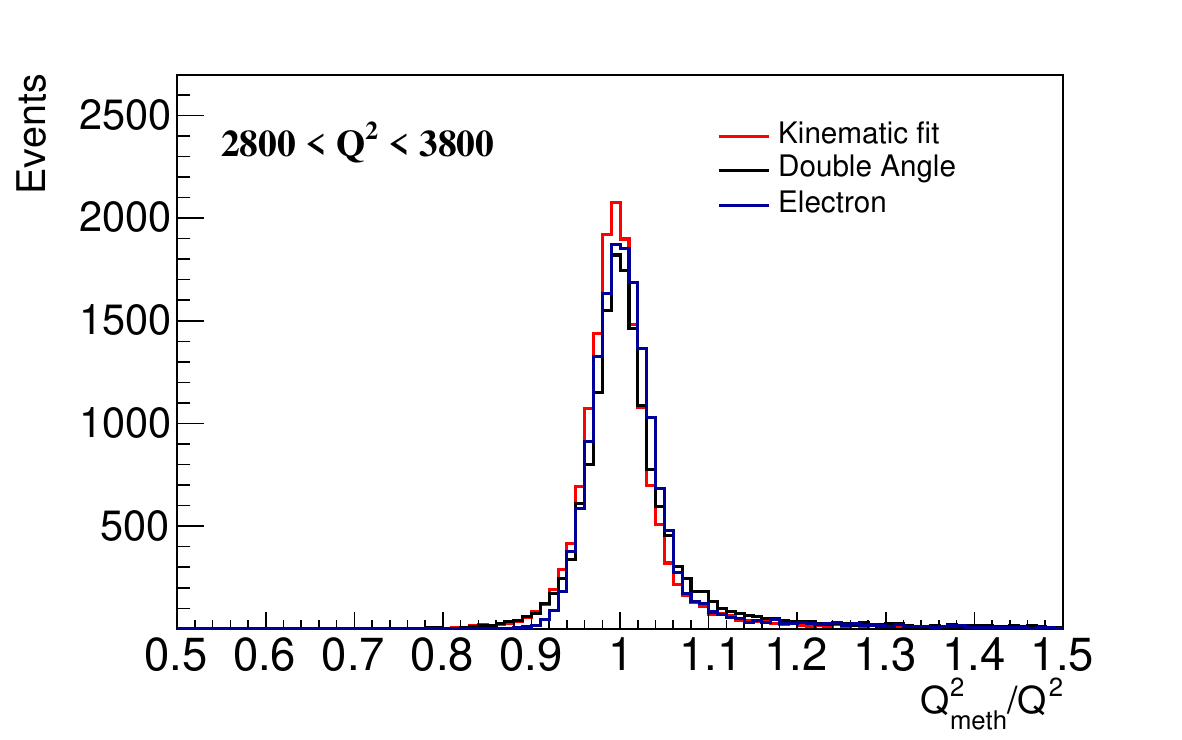}
\includegraphics[scale=.30]{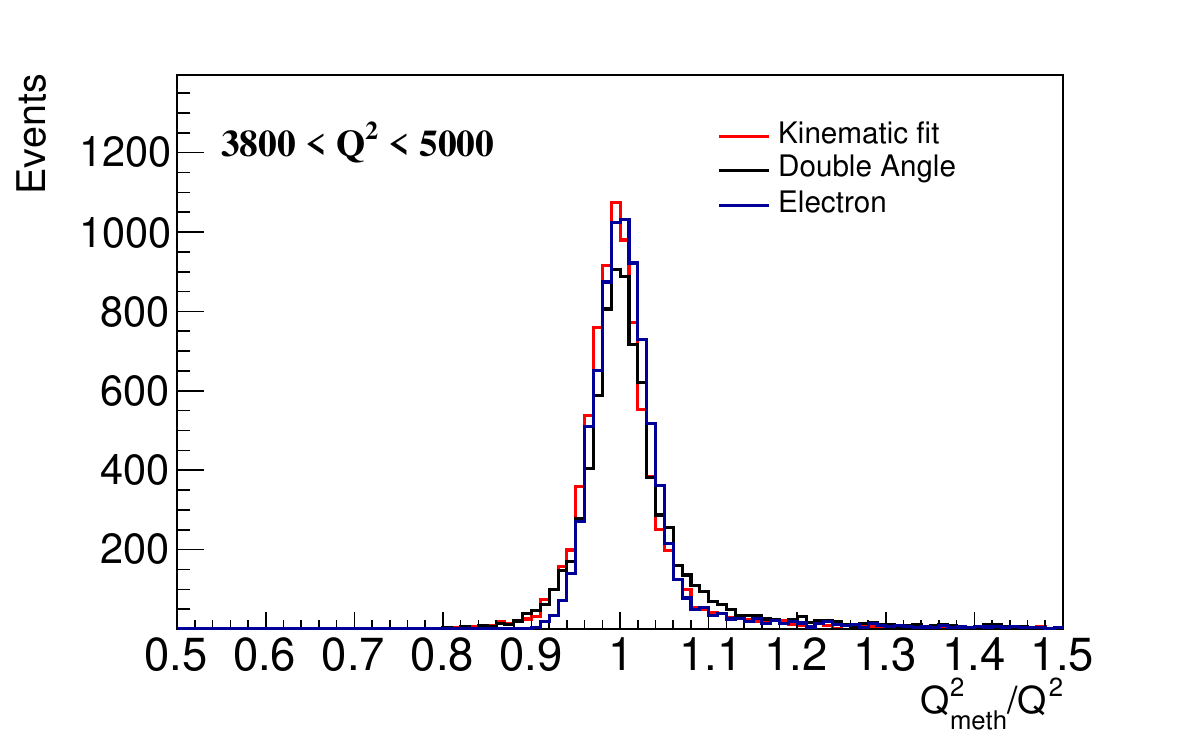}
\includegraphics[scale=.30]{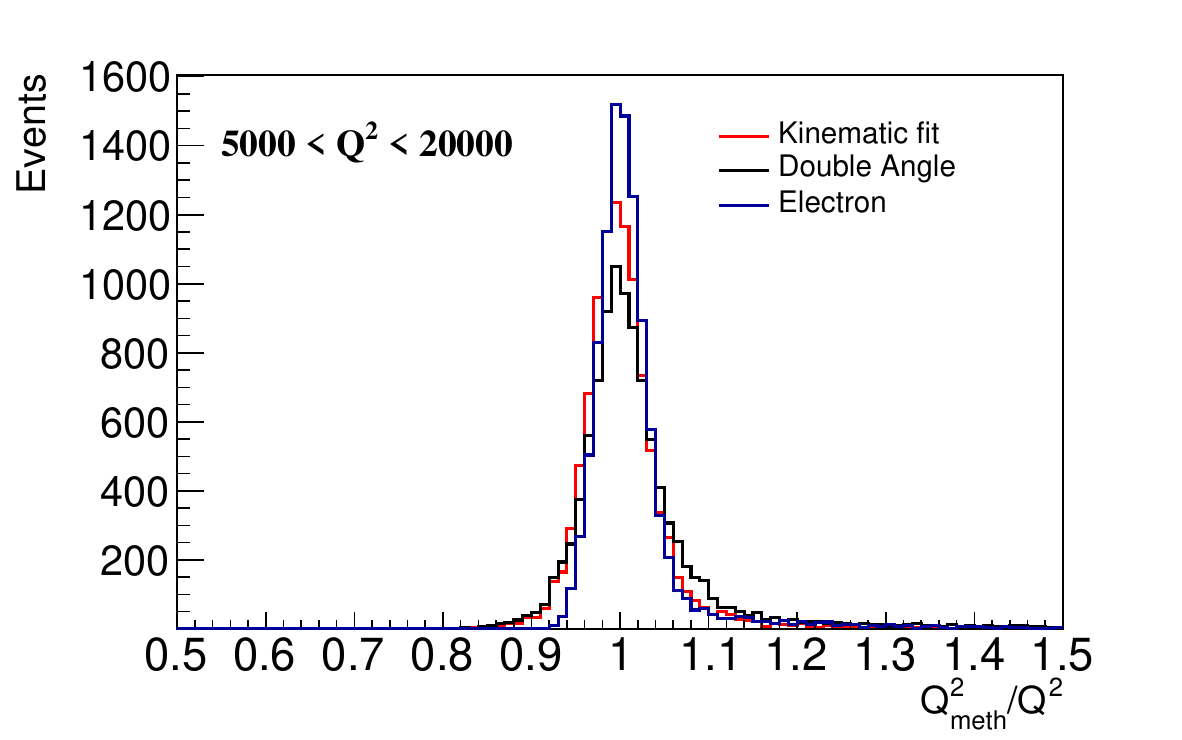}
\caption{Ratio of $Q^2$ from the kinematic fit, double angle and electron  reconstruction methods  to the true value in different bins of $Q^2$.}
\label{fig:q2_bins}

\end{figure}

\begin{figure}
\centering
\includegraphics[scale=.52]{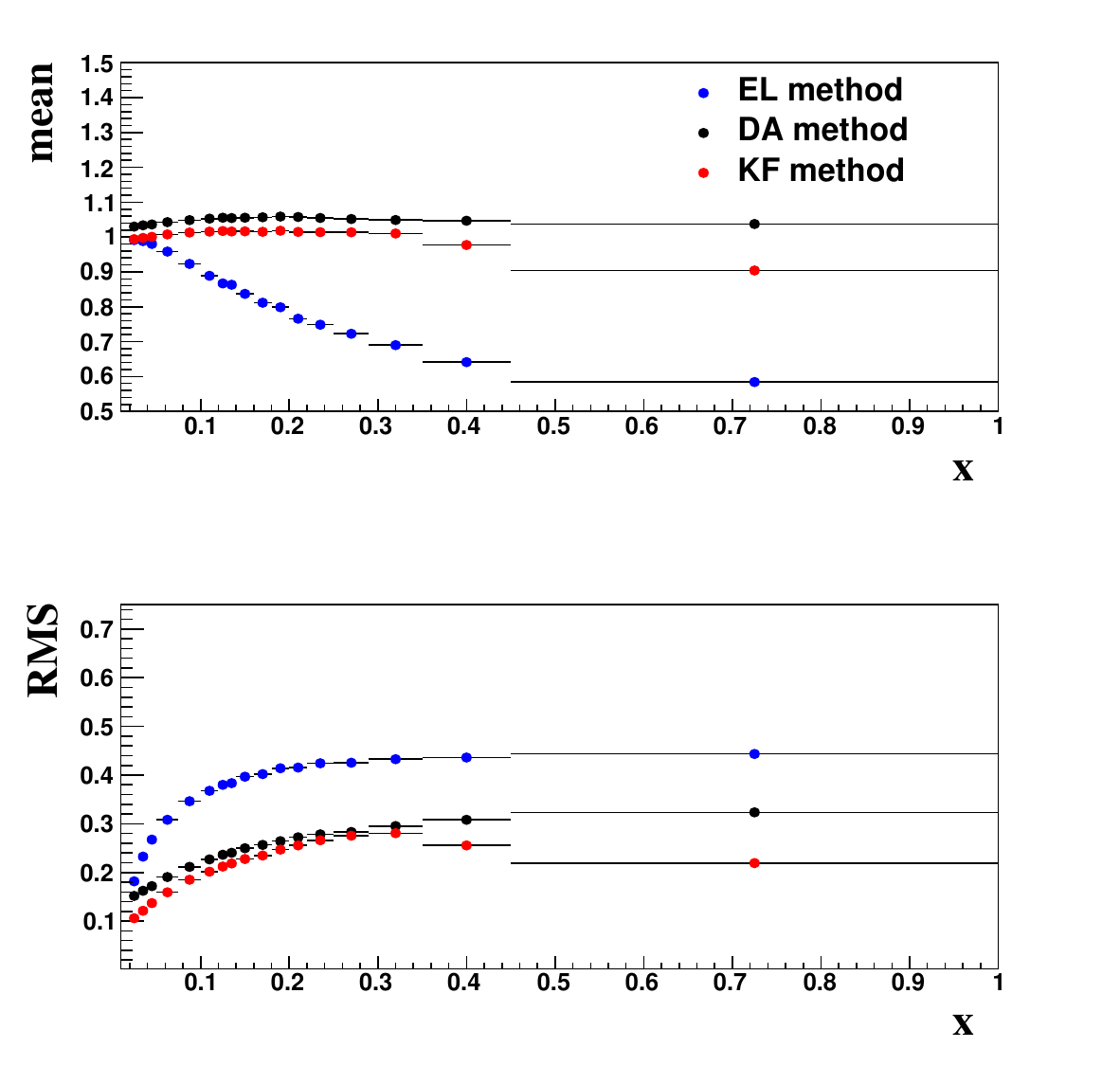}
\caption{Mean and standard deviation (RMS) of the ratio of  $x_{meth}/x$ from the kinematic fit, double angle and electron reconstruction methods. }
\label{fig:xres_bins}

\end{figure}

\begin{figure}
\centering
\includegraphics[scale=.52]{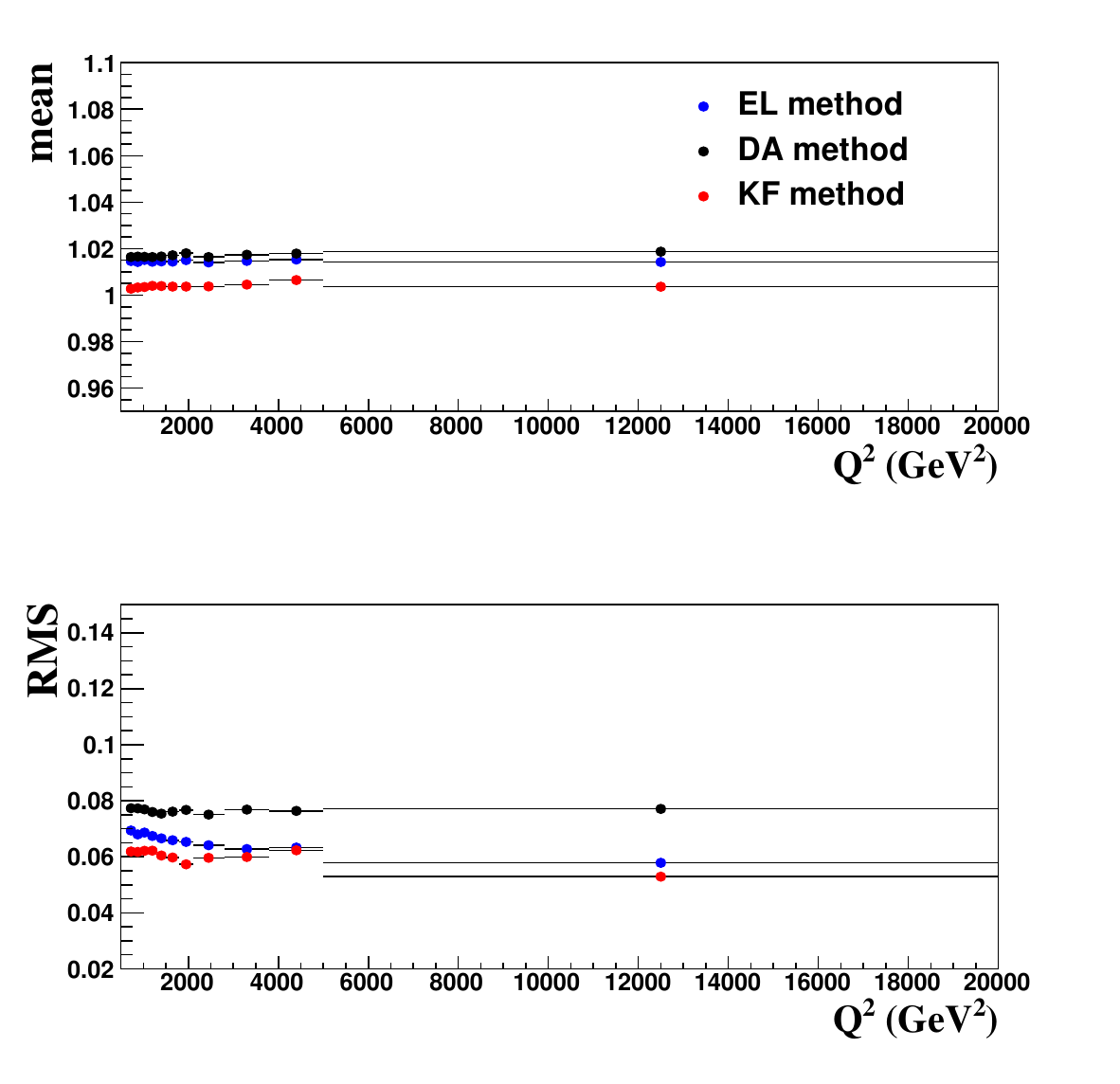}
\caption{Mean and standard deviation (RMS) of the ratio of  $Q^2_{meth}/Q^2$ from the kinematic fit, double angle and electron reconstruction methods.}
\label{fig:q2res_bins}

\end{figure}

The ratios  $ x_{meth}/x$  and $Q^2_{meth}/Q^2$  are studied in the bins of $x$ and $Q^2$ and are shown in Figures~\ref{fig:x_bins} and~\ref{fig:q2_bins} respectively. It is observed that the resolution of $x$ and $Q^2$ obtained from the kinematic fit are better than other methods in the whole $x$-$Q^2$ phase space scanned. \\
The mean and standard deviation (rms) of the ratios are collected in bins of $x$ and $Q^2$  and are shown in Figures~\ref{fig:xres_bins} and~\ref{fig:q2res_bins} respectively. Following detailed observations are made:

\begin{itemize}
	\item The ratio  $x_{meth}/x$ in bins of $x$ is found to have small bias in the kinematic fit and double angle methods. However, for the electron method the $x$ reconstruction becomes biased at higher $x$ values.
	\item The $x$ reconstruction from kinematic fit method is found to be the most precise one as seen from a relatively smaller RMS for the $x_{meth}/x$ ratios in bins of $x$. In the double angle reconstruction, the standard deviation is found to be better than the electron method with increasing $x$. For the electron method a large value of standard deviation is observed and therefore this method is conventionally not recommended at high $x$.
	\item The reconstruction  of $Q^2$ from all three methods has a very small bias, as the mean of the ratios $Q^2_{meth}/Q^2$ is centered at 1 with in 1-3$\%$, in the bins of $Q^2$. 
	\item The RMS of the ratios $Q^2_{meth}/Q^2$ from the kinematic fit method is found to have a least value, implying a better resolution in all of  the bins of $Q^2$.
	
	\end{itemize}

\begin{figure}
\centering
\includegraphics[width=0.48\linewidth]{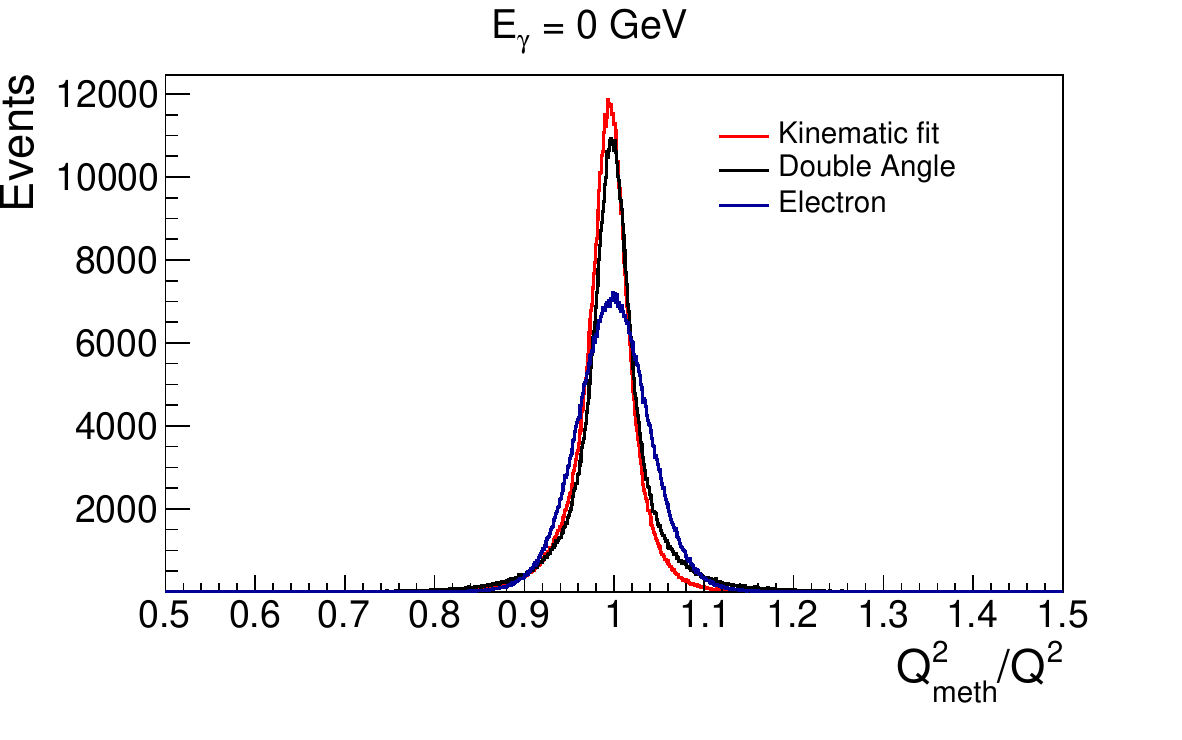}
\includegraphics[width=0.48\linewidth]{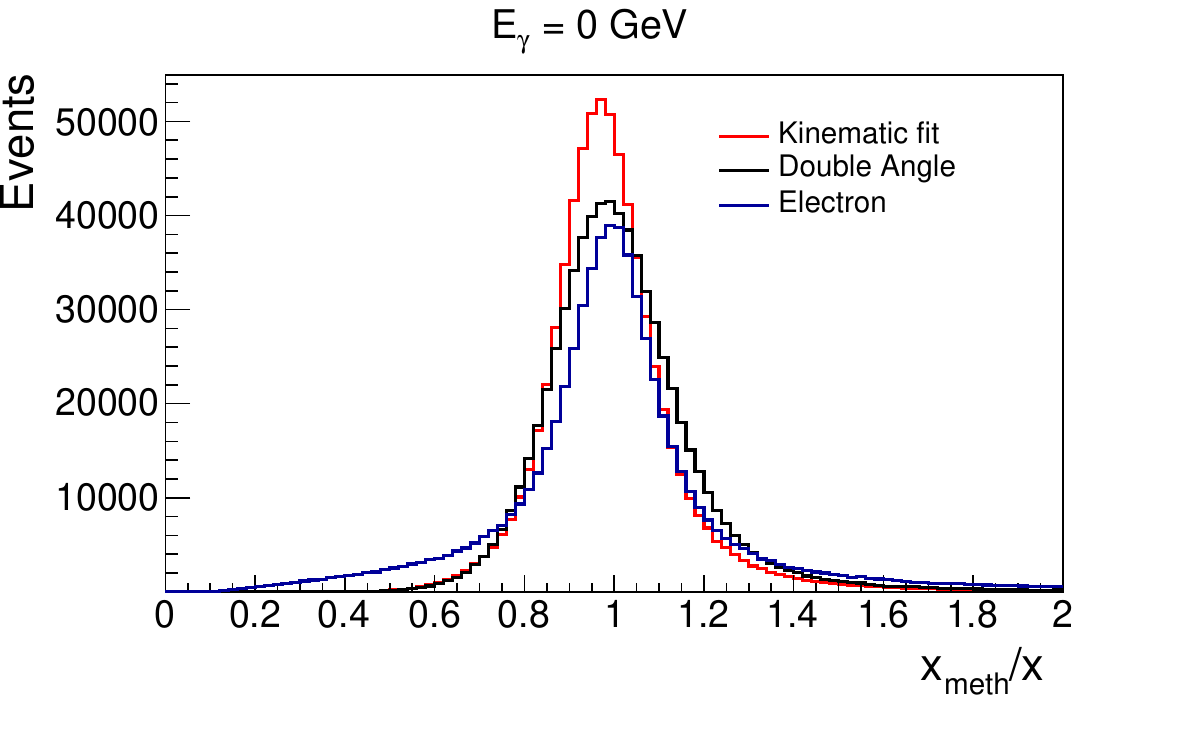}

\includegraphics[width=0.48\linewidth]{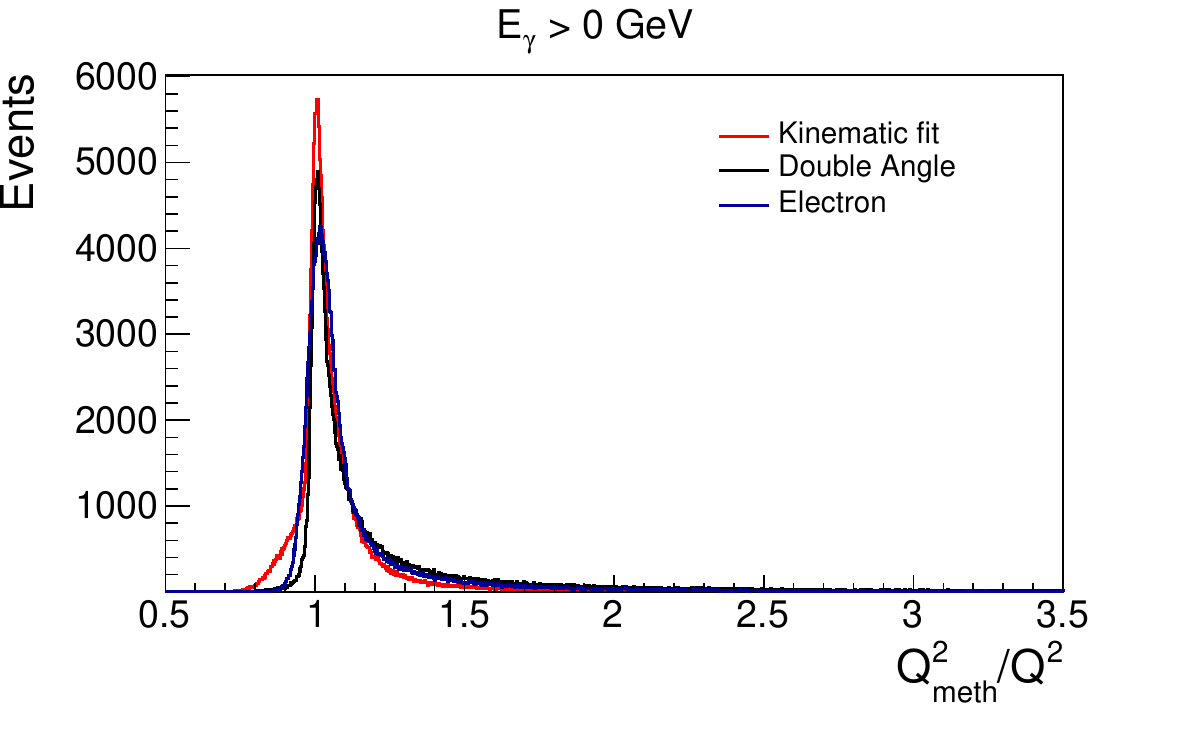}
\includegraphics[width=0.48\linewidth]{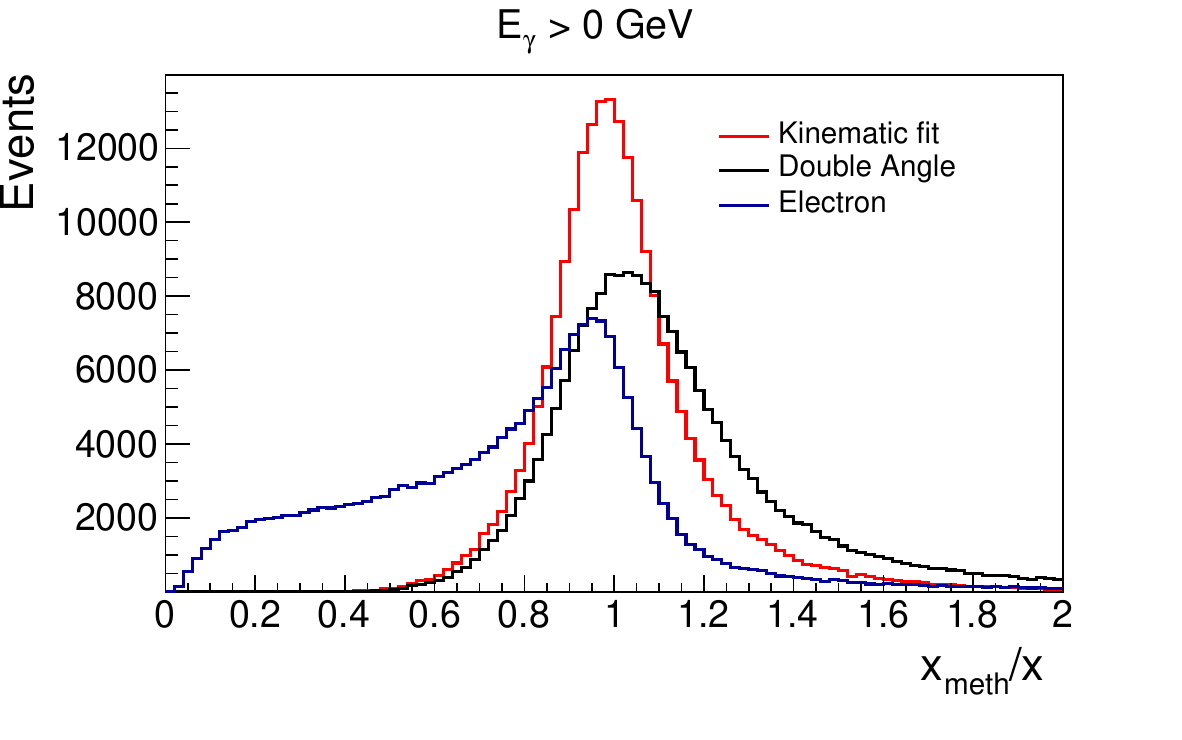}

\includegraphics[width=0.48\linewidth]{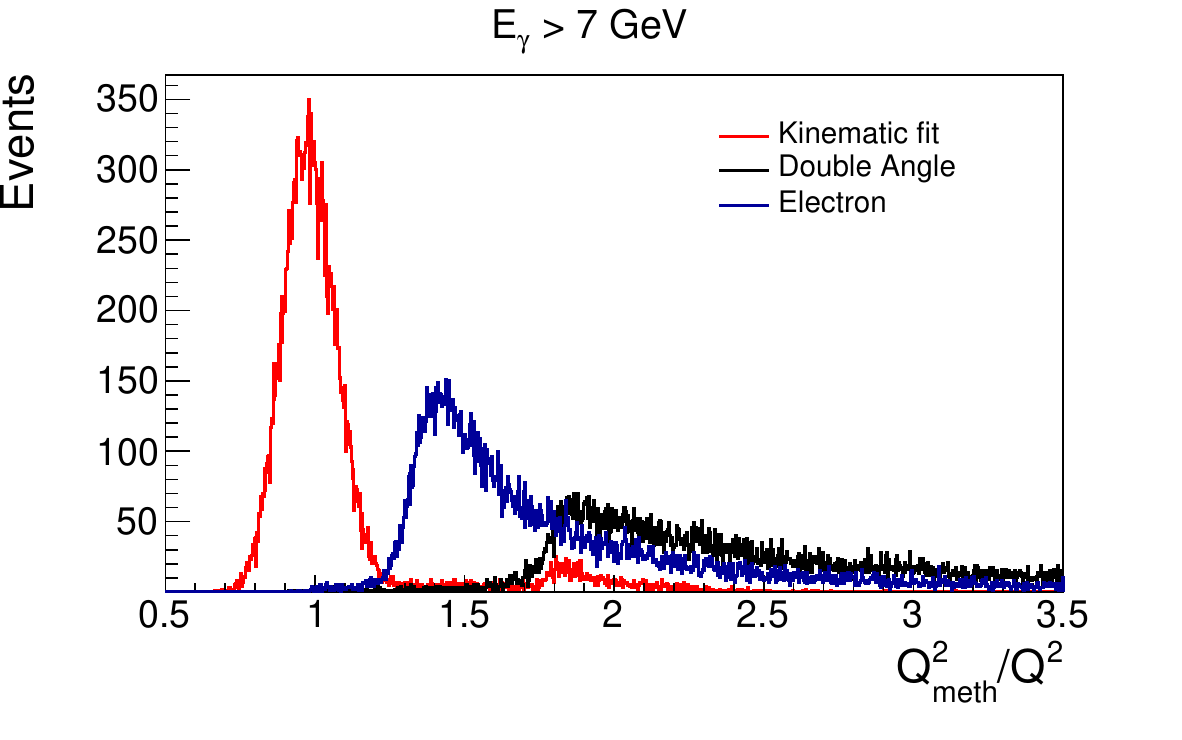}
\includegraphics[width=0.48\linewidth]{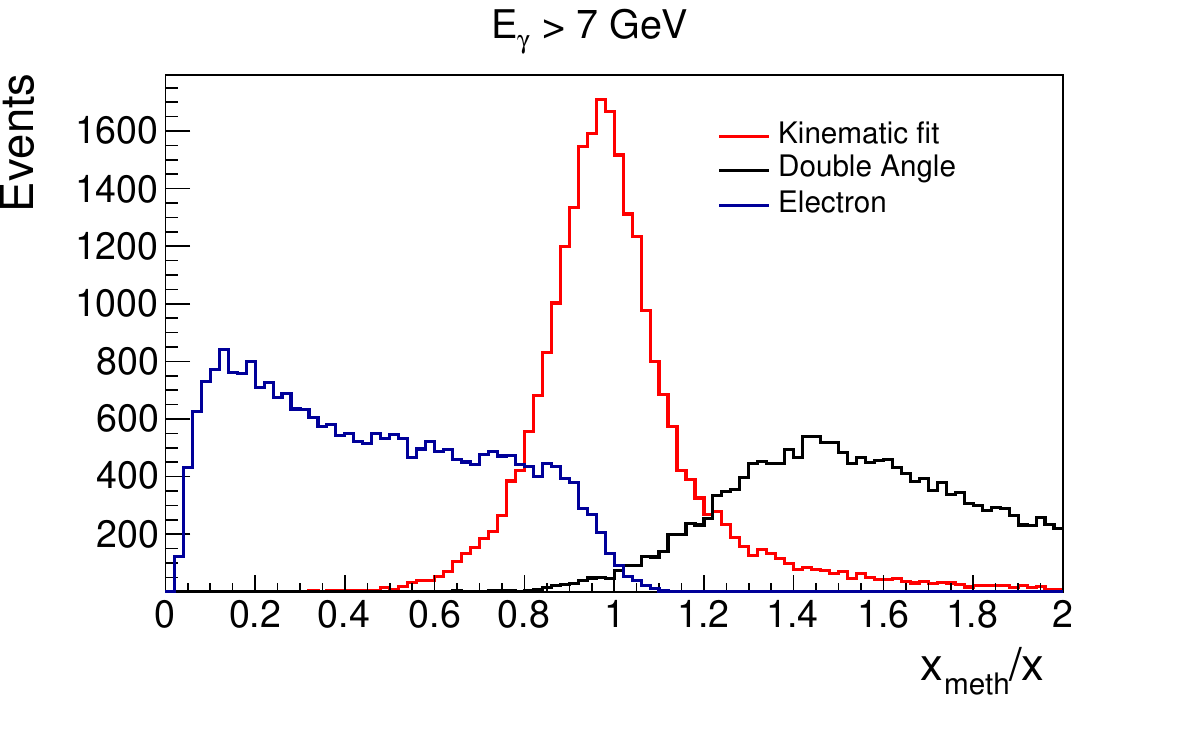}
\caption{(Left) \textbf{Ratio Plots} Ratios $Q^2_{meth}/Q^2$ and $x_{meth}/x$ are shown from electron, double angle and kinematic fit reconstruction methods. Different groups are made on the basis of energy of the ISR. Starting from the top row, the three rows are described as events with : no ISR, $E\gamma > 0$~GeV and $E\gamma > 7$~GeV.}
\label{fig:xq2_diffEgam}
\end{figure}

The presence of an ISR can lead to a very biased reconstruction of the scaling variables~\cite{ref:jbmk2},~\cite{ref:jbmk}. A comparison on the resolution of  $x$ and $Q^2$ reconstruction has been shown for the three different categories of events based upon the true ISR energies : $E_\gamma = 0$ (No ISR), $E_\gamma > 0 $ and $E_\gamma > 7 $.
Figure~\ref{fig:xq2_diffEgam}  shows the ratios $ x_{meth}/x$  and $Q^2_{meth}/Q^2$  for the three different categories of events.  For all the three cases, the kinematic fit method is observed to offer a robust reconstruction of $x$ and $Q^2$.

For the first case where  no ISR photon is present in the event, kinematic fit is doing better due to the full detector information taken into account.
For the second case, when an ISR photon is present, the electron and double angle methods have biased $x$ and $Q^2$ reconstruction as is visible from the  long tails in the  $ x_{meth}/x$  and $Q^2_{meth}/Q^2$ ratios. The kinematic fit method can estimate the ISR photon energy and take it into account in the $x$ and $Q^2$ reconstruction. The bias in $x$ and $Q^2$ reconstruction from the electron and double angle methods increases as the value of the ISR photon energy increases. 

For the cases where an ISR photon with  $E_\gamma >$ 7~GeV is emitted from the initial state electron, the reconstruction of $Q^2$ can be wrong by 50\% and 25\% from double angle and electron methods respectively. This is also observed from the ratios $ x_{meth}/x$  and $Q^2_{meth}/Q^2$ as shown in Figure ~\ref{fig:xq2_diffEgam} for the case with $E_\gamma >$ 7~GeV. 

The ISR events with large $E_\gamma$ can be discarded by putting a cut on total $E-Pz$ of the event: for a lower bound of 30~GeV of total $E-Pz$, all ISR with $E_\gamma$  > 12.5 GeV can be rejected.  For the events with  $E_\gamma$ below this value, the reconstruction of $x$ and $Q^2$ can still be wrong by more that 25\% from the conventional methods. For these events, the kinematic fit would play a vital role in the correct reconstruction of kinematic variables.

\section{Summary}
This paper successfully demonstrates the use of kinematic fit method to reconstruct the kinematic variables. The method is tested on the high $Q^2$ simulated NC scattering of electron on protons at HERA
energies. A kinematic fit is performed which uses the full detector potential in the form of all four directly measured quantities in the final state as input, namely energy and angle of the electron and $\delta_{h}$ and $P_{T,h}$ of the hadronic final state respectively. As a result of using the full event information, the scaling variables $x$ and $Q^2$ reconstructed from the kinematic fit are observed to have a good resolution which is better than double angle and electron methods. \\
The kinematic fit technique is found to be able to reconstruct the  energy of ISR (E$_\gamma$), which otherwise goes undetected down the
beam pipe.  For the events where an  ISR photon is reconstructed from the kinematic fit,  the resolution offered to the scaling variables $x$ and $Q^2$ is found to be preserved. \\
The results from the kinematic fit method, however, rely on the in depth knowledge of the detector response in collecting the event information. In the future, one may also try a more complex likelihood function instead of the one used in Equation~\ref{eq:Pos}, taking into account the correlations between different final state quantities. A further  improvement can be anticipated by using a different prior function for $E_\gamma$, which could reproduce the ISR spectrum for $E_\gamma$ < 1~GeV. 
Presenting this method, we  hope to use the full potential of the planned future lepton hadron experiments at very high energies.  In the future we plan to study this method for the lower $Q^2$ kinematic phase space at HERA and EIC energies and use this method for other experiments as well.  Recently Neural Networks have been used to reconstruct the scaling variables $x$, $y$ and $Q^2$ in the NC DIS events~\cite{ref:NN1}-~\cite{ref:NN2}. It will be very interesting to do a direct comparison of the two methods in the same kinematic phase space as a future study.

 \section*{Appendix A : Example Kinematic Fit Results}
 Two example fits of the kinematic variables using BAT~\cite{ref:bat}  are presented, with one event having a high energy ISR photon and the other with no ISR.
The $x$, $y$ and E$_\gamma$ values from the Kinematic Fit are given along with their standard deviation and compared to the true values. The values obtained from the double angle and electron methods are also shown.

 \subsection*{Example 1}
\begin{figure}
\centering
\includegraphics[scale=0.4]{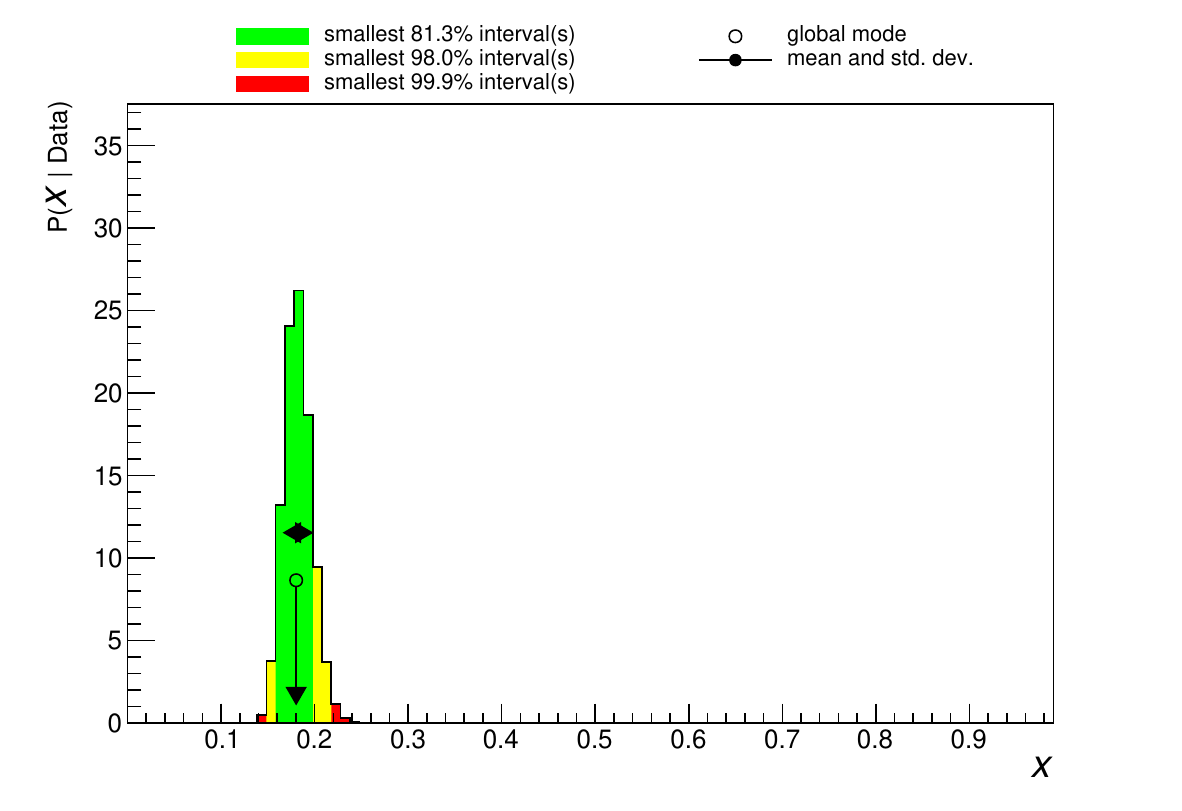}
\includegraphics[scale=0.4]{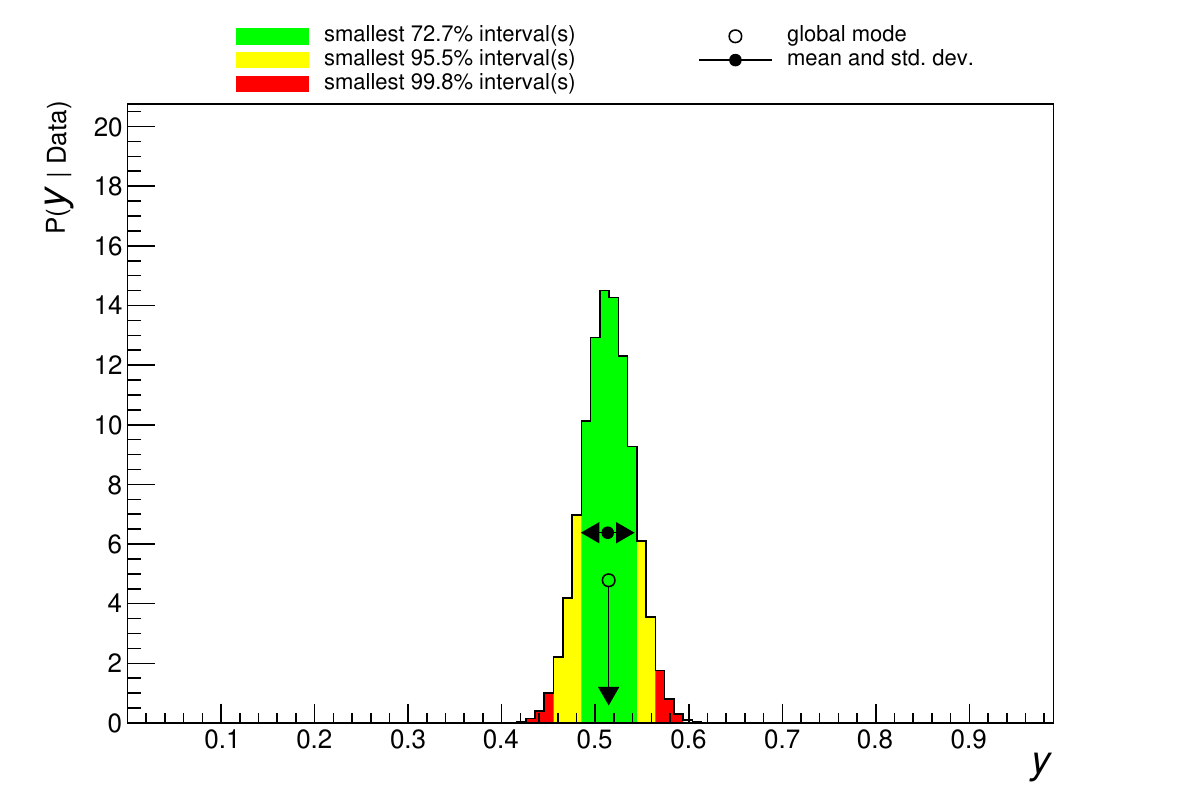}
\includegraphics[scale=0.4]{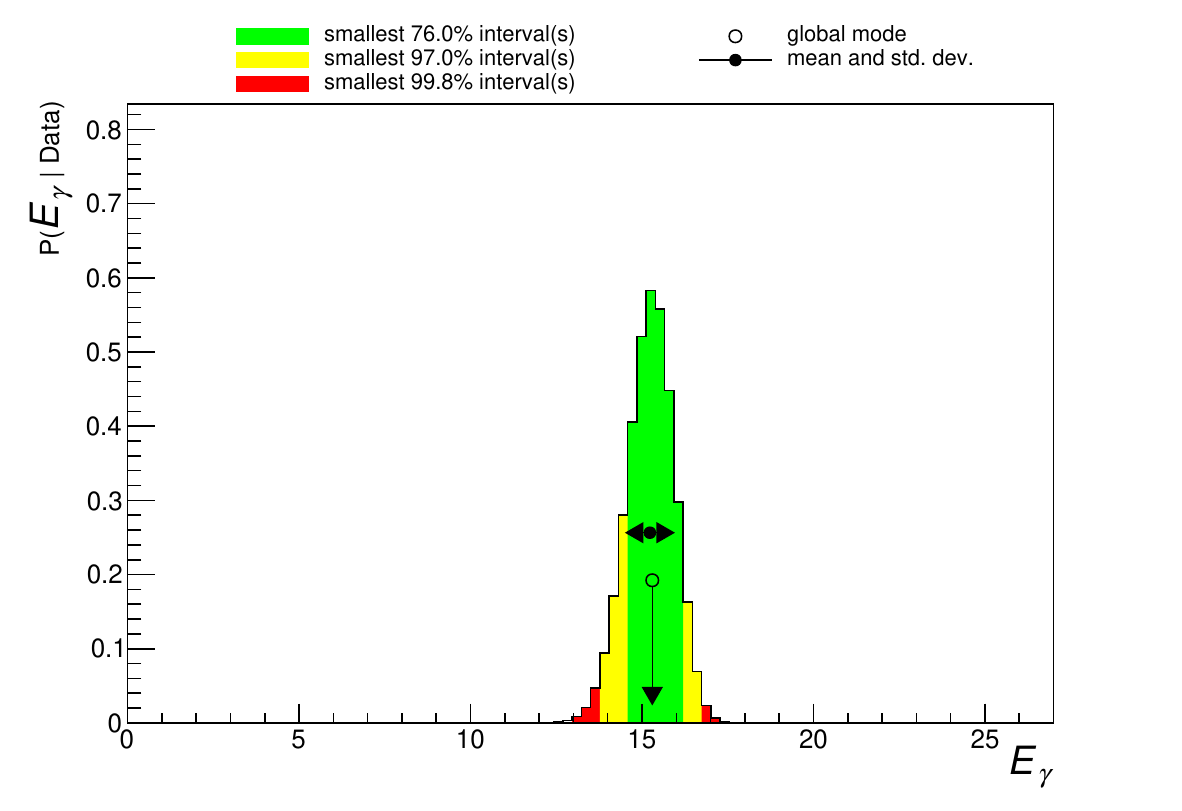}
\includegraphics[scale=0.4]{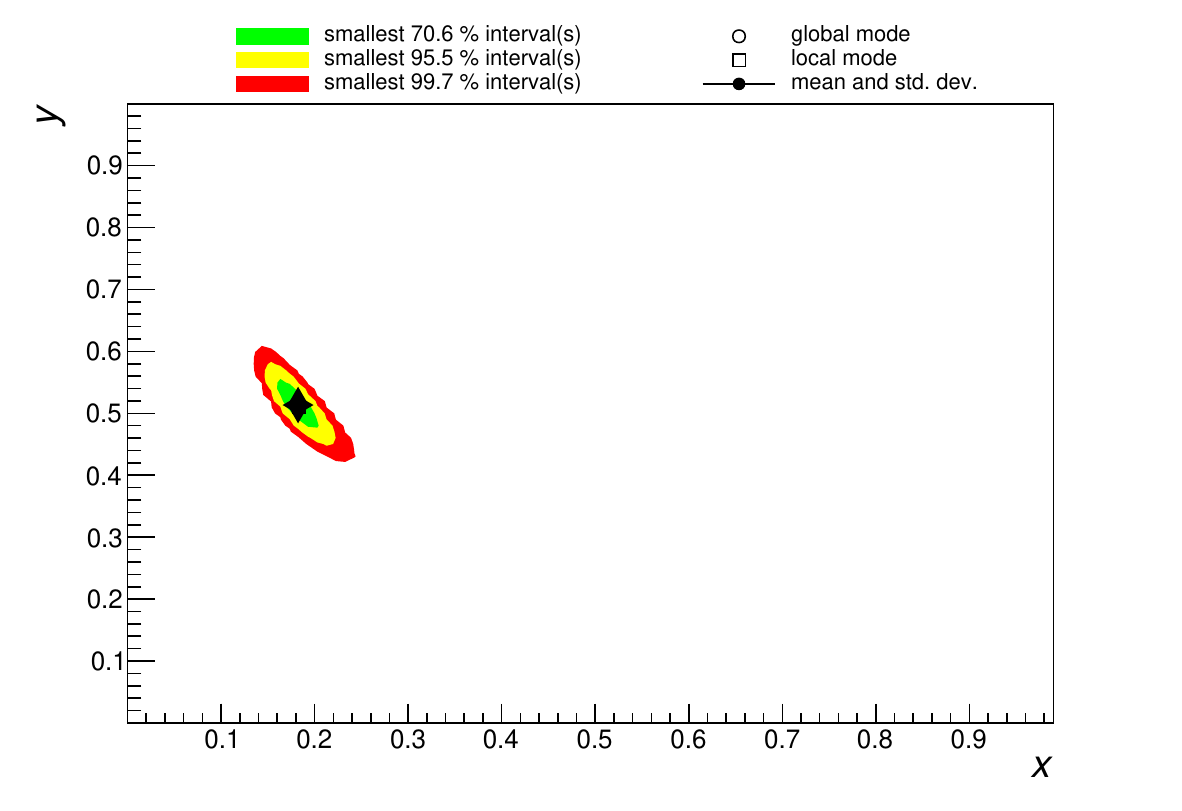}
\includegraphics[scale=0.4]{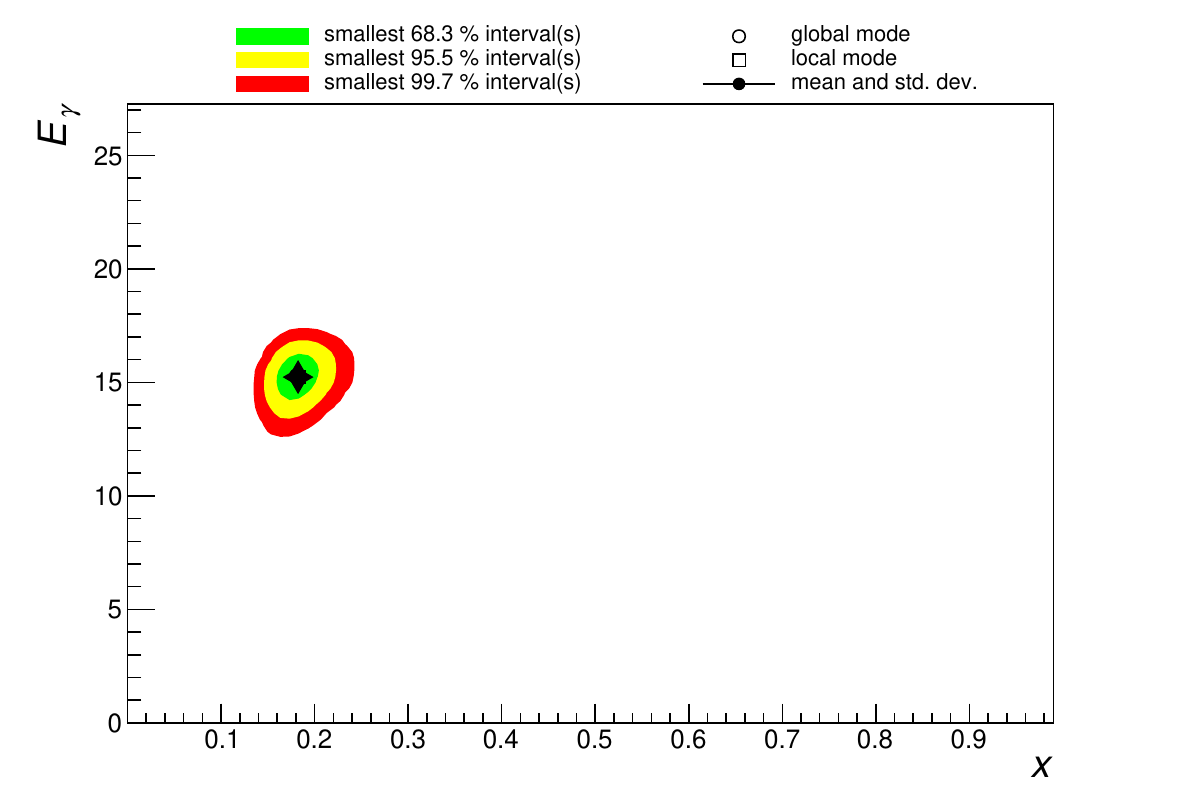}
\includegraphics[scale=0.4]{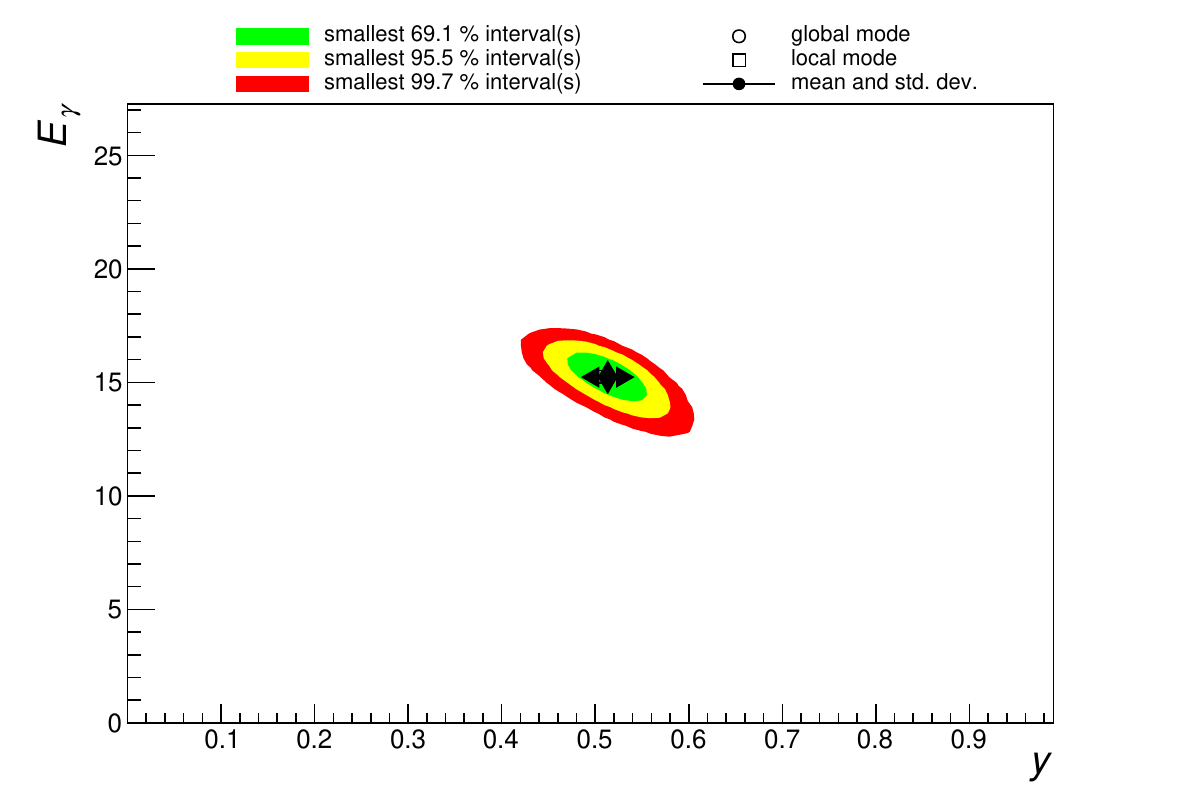}
\caption{Example 1 - Each plot shows marginalized distributions of the parameters $x$, $y$ and  E$_\gamma$ obtained from the Kinematic Fit. The global mode, local mode  and mean are also shown in the plots.}
\label{fig:Eg1}
\end{figure}

This event has a high energy ISR photon and is the type of event that is typically difficult to reconstruct with the standard reconstruction techniques. The calculated and generated values of  $Q^2$, $x$ and $y$ are shown in Table~\ref{tab:Eg1} as well as the photon energy. 
The values of $x$, $y$ and E$_\gamma$ at the global mode of the Kinematic Fit are given with uncertainty taken as the standard deviation of the marginalized distribution.
In addition, the one dimensional and two dimensional marginalized distributions of the parameters  $x$, $y$ and  E$_\gamma$ obtained from the Kinematic Fit are shown in Figure~\ref{fig:Eg1}. This Figure also shows the position of global mode, local mode and three different credible intervals of the marginalized distributions.   
 \begin{table}[h!]
\centering
 \begin{tabular}{|c|c| c| c|c|} 
 \hline
  & $Q^2$(GeV$^2$) & $x$ & $y$ & E$_\gamma$(GeV) \\ [0.5ex] 
 \hline
 Generated (with ISR  correction)  & 4126 & 0.212 &  0.459 &16.1\\
 \hline
 KF (Global mode)  & 4188 & 0.180 +- 0.015 & 0.514 +- 0.027 & 15.3 +- 0.7\\
 \hline
 EL &9830  &0.125  & 0.774&-\\
 \hline
 DA &  21135   &0.405  & 0.515&-\\
 \hline
 \end{tabular}
 
\caption{Generated $Q^2$, $x$ and $y$ for the first example event. The values of $x$, $y$ and  E$_\gamma$ from the Kinematic Fit are quoted at the global mode with their respective standard deviation from the marginalized distribution.}
\label{tab:Eg1}
\end{table}

 \subsection*{Example 2}
\begin{figure}
\centering
\includegraphics[scale=0.4]{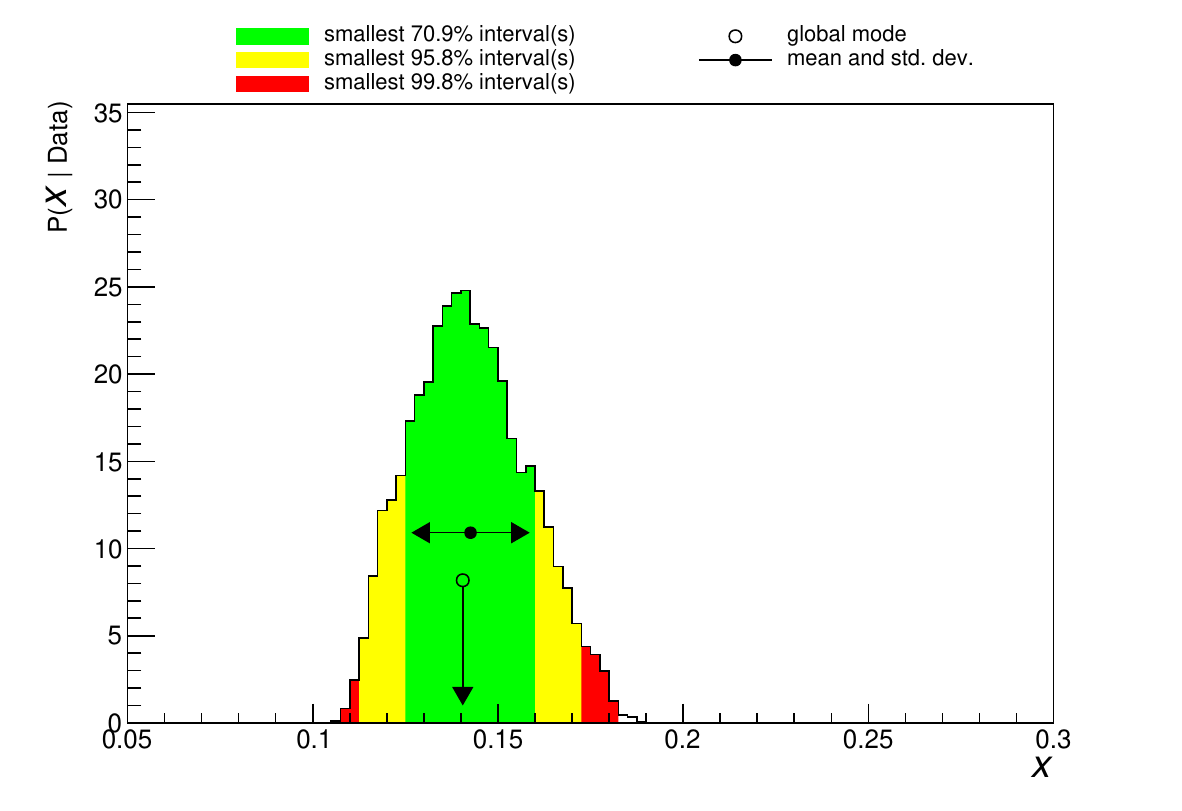}
\includegraphics[scale=0.4]{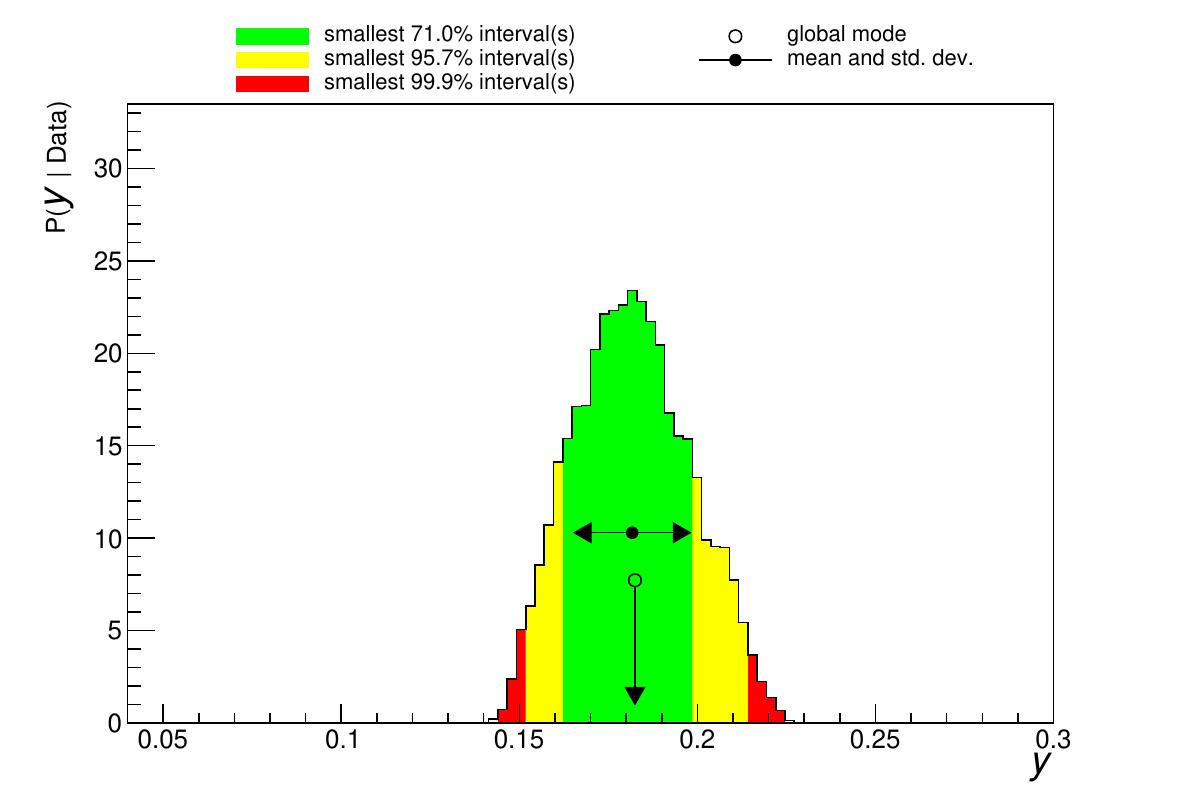}
\includegraphics[scale=0.4]{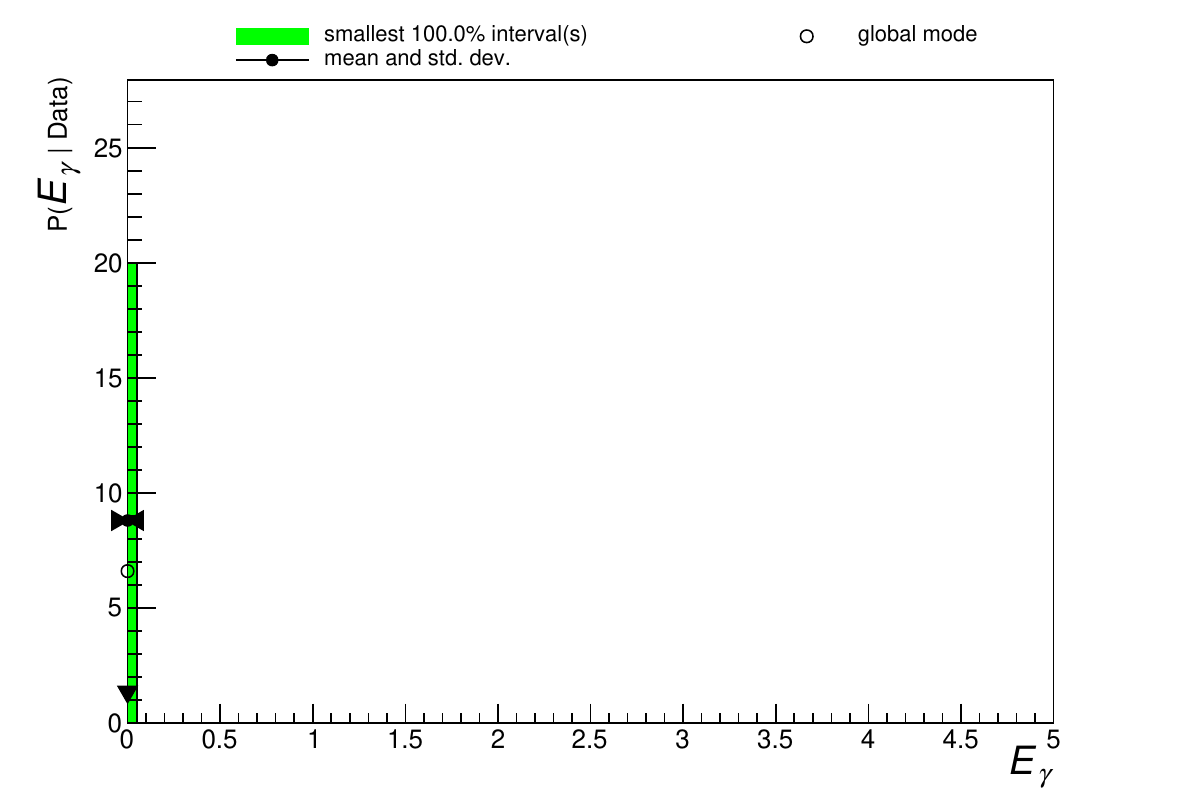}
\includegraphics[scale=0.4]{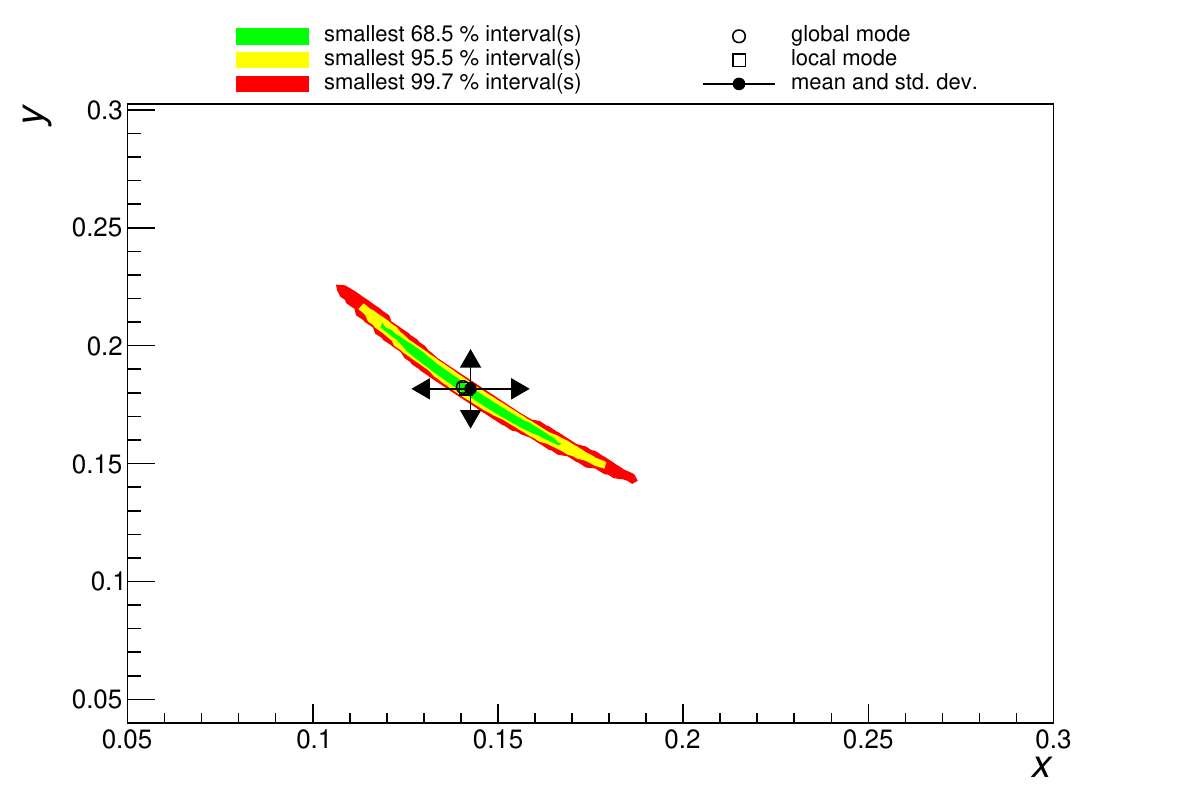}
\includegraphics[scale=0.4]{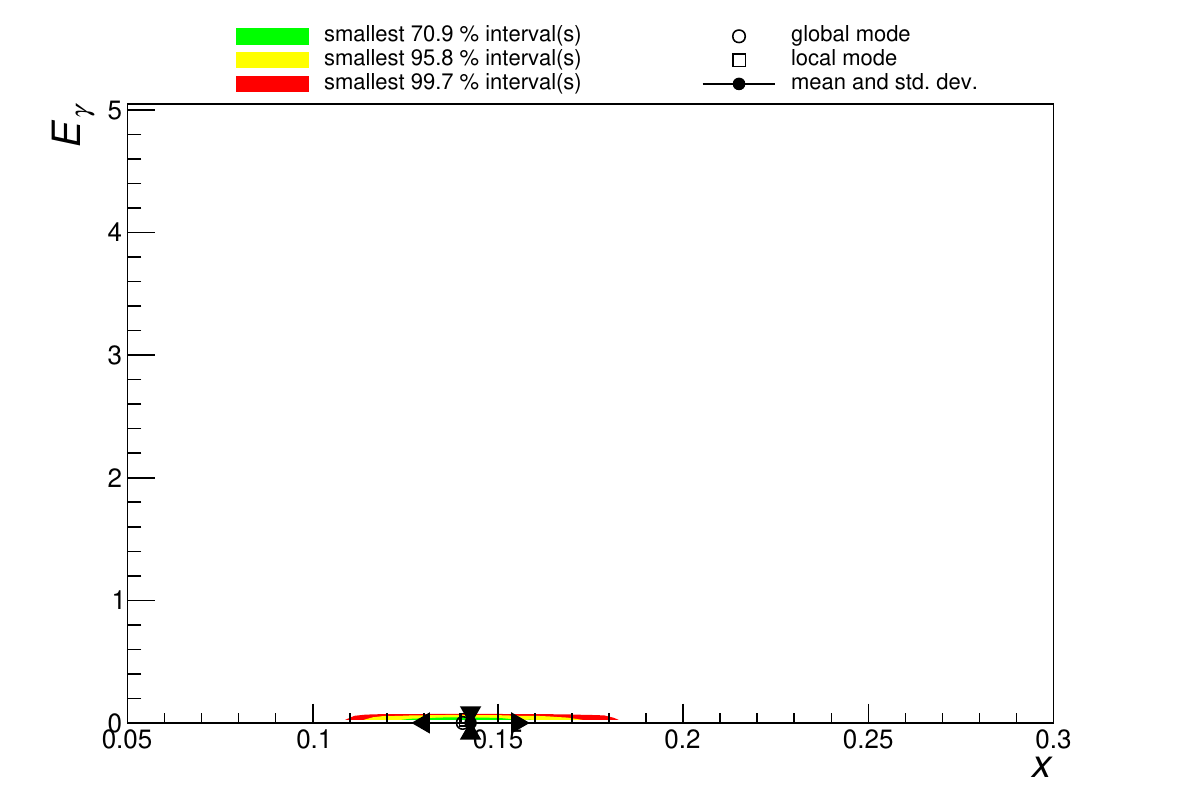}
\includegraphics[scale=0.4]{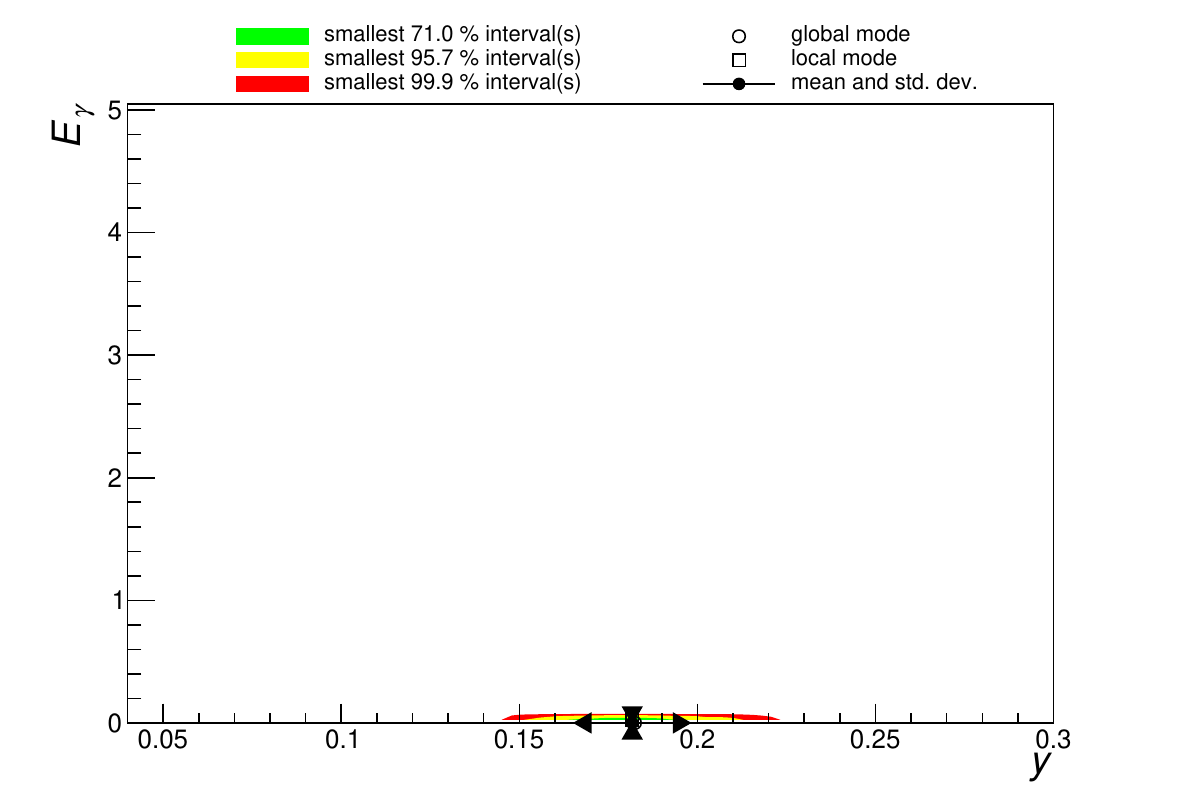}
\caption{Example 2 - Each plot shows marginalized distributions of the parameters $x$, $y$ and  E$_\gamma$ obtained from the Kinematic Fit. The global mode, local mode  and mean are also shown in the plots.}
\label{fig:Eg2}
\end{figure}

The second example is an event which has no ISR. The results of the Kinematic Fit for this event is given in Table~\ref{tab:Eg2}. The one dimensional and two dimensional marginalized distributions of the parameters  $x$, $y$ and  E$_\gamma$ obtained from the Kinematic Fit are shown in Figure~\ref{fig:Eg2}. 

 \begin{table}[h!]
\centering
 \begin{tabular}{|c|c| c| c|c|} 
 \hline
  & Q$^2$(GeV$^2$) & $x$ & $y$ & E$_\gamma$(GeV) \\ [0.5ex] 
 \hline
 True (with ISR  correction)  & 2558 &  0.128  & 0.197 &0\\
 \hline
 KF (Global mode)  & 2601 &   0.140 +- 0.017 &  0.182 +- 0.018 &  0\\
 \hline
EL &2648 &  0.155& 0.168&- \\
 \hline
DA & 2606 & 0.142  &0.181 &- \\
 \hline
 \end{tabular}
\caption{Generated $Q^2$, $x$ and $y$ for the second example event. The values of $x$, $y$ and  E$_\gamma$ from the Kinematic Fit are quoted at the global mode with their respective standard deviation from the marginalized distribution.}
\label{tab:Eg2}
\end{table}

\begin{thebibliography}{1}
\bibitem{ref:LHeC}P. Agostini et al., The Large Hadron-Electron Collider at the HL-LHC (April 2021). arXiv:2007.14491v2, CERN-ACC-Note-2020-0002.
\bibitem{ref:FCC}A. Abeda et. al., FCC physics opportunities. Eur. Phys. J. C 79 (2019) 474.
\bibitem{ref:Vhep}A. Caldwell, M. Wing,  VHEeP: A very high energy electron–proton collider. Eur. Phys. J. C 76 (2016) 463. 
\bibitem{ref:eic}A. Accardi  et. al., Electron-Ion Collider: The next QCD frontier. Eur. Phys. J. A  52 (2016) 268.
\bibitem{ref:Awake}M. Wing, Particle physics experiments based on the AWAKE acceleration scheme, Philosophical transactions. Ser. A Math. Phys. Eng. Sci. 377 (2019)  20180185.

\bibitem{ref:DIS}A. Cooper-Sarkar, R. Devenish, Deep Inelastic Scattering (Oxford University Press, Oxford (2004). 
\bibitem{ref:bayes1}A. Gelman, J. B. Carlin, H. S. Stern, D. B. Dunson,A. Vehtari, D. B. Rubin, Bayesian Data Analysis, Third Edition, Chapman and Hall/CRC  (2013).
\bibitem{ref:JB}F. Jacquet and A. Blondel, Proceedings on Study of an ep Facility for Europe, DESY 79/48 (1979) 391, editor U. Amaldi, DESY Publ., (November 1992).
\bibitem{ref:KF1}H. Chaves, R.J. Seyfert, G. Zech, Proceedings of the Workshop Physics at HERA, vol. 1, eds. W. Buchmüller, G. Ingelman, DESY (1992) 57-70.
\bibitem{ref:Recon1}U. Bassler, G. Bernardi, 
On the kinematic reconstruction of deep inelastic scattering at HERA: The sigma method
Nucl. Instrum. Methods A, 361 (1995), pp. 197-208.
\bibitem{ref:highx}H. Abramowicz et al. (ZEUS Collaboration), Phys. Rev. D 89 (2014)  072007 .
\bibitem{ref:Recon2}U. Bassler, G. Bernardi,
Structure function measurements and kinematic reconstruction at HERA, Nucl. Instrum. Methods A, 426 (1999) pp. 583-598.
\bibitem{ref:el}K.C. Hoeger. Measurement of x, y and $Q^2$ in neutral current events, editors W. Buchmüller and G. Ingelman, Proceedings of Workshop on Physics at HERA, vol. 1, DESY (1992) pages 43-55.
\bibitem{ref:DA}S. Bentvelsen et al., Proceedings of the Workshop Physics at HERA, editors W. Buchmüller and G. Ingelman, vol. 1,  DESY (1992) pages 23-40.
\bibitem{ref:rapgap}RAPGAP hepforge site, https://rapgap.hepforge.org; H. Jung, Comp. Phys. Comm. 86 (1995) 147.
\bibitem{ref:Heraceles}A. Kwiatkowski, H. Spiesberger and H.-J. Möhring, Comp. Phys. Comm. 69, 155 (1992). Also
in Proc. Workshop Physics at HERA. eds. W. Buchmüller and G. Ingelman, DESY, Hamburg (1991).
\bibitem{ref:aggarwal}R.Aggarwal, PhD thesis submitted to the Dept. of Physics, Panjab University (2013).
\bibitem{ref:tuning}N. Tuning, Proton Structure Functions at HERA, Ph.D. Thesis, Amsterdam University (2001).
\bibitem{ref:bat}Bayesian Analysis Toolkit :http://www.mppmu.mpg.de/bat/.
\bibitem{ref:jbmk2}J. Blümlein, $O(\alpha^2L^2)$ radiative corrections to deep inelastic ep scattering for different kinematical variables, Z. Phys. C - Particles and Fields 65 (1995) 293.
\bibitem{ref:jbmk}J. Blümlein and M. Klein, On the cross calibration of calorimeters at ep colliders, Nucl. Instr. Meth. A 329 (1993) 112.
\bibitem{ref:NN1}M. Diefenthaler, A. Farhat, A. Verbytskyi, Y. Xu, Deeply Learning Deep Inelastic Scattering Kinematics (Aug 2021). arXiv:2108.11638.
\bibitem{ref:NN2}M. Arratia, D. Britzger, O. Long, B. Nachman, Reconstructing the Kinematics of Deep Inelastic Scattering with Deep Learning, Nuclear Inst. and Methods in Physics Research, A 1025 (2022) 166164.


 \end{thebibliography}
\end{document}